\documentclass[%
reprint,
amsmat,amssymb,
aps,
prb,
]{revtex4-1}

\usepackage{hyperref}
\usepackage{graphicx}
\usepackage{dcolumn}
\usepackage{bm}
\usepackage{amsmath}
\usepackage{natbib}
\usepackage{color}

\begin{document}
\title{Theory of surface Andreev bound states  and  
tunneling spectroscopy in three-dimensional chiral superconductors }

\author{Shun Tamura$^{1}$, Shingo Kobayashi$^{1,2}$, Lu Bo$^{3,4}$, 
Yukio Tanaka$^{1}$}
\affiliation{%
   $^1$Department of Applied Physics, Nagoya University, Nagoya 464-8603, Japan\\
   $^2$Institute for Advanced Research, Nagoya University, Nagoya 464-8601, Japan\\
   $^3$National Graphene Institute, University of Manchester, Manchester, M13 9PL, UK\\
   $^4$School of Physics and Astronomy, University of Manchester, Manchester, M13 9PL, UK
}

\begin{abstract}
We study the surface Andreev bound states (SABSs) and quasiparticle 
tunneling spectroscopy of 
three-dimensional (3D) chiral superconductors 
by changing their surface (interface) misorientation angles. 
We obtain an analytical formula for the SABS energy dispersion of a general 
pair potential, for which an original $4\times 4$ 
BdG Hamiltonian can be reduced to two $2 \times 2$ blocks. 
The resulting SABS for 3D chiral superconductors 
with a pair potential given by $k_{z}(k_{x} + i k_{y})^{\nu}$ ($\nu=1,2$) 
has a complicated energy dispersion owing to the coexistence of 
both point and line nodes. 
We focus on the tunneling spectroscopy of this pairing 
in the presence of an applied magnetic field, which induces 
a Doppler shift in the quasiparticle spectra. 
In contrast to the previously known Doppler effect in unconventional superconductors, 
a zero bias conductance dip can change into a zero bias conductance peak 
owing to an external magnetic field. 
We also study SABSs and tunneling spectroscopy for possible 
pairing symmetries of UPt$_{3}$.  
For this purpose, we extend a standard formula for the tunneling conductance of  
unconventional superconductor junctions to treat spin-triplet 
non-unitary pairings. 
Magneto tunneling spectroscopy, $i.e$., tunneling spectroscopy in the 
presence of a magnetic field, can serve as a guide to determine the 
pairing symmetry of this material. 

\end{abstract}
\pacs{pacs}

\maketitle
\thispagestyle{empty}

\section{Introduction}
The surface Andreev bound state (SABS) is one of the key concepts
regarding unconventional superconductors \cite{kashiwaya00,Lofwander,Golubov_RMP}. 
To date, various types of SABSs have been revealed in two-dimensional (2D)
unconventional superconductors \cite{tanaka12,qi11}. 
It is known that a flat band SABS 
exists in a spin-singlet $d$-wave superconductor \cite{Hu} that is 
protected by a topological invariant defined 
in the bulk Hamiltonian \cite{RH02,index,tanaka12,Sato2016}. 
The ubiquitous presence of this zero energy 
SABS manifests itself as a zero bias conductance peak (ZBCP) 
in the tunneling spectroscopy of high-$T_{c}$ cuprates 
\cite{TK95,Experiment1,Experiment2,Experiment3,Experiment4,Experiment5,Experiment6}. 
There has been interest in the spin-triplet $p$-wave superconductor\cite{ABS,ABSb}
with a flat band SABS and sharp ZBCP, similar to the $d$-wave case
\cite{YTK98,Tanaka2002,Yakovenko}.
Apart from flat bands, it is known that chiral $p$-wave superconductors host 
an SABS with a linear dispersion as a function of momentum parallel to the edge, 
resulting in a much broader ZBCP
\cite{YTK97,Honerkamp,Kashiwaya11,Yakovenko,Samokhin}.

Magneto tunneling spectroscopy, $i$.$e$., tunneling 
spectroscopy in the presence of an applied magnetic field, 
is a powerful tool to make distinctions among
pairing symmetries.
Under an applied magnetic field, the shift of quasiparticle energy spectra, 
which is proportional to the transverse momentum, 
is generally known as the Doppler effect\cite{Fogel,Experiment3}.
It has been shown that the splitting of the ZBCP 
occurs in $d$-wave superconductor 
junctions owing to this\cite{Fogel,Experiment3}. 
In contrast, for spin-triplet $p$-wave cases,
the ZBCP does not split into two\cite{Tanaka2002}
since the perpendicular injection of quasiparticles dominantly contributes 
to the tunneling conductance. 
For perpendicular injection, the component of
the Fermi velocity parallel to the interface is zero and 
there is no energy shift of the quasiparticles. 
Therefore, we can distinguish between $d$- and $p$-wave pairing with 
magneto tunneling spectroscopy\cite{Tanuma2002b}.
For chiral $p$-wave superconductors, the magnitude of the ZBCP is enhanced
or suppressed depending on the direction of the applied
magnetic field\cite{Tanaka2002}. 
Chiral and helical superconductors also exhibit different features of magneto tunneling spectroscopy,
whereas both of these superconductors have similarly broad 
ZBCP without magnetic field\cite{TYBN08}. 
In any case, a ZBCP generated from a zero bias conductance dip by applying 
magnetic field has not been found.
 
For three-dimensional (3D) unconventional 
superconductors, the energy dispersion of 
SABS becomes more complicated \cite{ABS,Qi09,Chung,Murakawa,Asano2003,fu10,sasaki11,hao11,hsieh12,yamakage12,Hashimoto1,LuBo,Hashimoto2}. 
Recently, the SABSs of 3D chiral superconductors 
have been studied \cite{Kobayashi2015}, 
where a pair potential is given by 
$\Delta_{0}k_{z}(k_{x}+ik_{y})^{\nu}/k_\mathrm{F}^{\nu+1}$ with a nonzero integer $\nu$.  
These pairing symmetries are relevant to typical 
heavy fermion superconductors 
with $\nu=1$ and $\nu=2$, corresponding to the 
candidate pairing symmetries of URu$_{2}$Si$_{2}$ 
\cite{Schemm2015,Kasahara,Shibauchi,Schemm2015PRB}
and UPt$_{3}$ 
\cite{Sauls,Joynt,Schemm2014,Goswami2015,Tsutsumi2012,Tsutsumi2013}
respectively.
The simultaneous presence of line and point nodes gives rise to exotic SABS\@.
It has been shown that 
the flat band SABS is found to be fragile against the 
surface misorientation angle 
$\alpha$, as shown in Fig.~\ref{fig:pic_junction}. 
Although the topological natures of the flat band 
SABS have been clarified \cite{Kobayashi2015}, 
the overall features of the energy dispersion of the SABS 
have not been systematically analyzed.
Thus, it is a challenging issue to 
identify 3D pairing states theoretically in terms of magneto tunneling 
spectroscopy.

In this work, we study the SABS 
and quasiparticle tunneling spectroscopy of 
3D chiral superconductors 
by changing the surface (interface) misorientation angle $\alpha$. 
For this purpose, we analytically derive a  
formula for the energy dispersion of SABSs available of a general 
pair potential, for which an original $4\times 4$ matrix of 
a Bogoliubov-de Gennes (BdG) Hamiltonian can be 
decomposed into two blocks of 
$2 \times 2$ matrices. 
We apply this formula to 3D chiral superconductors with a 
pair potential given by $\Delta_{0}k_{z}(k_{x} + i k_{y})^{\nu}/k_\mathrm{F}^{\nu+1}$ 
($\nu=1,2$). 
The resulting SABS has a complex momentum dependence 
due to the coexistence of point and line nodes. 
SABSs arising from topological and nontopological origins are found to coexist.
The number of branches of the energy dispersion of 
SABSs with topological origin can be classified 
by $\nu$ for various $\alpha$. 
On the other hand, if we apply our formula 
to 2D-like chiral superconductors 
with a pair potential given by 
$\Delta_{0}(k_{x} + i k_{y})^{\nu}/k_\mathrm{F}^{\nu}$ ($\nu=1,2$),  
the number of branches of SABSs is equal to 
$2\nu$, where the pair potential has only point nodes. 

In order to distinguish between two 3D chiral superconductors with different $\nu$, 
we calculate the tunneling conductance of 
normal metal / insulator / chiral superconductor junctions 
in the presence of an applied magnetic field, which induces a Doppler shift. 
The obtained angle-resolved conductance has a complicated 
momentum dependence reflecting on the dispersion of 
SABS for nonzero $\alpha$.
In contrast to previous studies of the Doppler effect on 
tunneling conductance, a zero bias conductance dip can evolve into 
a ZBCP by applying magnetic field. 
This unique feature stems from the complex 
nodal structures of the pair potential where both 
line and point nodes coexist. 
Furthermore, we focus on four possible candidates of the 
pairing symmetry of UPt$_{3}$, where the momentum dependences of the 
pair potentials are proportional to 
$k_{z}(k_{x}+ik_{y})^{2}d_{z}$, 
$(5k_{z}^{2}-k_\mathrm{F}^2)(k_{x}+ik_{y})d_{z}$, 
$(5k_{z}^{2}-k_\mathrm{F}^2)
(d_{y}k_{x} +d_{z}k_{y})$, 
and $f+p$-wave belonging to the $E_{2u}$ representation. 
Here, we derive a general formula for tunneling conductance, 
which is available even for non-unitary spin-triplet 
superconductors. 
We show that these four pairings can be classified by 
using magneto tunneling spectroscopy. 
Thus, our theory serves as a guide to determine the 
pairing symmetry of UPt$_{3}$. 

The remainder of this paper is organized as follows. 
In section~\ref{sec:MM}, we explain the model and formulation. 
We analytically derive a  
formula for the energy dispersion of SABSs available for a general 
pair potential for which an original $4\times 4$ matrix of 
BdG Hamiltonian is decomposed into two blocks of 
$2 \times 2$ matrices. 
We also derive a general conductance formula, 
available even for non-unitary spin-triplet pairing cases. 
Besides these, to understand the topological origin of SABS, 
we calculate winding number. 
In section~\ref{sec:H0}, based on above formula, 
we calculate the SABS and tunneling conductance   
for 3D chiral superconductors for various $\alpha$. 
As a reference, we also calculate the SABS for 2D-like chiral 
superconductors.  
In section~\ref{sec:H}, we calculate the tunneling conductance in the presence of 
an external magnetic field using so-called magneto tunneling spectroscopy. 
In section~\ref{sec:UPt3}, we study the SABS and tunneling conductance for  
promising pairing symmetries of UPt$_{3}$. 
In section~\ref{sec:conclusion}, we summarize our results. 

\section{\label{sec:MM}Model and Method}
\begin{figure}[htbp]
   \centering
   \includegraphics[width=6.5cm,bb=0 0 1432 1009]{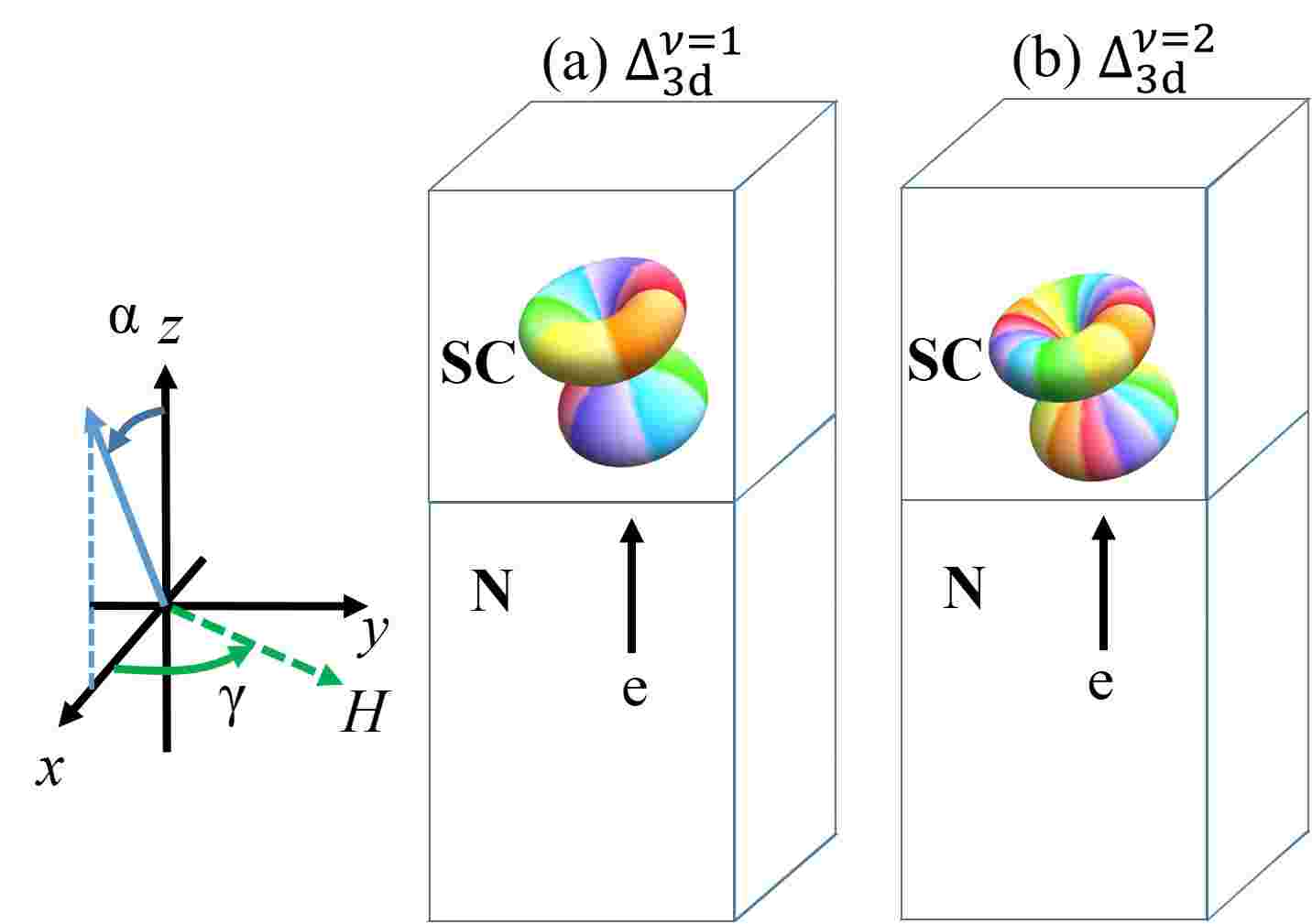}
   \caption{%
      We mainly consider above two types of normal (N) 
      --superconductor (SC) junctions. 
      Here, $\alpha$ is the misorientation angle from the $k_{z}$-axis.
      %
      Magnetic field is applied in the $x-y$ plane, rotated from the $x$-axis 
      by $\gamma$.
   }
   \label{fig:pic_junction}
\end{figure}

In this section, we introduce the mean field Hamiltonian of 
3D chiral superconductors. 
We derive an analytical formula for SABS for which the 
original 4 $\times$ 4 BdG Hamiltonian can be reduced to be 
two blocks of 2 $\times$ 2 matrices. 
We calculate the tunneling conductance in the presence of 
an external applied magnetic field.
In order to study the case of non-unitary spin-triplet pairings, 
we derive an analytical formula for the tunneling conductance.    
The bulk BdG Hamiltonian is given as follows,
\begin{align}
   {\cal H}
   =&
   \frac{1}{2}
   \sum_\mathbf{k} \Psi^\dag(\mathbf{k})
   H(\mathbf{k})
   \Psi(\mathbf{k}),
   \nonumber
   \\
   H(\mathbf{k})
   =&
   \begin{pmatrix}
      \hat{\varepsilon}(\mathbf{k}) & \Delta(\mathbf{k})\\
      \Delta^\dag(\mathbf{k}) & -\hat{\varepsilon}(-\mathbf{k})
   \end{pmatrix},
   \label{eq:BdG_H}
   \\
   \Psi(\mathbf{k})
   =&
   (%
      c_{\mathbf{k},\uparrow},c_{\mathbf{k},\downarrow},
      c_{-\mathbf{k},\uparrow}^\dag,c_{-\mathbf{k},\downarrow}^\dag
	)^\mathrm{T},
   \nonumber
\end{align}
where 
$\hat{\varepsilon}(\mathbf{k})
=\mathrm{diag}[\varepsilon(\mathbf{k}),\varepsilon(\mathbf{k})]$ 
and $\Delta(\mathbf{k})$ are $2 \times 2$ matrices.
Here, $\varepsilon(\mathbf{k})$ denotes the energy dispersion  
$\hbar^2\mathbf{k}^2/(2m)-\mu$ in the normal state.
For spin-triplet pairing, the pair potential is given by 
\begin{align}
	\Delta=\mathbf{d}\cdot\boldsymbol{\sigma}i\sigma_2,
   \nonumber
\end{align}
by using a $\mathbf{d}$-vector, 
where $(\sigma_1,\sigma_2,\sigma_3)=\boldsymbol{\sigma}$ 
are the Pauli matrices.
We consider normal metal ($z<0$)/superconductor 
($z>0$) junctions with a flat interface, as shown in 
Fig.~\ref{fig:pic_junction}. 
The momentum parallel to the interface  
$\mathbf{k}_{\parallel}=(k_x,k_y)$ becomes a good quantum number.

In the following, we explain 
eight types of pair potentials and corresponding formulae for tunneling 
conductance and SABS.
In Sec.~\ref{sec:22},
we explain the case for which the BdG Hamiltonian can be reduced 
to a $2\times2$ form [$\Delta_\mathrm{3d}^{\nu=0,1,2}$,
$\Delta_\mathrm{2d}^{\nu=1,2}$, and $\Delta_{E_{1u}}^\mathrm{chiral}$].  
In Sec.~\ref{sec:44}, we explain the case for which the $4\times4$ BdG Hamiltonian 
cannot be reduced to two $2 \times 2$ blocks 
[$\Delta_{E_{1u}}^\mathrm{planar}$ and $\Delta_{E_{2u}}^{f+p}$].
In Sec.~\ref{sec:topological_n}, we briefly summarize the zero-energy SABS (ZESABS)
stemming from topological numbers.

\subsection{\label{sec:22}$2\times2$ BdG Hamiltonian}
\begin{figure}[htbp]
   \centering
   \includegraphics[width=8.5cm,bb=0 0 1886 507]{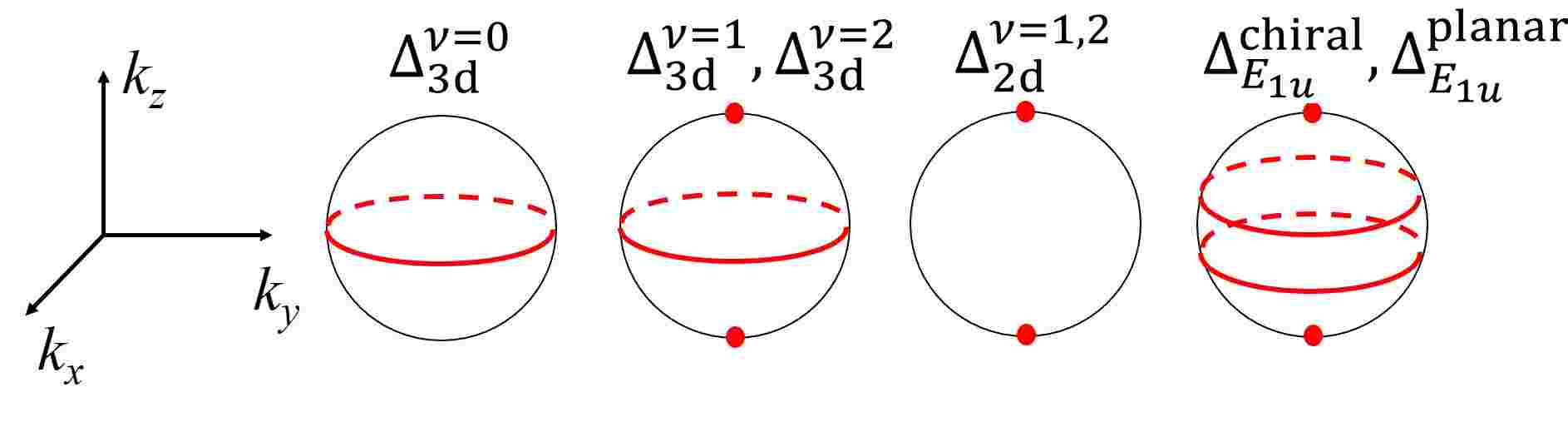}
   \caption{%
   Schematic illustration of nodal structures of $\Delta(\mathbf{k})$ with
   $\alpha=0$. 
   Red points and lines indicate the positions of nodes.
   }
   \label{fig:nodes}
\end{figure}

In this subsection, we introduce pair potentials for
3D and 2D-like chiral superconductors. 
We derive a formula for the SABS    
for which the BdG Hamiltonian can be reduced to two $2\times2$ matrices.
$\Delta(\mathbf{k})$ takes
\begin{align}
	\Delta_{3\mathrm{d}}^\nu(\mathbf{k})
	=&
   \left\{
      \begin{aligned}
         & \tilde{\Delta}_\mathrm{3d}^\nu i\sigma_2
         & \nu:\mathrm{odd},
         \\
         & \tilde{\Delta}_\mathrm{3d}^\nu \sigma_3i\sigma_2
         & \nu:\mathrm{even},
         \\
      \end{aligned}
      \label{eq:Delta_3d_nu}
   \right.\\
	 \Delta_{2\mathrm{d}}^\nu(\mathbf{k})
	 =&
   \left\{
      \begin{aligned}
         &\tilde{\Delta}_\mathrm{2d}^\nu\sigma_3 i\sigma_2
         & \nu:\mathrm{odd},\\
         &\tilde{\Delta}_\mathrm{2d}^\nu i\sigma_2
         & \nu:\mathrm{even},
      \end{aligned}
      \label{eq:Delta_2d_nu}
   \right.\\
   \Delta_{E_{1u}}^\mathrm{chiral}(\mathbf{k})
	 =&
    \tilde{\Delta}_{E_{1u}}^\mathrm{chiral}\sigma_3i\sigma_2,
      \label{eq:Delta_E1u_c}
   \\
   \tilde{\Delta}_\mathrm{3d}^\nu
   =&
   \frac{\Delta_0}{r_{\mathrm{3d},\nu}k_\mathrm{F}^{\nu+1}}
   k'_z(k'_x+ik'_y)^\nu,
   \nonumber
   \\
   \tilde{\Delta}_\mathrm{2d}^\nu
   =&
   \frac{\Delta_0}{k_\mathrm{F}^{\nu}}
   (k'_x+ik'_y)^\nu,
   \nonumber
   \\
   \tilde{\Delta}_{E_{1u}}^\mathrm{chiral}
   =&
   \frac{\Delta_0}{r_{E_{1u}}k_\mathrm{F}^3}
   \left(
      {5k'_z}^2 - k_\mathrm{F}^2
   \right)
   (k'_x+ik'_y),
   \nonumber
   \\
   k'_x=&k_x\cos\alpha-k_z\sin\alpha,
   \nonumber
   \\
   k'_y=&k_y,
   \nonumber
   \\
   k'_z=&k_x\sin\alpha+k_z\cos\alpha,
   \nonumber
\end{align}
where $\alpha$ is the misorientation angle from the $k_z$ axis and
$(r_{\mathrm{3d},\nu=0}, r_{\mathrm{3d},\nu=1}, r_{\mathrm{3d},\nu=2}, r_{E_{1u}})
=(1,\:1/2,\:2/\sqrt{27},\:16/(3\sqrt{15}))$ 
are the normalization factors so that the maximum value of the
pair potential becomes $\Delta_0$.
Because the direction of the $\mathbf{d}$-vector in spin-space does not 
affect conductance, $i.e$., conductance is invariant under spin rotation 
(Appendix~\ref{sec:App_cond4}), 
we fix the direction of the $\mathbf{d}$-vector given in Eqs.~(\ref{eq:Delta_3d_nu}),
(\ref{eq:Delta_2d_nu}), and (\ref{eq:Delta_E1u_c})
for the spin-triplet cases.
$\Delta_\mathrm{3d}^\nu$ and $\Delta_\mathrm{2d}^{\nu}$ are chosen in
Sec.~\ref{sec:H0} and Sec:~\ref{sec:H}.
We study $\Delta_\mathrm{3d}^{\nu=2}$ and $\Delta_{E_{1u}}^\mathrm{chiral}$
in Sec.~\ref{sec:UPt3}.
Under the quasiclassical approximation, the magnitude of the wave vector is pinned 
to the value on the Fermi surface, 
$k_z=\sqrt{k_\mathrm{F}^2-k_x^2-k_y^2}$.
As shown in Fig.~\ref{fig:nodes},
$\Delta_{3\mathrm{d}}^{\nu=0}$ has a line node,
$\Delta_{3\mathrm{d}}^\nu$($\nu=1,2$) has two point nodes and one line node, and
$\Delta_{2\mathrm{d}}^\nu$ ($\nu>0$) has two point nodes.
$\Delta_{E_{1u}}^\mathrm{chiral}$ has two point nodes and two line nodes.
Then, we show that the BdG Hamiltonian can be reduced to a 
$2\times2$ form for all the cases.
For the spin-singlet cases ($\Delta_\mathrm{3d}^{\nu=1}$ and 
$\Delta_\mathrm{2d}^{\nu=2}$), Eq.~(\ref{eq:BdG_H}) is reduced to
\begin{align}
	{\cal H}(\mathbf{k})
	=&
   \frac{1}{2}
	\sum_\mathbf{k}
	\Psi^\dag(\mathbf{k})
	\begin{pmatrix}
		\varepsilon(\mathbf{k}) & 0 & 0 & D \\	
		0 & \varepsilon(\mathbf{k}) & -D & 0 \\
		0 & -D^* & -\varepsilon(\mathbf{k}) & 0 \\
		D^* & 0 & 0 & -\varepsilon(\mathbf{k})
	\end{pmatrix}
	\Psi(\mathbf{k})
	\nonumber\\
	=&
   \frac{1}{2}
	\sum_\mathbf{k}
	(c_{\mathbf{k},\uparrow}^\dag\:c_{-\mathbf{k},\downarrow})
	\begin{pmatrix}
		\varepsilon(\mathbf{k}) & D \\	
		D^* & -\varepsilon(\mathbf{k})
	\end{pmatrix}
	\begin{pmatrix}
	c_{\mathbf{k},\uparrow}\\
		c_{-\mathbf{k}\downarrow}^\dag
	\end{pmatrix}
	\nonumber\\
	&+
   \frac{1}{2}
	\sum_\mathbf{k}
	(c_{\mathbf{k},\downarrow}^\dag\:c_{-\mathbf{k},\uparrow})
	\begin{pmatrix}
		\varepsilon(\mathbf{k}) & -D \\	
		-D^* & -\varepsilon(\mathbf{k})
	\end{pmatrix}
	\begin{pmatrix}
	c_{\mathbf{k},\downarrow}\\
		c_{-\mathbf{k}\uparrow}^\dag
	\end{pmatrix},
   \nonumber
\end{align}
with $Di\sigma_2=\Delta_\mathrm{3d}^{\nu=1}$ or 
$\Delta_\mathrm{2d}^{\nu=2}$.
For the spin-triplet cases ($\Delta_\mathrm{3d}^{\nu=0,2}$, 
$\Delta_\mathrm{2d}^{\nu=1}$, and $\Delta_{E_{1u}}^\mathrm{chiral}$), 
with a $\mathbf{d}$-vector, Eq.~(\ref{eq:BdG_H}) becomes
\begin{align}
	{\cal H}(\mathbf{k})
	=&
   \frac{1}{2}
	\sum_\mathbf{k}
	\Psi^\dag(\mathbf{k})
	\begin{pmatrix}
		\varepsilon(\mathbf{k}) & 0 & 0 & d_3 \\	
		0 & \varepsilon(\mathbf{k}) & d_3 & 0 \\
		0 & d_3^* & -\varepsilon(\mathbf{k}) & 0 \\
		d_3^* & 0 & 0 & -\varepsilon(\mathbf{k})
	\end{pmatrix}
	\Psi(\mathbf{k})
	\nonumber\\
	=&
   \frac{1}{2}
	\sum_\mathbf{k}
	(c_{\mathbf{k},\uparrow}^\dag\:c_{-\mathbf{k},\downarrow})
	\begin{pmatrix}
		\varepsilon(\mathbf{k}) & d_3 \\	
		d_3^* & -\varepsilon(\mathbf{k})
	\end{pmatrix}
	\begin{pmatrix}
	c_{\mathbf{k},\uparrow}\\
		c_{-\mathbf{k}\downarrow}^\dag
	\end{pmatrix}
	\nonumber\\
	&+
   \frac{1}{2}
	\sum_\mathbf{k}
	(c_{\mathbf{k},\downarrow}^\dag\:c_{-\mathbf{k},\uparrow})
	\begin{pmatrix}
		\varepsilon(\mathbf{k}) & d_3 \\	
		d_3^* & -\varepsilon(\mathbf{k})
	\end{pmatrix}
	\begin{pmatrix}
	c_{\mathbf{k},\downarrow}\\
		c_{-\mathbf{k}\uparrow}^\dag
	\end{pmatrix},
   \nonumber
\end{align}
with $(d_3\sigma_3)(i\sigma_2)=\Delta_\mathrm{3d}^{\nu=2}$,
$\Delta_\mathrm{2d}^{\nu=1}$, or $\Delta_{E_{1u}}^\mathrm{chiral}$.

In the following, we calculate 
the charge conductance 
by using the extended version \cite{TK95,kashiwaya00,KT96} 
of the Blonder--Tinkham--Klapwijk (BTK) formula \cite{BTK} 
for unconventional superconductors \cite{Bruder90}. 

Since we assume that the penetration depth of 
the magnetic field $\lambda$ is much larger
than the coherence length of the pair potential 
\cite{Fogel,Tanuma2002a}, 
we can neglect the spatial dependence of the magnetic field.
Therefore, we can take the vector potential as
\begin{align}
   \mathbf{A}=(-H\lambda\sin\gamma,H\lambda\cos\gamma,0).
   \label{eq:vector_potential}
\end{align}

Solving the BdG equation with the quasiclassical approximation, 
where $\mu$ is much larger than the energy of an injected electron
$|eV|$ and $|\Delta(\mathbf{k}_\parallel)|$, 
the wave function in the normal metal (N) and 
superconductor (S) are obtained as
\begin{align}
   \Psi_\sigma^\mathrm{N}(z<0,\mathbf{k}_\parallel)
   =&\psi_{e,\sigma}^\mathrm{N}e^{i\mathbf{k}\cdot\mathbf{r}}
   \nonumber\\
   &+\sum_{\sigma'=\pm}
   \left(
      a_{\sigma,\sigma'}\psi_{h,\sigma'}^\mathrm{N}e^{i\mathbf{k}\cdot\mathbf{r}}
      +
      b_{\sigma,\sigma'}\psi_{e,\sigma'}^\mathrm{N}e^{i\tilde{\mathbf{k}}\cdot\mathbf{r}}
   \right),
   \label{eq:Psi_N}
   \\
   \Psi^\mathrm{S}(z>0,\mathbf{k}_\parallel)
   =&\sum_{\sigma'=\pm}
   \left(
      c_{\sigma'}\psi_{e,\sigma'}^\mathrm{S}e^{i\mathbf{k}\cdot\mathbf{r}}
      +
      d_{\sigma'}\psi_{h,\sigma'}^\mathrm{S}e^{i\tilde{\mathbf{k}}\cdot\mathbf{r}}
   \right),
   \label{eq:Psi_S}
\end{align}
and
\begin{align}
   \psi_{e,\sigma}^\mathrm{N}=&(1+\sigma,1-\sigma,0,0)^\mathrm{T}/2,
	 \label{eq:WF_N1}
	 \\
   \psi_{h,\sigma}^\mathrm{N}=&(0,0,1+\sigma,1-\sigma)^\mathrm{T}/2,
	 \label{eq:WF_N2}
	 \\
   \psi_{e,\sigma}^\mathrm{S}
	 =&
	 [1+\sigma,1-\sigma,(1-\sigma)\rho\Gamma_+,
	 (1+\sigma)\Gamma_+]^\mathrm{T}/2,
    \nonumber
	 \\
   \psi_{h,\sigma}^\mathrm{S}
	 =&
	 [(1+\sigma)\Gamma_-,(1-\sigma)\rho\Gamma_-,
	 1-\sigma,1+\sigma]^\mathrm{T}/2,
    \nonumber
	 \\
   \Gamma_+
	 =&
	 \frac{\Delta^*(\mathbf{k})}
	 {\tilde{E}+\sqrt{\tilde{E}^2-|\Delta(\mathbf{k})|^2}},
   \label{eq:Gamma_p}
   \\
   \Gamma_-
	 =&
	 \frac{\Delta(\tilde{\mathbf{k}})}
	 {\tilde{E}+\sqrt{\tilde{E}^2-|\Delta(\tilde{\mathbf{k}})|^2}},
   \label{eq:Gamma_m}
   \\
   \tilde{E}
   =&
   eV-\frac{H}{H_0}\Delta_0
   \left(
      \frac{k_y}{k_\mathrm{F}}\cos\gamma
      -
      \frac{k_x}{k_\mathrm{F}}\sin\gamma
   \right),
   \label{eq:Doppler}
\end{align}
with $\tilde{\mathbf{k}}=(k_x,k_y,-k_z)$ and
$H_0=\Delta_0/(e\lambda v_\mathrm{F})$.
$\Delta(\mathbf{k})=\tilde{\Delta}_\mathrm{3d}^\nu(\mathbf{k})$, 
$\tilde{\Delta}_\mathrm{2d}^\nu(\mathbf{k})$, or 
$\tilde{\Delta}_{E_{1u}}^\mathrm{chiral}(\mathbf{k})$.
Here, $\rho=-1$ for $\Delta_{3\mathrm{d}}^{\nu=1}$ and 
$\Delta_{2\mathrm{d}}^{\nu=2}$ (even parity) and 
$\rho=1$ for $\Delta_{3\mathrm{d}}^{\nu=0,2}$,
$\Delta_{2\mathrm{d}}^{\nu=1}$,
and $\Delta_{E_{1u}}^\mathrm{chiral}$
(odd parity). 
The coefficients ($a_{\sigma,\sigma'},b_{\sigma,\sigma'},c_{\sigma},d_{\sigma}$)
are determined by the boundary conditions:
\begin{align}
   \Psi_\sigma^\mathrm{N}(0_-,\mathbf{k}_\parallel)
   =&
   \Psi^\mathrm{S}(0_+,\mathbf{k}_\parallel),
   \label{eq:bound1}
   \\
   \left.
      \frac{d\Psi^\mathrm{S}}{dz}
   \right|_{z=0_+}
   -
   \left.
   \frac{d\Psi^\mathrm{N}_\sigma}{dz}
   \right|_{z=0_-}
   =&\frac{2mU_0}{\hbar^2}\Psi^\mathrm{N}_\sigma(0_-,\mathbf{k}_\parallel),
   \label{eq:bound2}
\end{align}
where the insulating barrier at $z=0$ is simplified as $V(z)=U_0\delta(z)$.
The angle-resolved conductance is given by \cite{TK95,KT96}
\begin{align}
   \sigma_\mathrm{S}(eV,\mathbf{k}_\parallel)
   =&1+\frac{1}{2}\sum_{\sigma,\sigma'=\pm}
   \left[
      |a_{\sigma,\sigma'}|^2-|b_{\sigma,\sigma'}|^2
   \right]
   \nonumber\\
   =&\sigma_\mathrm{N}\frac{1+\sigma_\mathrm{N}|\Gamma_+|^2
   +(\sigma_\mathrm{N}-1)|\Gamma_+\Gamma_-|^2}
   {|1+(\sigma_\mathrm{N}-1)\Gamma_+\Gamma_-|^2},
   \label{eq:conductance}\\
   \sigma_\mathrm{N}(\mathbf{k}_\parallel)=&\frac{4\cos^2\theta}{4\cos^2\theta+Z^2},
\end{align}
with $\cos\theta=k_z/k_\mathrm{F}$ 
and $Z=2mU_0/(\hbar^2k_\mathrm{F})$.

      In the procedure of obtaining conductance with magnetic field, 
      we neglect Zeeman effect.
      For UPt$_3$, the order of $\lambda \sim 10^4$\AA\cite{Broholm90} $k_\mathrm{F}\sim1/$\AA
      \cite{Marabelli86}.
      Here the order of the 
      energy of Doppler shift is $H \Delta/H_{0}$ with
      $H_{0}=h/(2e\pi^{2}\xi \lambda)$
      and $\xi=\hbar^{2} k_\mathrm{F}/(\pi m \Delta)$ \cite{Fogel}. 
      Since the Zeeman energy is given by 
      $\mu_{B}H$, the ratio of the energy of Doppler shift to Zeeman effect is
      $2\lambda k_\mathrm{F}\sim10^4$
      times larger than that of Zeeman energy for UPt$_3$.  
      Thus, neglecting Zeeman effect is a good approximation in present case.

In Sec.~\ref{sec:H0}, we discuss SABSs with $H=0$, which is determined by
requiring the condition that 
the denominator of Eq.~(\ref{eq:conductance}) 
is zero for $Z\rightarrow\infty$ ($\sigma_\mathrm{N}\rightarrow0$).
Then, at the energy dispersion of the SABS, 
the denominator of Eq.~(\ref{eq:conductance}) must satisfy
following conditions:
\begin{align}
   \mathrm{Re}(\Gamma_+\Gamma_-)=&1,
   \label{eq:Re1}\\ 
   \mathrm{Im}(\Gamma_+\Gamma_-)=&0.
   \label{eq:Im0}
\end{align}
In this case, $\sigma_\mathrm{S}$ becomes two, which is the maximum value of
angle-resolved conductance owing to the perfect resonance.
We define $E(\mathbf{k}_\parallel)=\tilde{E}$, which satisfies 
Eq.~(\ref{eq:Re1}) and Eq.~(\ref{eq:Im0}).

Here, we derive a general formula of 
SABS for pair potentials with arbitrary momentum dependence
(details are explained in Appendix~\ref{sec:App_SABS}).
There are two cases. 
For $\mathrm{Im}\left[\Delta^*(\mathbf{k})\Delta(\tilde{\mathbf{k}})\right]\neq0$,
the energy dispersion of SABS $E(\mathbf{k}_\parallel)$ is given by 
\begin{align}
   E(\mathbf{k}_\parallel)&=
   \frac{\mathrm{Im}\left[\Delta^*(\mathbf{k})\Delta(\tilde{\mathbf{k}})\right]}
   {\left|\Delta(\mathbf{k}) - \Delta(\tilde{\mathbf{k}})\right|}
   \label{eq:SABS1}
\end{align}
where 
$\Delta(\mathbf{k})$ and $\Delta(\tilde{\mathbf{k}})$ 
must satisfy
\begin{align}
   &
   \left\{
      \left|\Delta(\mathbf{k})\right|^2
      -
      \mathrm{Re}
      \left[
         \Delta^*(\mathbf{k})\Delta(\tilde{\mathbf{k}})
      \right]
   \right\}
   \nonumber\\
   &
   \times
   \left\{
      \left|\Delta(\tilde{\mathbf{k}})\right|^2
      -
      \mathrm{Re}
      \left[
         \Delta^*(\mathbf{k})\Delta(\tilde{\mathbf{k}})
      \right]
   \right\}
   \geq0.
   \label{eq:det_SABS}
\end{align}
For $\mathrm{Im}\left[\Delta^*(\mathbf{k})\Delta(\tilde{\mathbf{k}})\right]=0$,
\begin{align}
   E(\mathbf{k}_\parallel)
   =&0,
   \label{eq:SABS2}
   \\
   \mathrm{with}\:
   \frac{\Delta(\mathbf{k})}{|\Delta(\mathbf{k})|}
   =&-
   \frac{\Delta(\tilde{\mathbf{k}})}{|\Delta(\tilde{\mathbf{k}})|}.
   \label{eq:det_SABS2}
\end{align}
This formula reproduces all of the known results of SABSs 
in 2D unconventional superconductors, 
such as $d$-wave \cite{Hu,kashiwaya00}, $p$-wave \cite{ABSb,Yakovenko}, 
$d+is$-wave \cite{Matsumoto95,Kashiwayadis}, 
chiral $p$-wave \cite{FMS01}, and chiral $d$-wave. \par

In Sec.~\ref{sec:H} and Sec.~\ref{sec:UPt3}, 
we discuss the normalized conductance given by 
\begin{align}
   \sigma(eV)
	 =
	 \frac{\int_{|k_x^2+k_y^2|<k_\mathrm{F}} dk_xdk_y \sigma_\mathrm{S}(eV,k_x,k_y)}
   {\int_{|k_x^2+k_y^2|<k_\mathrm{F}} dk_xdk_y \sigma_\mathrm{N}(k_x,k_y)},
	 \label{eq:cond_norm}
\end{align}
which can be measured experimentally \cite{kashiwaya00}.
\subsection{\label{sec:44}
$4\times4$ BdG Hamiltonian}
In this subsection, we explain the case in which the BdG Hamiltonian is 
in the $4 \times 4$ form.
In Sec.~\ref{sec:UPt3}, in addition to $\Delta_\mathrm{3d}^{\nu=2}$
and 
$\Delta_{E_{1u}}^\mathrm{chiral}$, we choose 
$\Delta_{E_{1u}}^\mathrm{planar}$
(Fig.~\ref{fig:nodes})
and $\Delta_{E_{2u}}^{f+p}$.
$\Delta_{E_{1u}}^\mathrm{planar}$ is a spin-triplet pair potential in  
a unitary state 
$i.e$.\ $\mathbf{q}=i\mathbf{d}\times\mathbf{d}^*=0$,
while $\Delta_{E_{2u}}^{f+p}$ can incorporate non-unitary 
case ($\mathbf{q}\neq0$).
$\Delta_{E_{1u}}^\mathrm{planar}$ is given by 
\begin{align}
   \Delta_{E_{1u}}^\mathrm{planar}(\mathbf{k})=&
   \frac{\Delta_0}{r_{E_{1u}}k_\mathrm{F}^3}
   \left(
      {5k'_z}^2-k_\mathrm{F}^2
   \right)
   (ik'_x\sigma_0+k'_y\sigma_1).
\end{align}
It is noted that the time reversal
symmetry is not broken for $\Delta_{E_{1u}}^\mathrm{planar}$ 
but $\Delta_{E_{1u}}^\mathrm{chiral}$ does not have time reversal symmetry.

\begin{figure}[htbp]
   \centering
   \includegraphics[width=8.5cm,bb=0 0 360 273]{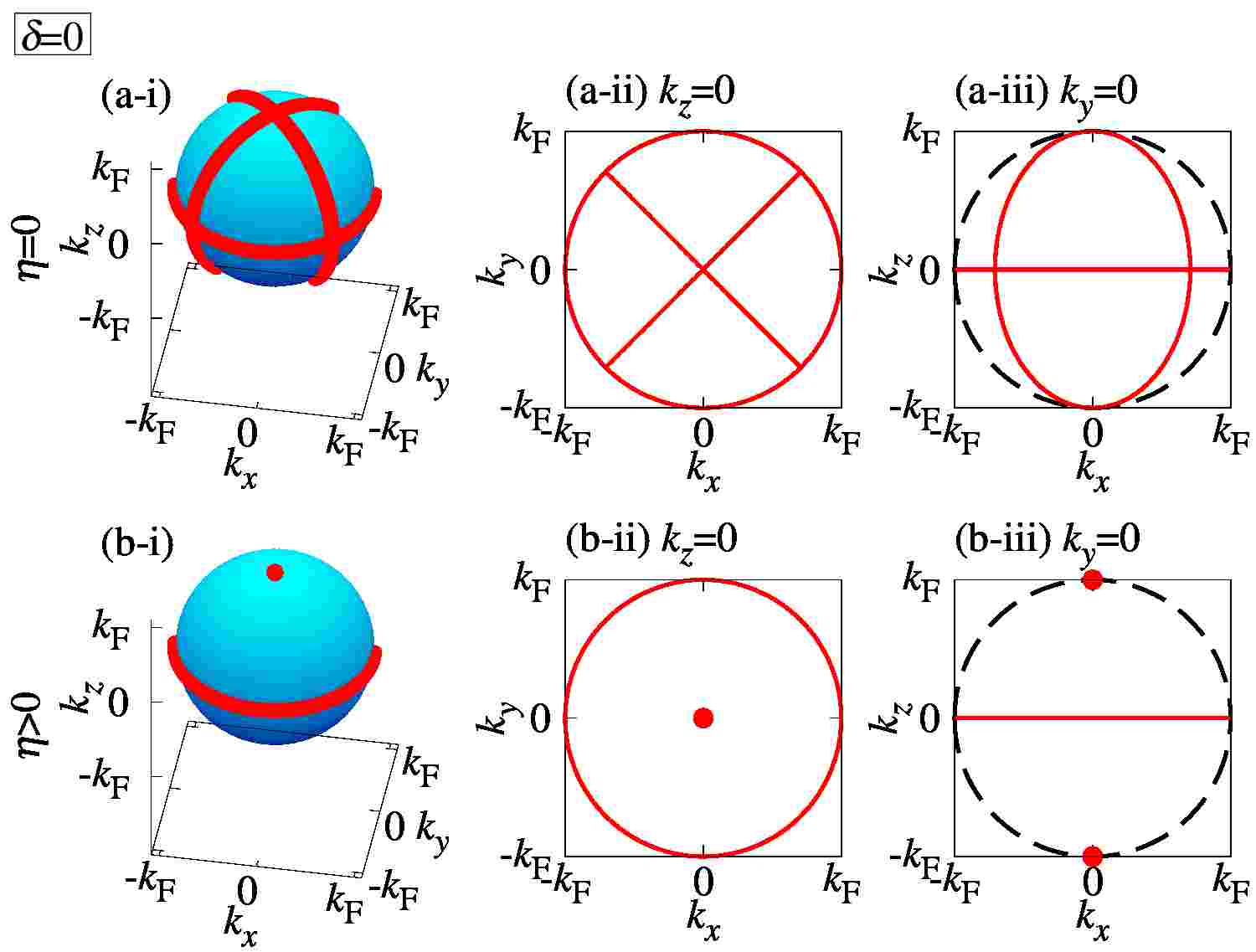}
   \caption{%
      Red lines and dots indicate the positions of nodes of the 
      pair potentials $\Delta_{E_{2u}}^{f+p}$ for $\delta=0$.
      $\eta=0$ for (a-i)--(a-iii).
      $\eta>0$ for (b-i)--(b-iii).
      (a-ii) and (b-ii) are nodes projected on the $k_z=0$ plane.
      (a-iii) and (b-iii) are nodes projected on the $k_y=0$ plane.
      Dashed lines in (a-iii) and (b-iii) indicate projected Fermi surface on $k_y=0$ plane.
   }
   \label{fig:nodes_f_p_d0}
\end{figure}

\begin{figure}[htbp]
   \centering
   \includegraphics[width=8.5cm,bb=0 0 360 396]{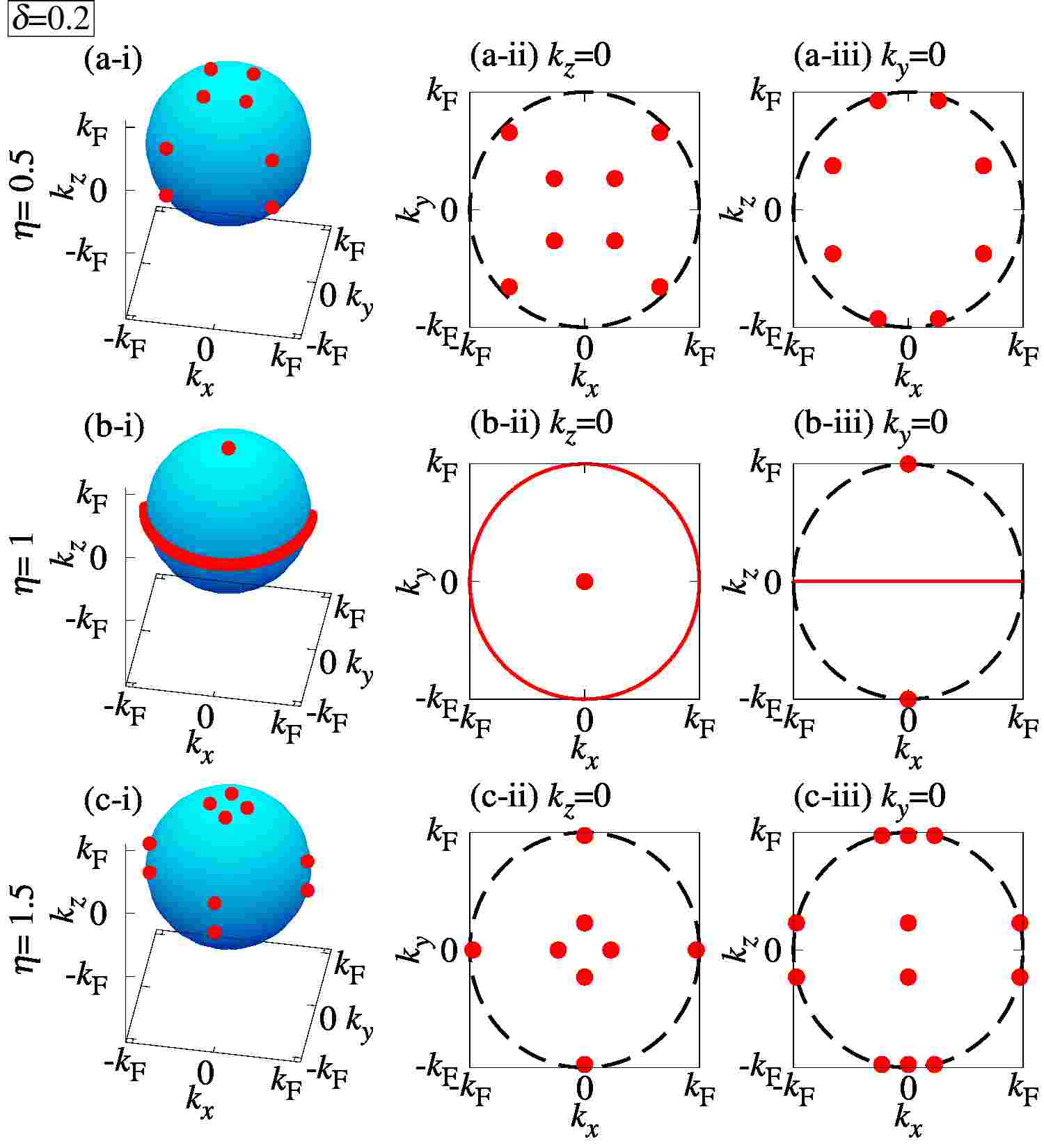}
   \caption{%
      Red lines and dots indicate the positions of nodes of the 
      pair potential $\Delta_{E_{2u}}^{f+p}$ for $\delta=0.2$.
      $\eta=0.5$ for (a-i)--(a-iii).
      $\eta=1$ for (b-i)--(b-iii).
      $\eta=1.5$ for (c-i)--(c-iii).
      (a-ii), (b-ii), and (c-ii) are nodes projected on the $k_z=0$ plane.
      (a-iii), (b-iii), and (c-iii) are nodes projected on the $k_y=0$ plane.
      Dashed lines in (a-ii), (a-iii), (b-iii), (c-ii), and (c-iii) indicate the 
      projected Fermi surface on the $k_z=0$ or $k_y=0$ plane.
   }
   \label{fig:nodes_f_p}
\end{figure}

$\Delta_{E_{2u}}^{f+p}$ is the combination of 
chiral $p$-wave and $f$-wave pairings. 
The $\mathbf{d}$-vector of $\Delta_{E_{2u}}^{f+p}$ is given by 
\begin{align}
   \mathbf{d}
   =&
   \frac{\Delta_0}{r^{f+p}}
   \left\{
      \delta\frac{1}{k_\mathrm{F}}
      \left[
         \left(
            k'_x+i\eta k'_y
         \right)
         d_x
         +i
         \left(
            \eta k'_x+ik'_y
         \right)
         d_y
      \right]
   \right.
   \nonumber\\
   &+
   \left.
      \frac{1}{k_\mathrm{F}^3}
      \left[
         k'_z({k'_x}^2-{k'_y}^2)
         +
         2i\eta k'_zk'_xk'_y
      \right]
      d_z
   \right\},
   \label{eq:d_nonu}
\end{align}
where $\delta$ is considered to be small \cite{Yanase2016} 
and $r^{f+p}$ is a normalized factor 
that is determined numerically so that the maximum value of the pair potential  
becomes $\Delta_0$.
If $(\eta,\delta)=(1,0)$ is satisfied, 
we obtain $\Delta_{E_{2u}}^{f+p}=\Delta_\mathrm{3d}^{\nu=2}$.
The position of nodes of this $f$+$p$-wave pairing 
depends on the values of $\eta$ and $\delta$.
If $\delta=0$ is satisfied (Fig.~\ref{fig:nodes_f_p_d0}), there are two cases.
In the case of $\eta=0$ [Fig.~\ref{fig:nodes_f_p_d0} (a-i)--(a-iii)], 
there are three line nodes.
For $\eta>0$ [Fig.~\ref{fig:nodes_f_p_d0} (b-i)--(b-iii)], 
there is a line node and two point nodes.
For $\delta>0$ (Fig.~\ref{fig:nodes_f_p}),
there are three cases.
For $\eta<1$ [Fig.~\ref{fig:nodes_f_p} (a-i)--(a-iii)],
there are 16 point nodes on $k_y=\pm k_x$ lines 
[Fig.~\ref{fig:nodes_f_p} (a-ii)].
In the case of $\eta=1$ [Fig.~\ref{fig:nodes_f_p} (b-i)--(b-iii)],
the positions of nodes are the same as that in 
the case of $\eta>0$ and $\delta=0$
[Fig.~\ref{fig:nodes_f_p_d0} (b-i)--(b-iii)].
For $\eta>1$ [Fig.~\ref{fig:nodes_f_p} (c-i)--(c-iii)],
there are 16 point nodes on lines $k_xk_y=0$, as shown in 
Fig.~\ref{fig:nodes_f_p} (c-ii).

The wave function for the normal metal side is shown in Eq.~(\ref{eq:Psi_N}) with
Eq.~(\ref{eq:WF_N1}) and Eq.~(\ref{eq:WF_N2})
and that for the superconducting side is given in Eq.~(\ref{eq:Psi_S}) 
with
\begin{align}
	\begin{pmatrix}
		\psi_{e,\uparrow}^\mathrm{S} & \psi_{e,\downarrow}^\mathrm{S} 
		&
		\psi_{h,\uparrow}^\mathrm{S} & \psi_{h,\downarrow}^\mathrm{S} 
	\end{pmatrix}
	=&
	\begin{pmatrix}
		u_e & v_h \\
		v_e & u_h
	\end{pmatrix},
   \nonumber
\end{align}
with
\begin{align}
	u_{e(h)}
	=&
	\frac{[\tilde{E}+\omega_{+(-)}]\sigma_0}
   {\sqrt{(\tilde{E}+\omega_{+(-)})^2+\frac{1}{2}\mathrm{Tr}
   \Delta_{+(-)}\Delta_{+(-)}^\dag}},
   \nonumber
		\\
	v_{e}
	=&
	\frac{\Delta_+^\dag}
	{\sqrt{(\tilde{E}+\omega_+)^2+\frac{1}{2}\mathrm{Tr}\Delta_+\Delta_+^\dag}},
   \nonumber
	\\
	v_{h}
	=&
	\frac{\Delta_-}
	{\sqrt{(\tilde{E}+\omega_-)^2+\frac{1}{2}\mathrm{Tr}\Delta_-\Delta_-^\dag}},
   \nonumber
	\\
	\omega_\pm
	=&
	\sqrt{\tilde{E}^2-\frac{1}{2}\mathrm{Tr}\Delta_\pm\Delta_\pm^\dag},
   \nonumber
	\\
	\Delta_+
	=&
\Delta_{E_{1u}}^\mathrm{planar}(\mathbf{k}),
   \nonumber
	\\
	\Delta_-
	=&
	\Delta_{E_{1u}}^\mathrm{planar}(\tilde{\mathbf{k}}),
   \nonumber
\end{align}
for $\Delta_{E_{1u}}^\mathrm{planar}(\mathbf{k})$ and
\begin{align}
   u_{e}
   =&
   a_{+}
   [%
      |\mathbf{q}_{+}|\sigma_0
      +\mathbf{q}_{+}\cdot\boldsymbol{\sigma}
   ](\sigma_0+\sigma_3)
   \nonumber\\
   &+
   b_{+}
   [%
      |\mathbf{q}_{+}|\sigma_0
      -\mathbf{q}_{+}\cdot\boldsymbol{\sigma}
   ](\sigma_0-\sigma_3),
   \nonumber
   \\
   u_{h}
   =&
   a_{-}
   [%
      |\mathbf{q}_{-}|\sigma_0
      +\mathbf{q}_{-}\cdot\boldsymbol{\sigma}^*
   ](\sigma_0+\sigma_3)
   \nonumber\\
   &+
   b_{-}
   [%
      |\mathbf{q}_{-}|\sigma_0
      -\mathbf{q}_{-}\cdot\boldsymbol{\sigma}^*
   ](\sigma_0-\sigma_3),
   \nonumber
   \\
   a_{\pm}
   =&
   \frac{1}
   {\sqrt{%
   16|\mathbf{q}_{\pm}|[|\mathbf{q}_{\pm}|+(\mathbf{q}_{\pm})_3]
   }}
   \sqrt{\frac{\tilde{E}+\omega_{\pm,p}}{\tilde{E}}},
   \nonumber
   \\
   b_{\pm}
   =&
   \frac{1}
   {\sqrt{%
   16|\mathbf{q}_{\pm}|[|\mathbf{q}_{\pm}|+(\mathbf{q}_{\pm})_3]
   }}
   \sqrt{\frac{\tilde{E}+\omega_{\pm,m}}{\tilde{E}}},
   \nonumber
   \\
   v_{e}
   =&
   \Delta_+^\dag u_e (\tilde{E}\sigma_0+\hat{\omega}_{+,pm})^{-1},
   \nonumber
   \\
   v_{h}
   =&
   \Delta_- u_h (\tilde{E}\sigma_0+\hat{\omega}_{-,pm})^{-1},
   \nonumber
   \\
   \hat{\omega}_{\pm,pm}
   =&
   \begin{pmatrix}
      \omega_{\pm,p} & 0\\
      0 & \omega_{\pm,m}
   \end{pmatrix},
   \nonumber
   \\
   \omega_{\pm,p}
   =&
   \sqrt{\tilde{E}^2-(|\mathbf{d}_\pm|^2+|\mathbf{q_\pm}|)},
   \nonumber
   \\
   \omega_{\pm,m}
   =&
   \sqrt{\tilde{E}^2-(|\mathbf{d}_\pm|^2-|\mathbf{q_\pm}|)},
   \nonumber
   \\
   \mathbf{q}_\pm
   =&
   i\mathbf{d}_\pm\times\mathbf{d}^*_\pm,
   \nonumber
   \\
   \mathbf{d}_+
   =&
   \mathbf{d}(\mathbf{k}),
   \nonumber
   \\
   \mathbf{d}_-
   =&
   \mathbf{d}(\tilde{\mathbf{k}}),
   \nonumber
\end{align}
for
$\Delta_{E_{2u}}^{f+p}$.
The boundary conditions are given in Eq.~(\ref{eq:bound1}) 
and Eq.~(\ref{eq:bound2}).
We derive a general formula for conductance, which includes the non-unitary 
case. This formula is similar to that derived in the context of 
doped topological insulators \cite{Takami2014}. 
In the present case, $\Gamma_{\pm}$ is available for a general pair potential, 
including non-unitary spin-triplet pairing. 
The derivation of conductance for a general pair potential is given in 
Appendix~\ref{sec:App_cond}.
\begin{align}
   \sigma_\mathrm{S}
   =&
   \frac{\sigma_\mathrm{N}}{2}\mathrm{Tr}
   \left[
      \sigma_0-(1-\sigma_\mathrm{N})
      \hat{\Gamma}_+^\dag\hat{\Gamma}_-^\dag
   \right]^{-1}
   \nonumber\\
   &\times
   \left[
      1
      +
      \sigma_\mathrm{N}\hat{\Gamma}_+^\dag\hat{\Gamma}_+
      +
      (\sigma_\mathrm{N}-1)\
      \hat{\Gamma}_+^\dag\hat{\Gamma}_-^\dag
      \hat{\Gamma}_-\hat{\Gamma}_+
   \right]
   \nonumber\\
   &\times
   \left[
      \sigma_0-(1-\sigma_\mathrm{N})\hat{\Gamma}_-\hat{\Gamma}_+
   \right]^{-1},
   \label{eq:cond_gen1}
\end{align}
with
\begin{align}
   \hat{\Gamma}_+
   =&
   \frac{\Delta_+^\dag}{\tilde{E}+\omega_{+}},
   \nonumber
   \\
   \hat{\Gamma}_-
   =&
   \frac{\Delta_-}{\tilde{E}+\omega_{-}},
   \nonumber
\end{align}
for $\Delta_{E_{1u}}^\mathrm{planar}$, and
\begin{align}
   \hat{\Gamma}_+
   =&
   \frac{\Delta_+^\dag}{2}
   \left[
      \left(
         \frac{1}{\tilde{E}+\omega_{+,p}}
         +
         \frac{1}{\tilde{E}+\omega_{+,m}}
      \right)
   \right.
   \nonumber\\
   &
   \left.
      +
      \left(
         \frac{1}{\tilde{E}+\omega_{+,p}}
         -
         \frac{1}{\tilde{E}+\omega_{+,m}}
      \right)
      \frac{\mathbf{q_+}}{|\mathbf{q_+}|}\cdot\boldsymbol{\sigma}
   \right],
   \nonumber
   \\
   \hat{\Gamma}_-
   =&
   \frac{\Delta_-}{2}
   \left[
      \left(
         \frac{1}{\tilde{E}+\omega_{-,p}}
         +
         \frac{1}{\tilde{E}+\omega_{-,m}}
      \right)
   \right.
   \nonumber\\
   &
   \left.
      +
      \left(
         \frac{1}{\tilde{E}+\omega_{-,p}}
         -
         \frac{1}{\tilde{E}+\omega_{-,m}}
      \right)
      \frac{\mathbf{q_-}}{|\mathbf{q_-}|}\cdot\boldsymbol{\sigma}^*
   \right],
   \nonumber
\end{align}
for $\Delta_{E_{2u}}^{f+p}$.
It is noted that the 
charge conductance 
for $\Delta_{E_{1u}}^\mathrm{planar}$ can be written by using
$\Delta_{E_{1u}}^\mathrm{chiral}$.
\begin{align}
   \sigma_\mathrm{S}(eV,\mathbf{k}_\parallel)
   =&\frac{\sigma_\mathrm{N}}{2}
   \left[
      1+\sigma_\mathrm{N}|\Gamma_+|^2+(\sigma_\mathrm{N}-1)|\Gamma_+\Gamma_-|^2
   \right]\nonumber\\
   &\times\left(\frac{1}{S_1}+\frac{1}{S_2}\right),
   \label{eq:sigma_planar}
	 \\
	 S_1
	 =&
	 \frac{1}{|1+(\sigma_\mathrm{N}-1)\Gamma_+\Gamma_-|^2},
   \label{eq:A}
	 \\
	 S_2
   =&
   \frac{1}{|1+(\sigma_\mathrm{N}-1)\Omega_+\Omega_-|^2},
   \label{eq:B}
   \\
   \Omega_+
   =&
   \frac{\tilde{\Delta}_{E_{1u}}^\mathrm{chiral}(\mathbf{k})}
   {\tilde{E}+\sqrt{\tilde{E}^2-|\tilde{\Delta}_{E_{1u}}^\mathrm{chiral}(\mathbf{k})|^2}},
   \nonumber
   \\
   \Omega_-
   =&
   \frac{[\tilde{\Delta}_{E_{1u}}^\mathrm{chiral}(\tilde{\mathbf{k}})]^*}
   {\tilde{E}+\sqrt{\tilde{E}^2
   -
   |\tilde{\Delta}_{E_{1u}}^\mathrm{chiral}(\tilde{\mathbf{k}})|^2}},
   \nonumber
\end{align}
where $\Gamma_\pm$ in Eq.~(\ref{eq:A}) is the same as in
Eq.~(\ref{eq:Gamma_p}) and Eq.~(\ref{eq:Gamma_m}) if 
$\Delta$ is replaced by $\tilde{\Delta}_{E_{1u}}^\mathrm{chiral}$.
SABS is given by Eq.~(\ref{eq:Re1}), Eq.~(\ref{eq:Im0}), and 
\begin{align}
   \mathrm{Re}(\Omega_+\Omega_-)=&1,
   \nonumber
   \\
   \mathrm{Im}(\Omega_+\Omega_-)=&0.
   \nonumber
\end{align}
Owing to the presence of time reversal symmetry, the energy dispersion of the SABS is
given by 
\begin{align}
   E(\mathbf{k}_\parallel) = \pm E^\mathrm{chiral}(\mathbf{k}_\parallel),
   \nonumber
\end{align}
where $E^\mathrm{chiral}(\mathbf{k}_\parallel)$ is the energy dispersion of the SABS for 
$\Delta_{E_{1u}}^\mathrm{chiral}$. 
The normalized conductance is given in Eq.~(\ref{eq:cond_norm}).

\subsection{\label{sec:topological_n}Topological number}
In this subsection, 
we briefly summarize the main discussion about the number of 
the ZESABSs for $\Delta_\mathrm{2d}^\nu$ and
$\Delta_\mathrm{3d}^{\nu}$ ($\nu\leq2$)
\cite{Kobayashi2015}.
This result is used in Sec.~\ref{sec:H0}.
The ZESABSs for $\Delta_\mathrm{2d}^\nu$ are understood only from  
a 2D topological number (Chern number) and 
those for $\Delta_\mathrm{3d}^\nu$ are understood from a one-dimensional
topological number (winding number) and the Chern number.
Similar discussions for $\Delta_\mathrm{3d}^{\nu>2}$ and 
$\Delta_{E_{1u}}^\mathrm{chiral}$ are given in Appendix~\ref{sec:App_top}.
For $\Delta_\mathrm{3d}^{\nu>2}$ with $0<\alpha<\pi/4$, 
cylindrical cuts must be used to calculate the Chern number.

Generally, if a Hamiltonian possesses time reversal symmetry, 
the winding number can be defined by using a chiral operator 
$\Gamma=-iCT$ ($C=\sigma_0\tau_1K$: Charge conjugation, 
$T=i\sigma_2\tau_0K$: Time reversal.  
$\sigma_i$ and $\tau_i$ are the Pauli 
matrices in spin and Nambu spaces, respectively,)
which anticommutes with the Hamiltonian.
The winding number is given by \cite{STYY11,Schnyder2011},
\begin{align}
   W(\mathbf{k}_\parallel,\Gamma)
   =
   \frac{-1}{4\pi i}
   \int_{-\infty}^\infty dk_\perp 
   \mathrm{Tr}
   \left[
      \Gamma H^{-1}(\mathbf{k})\partial_{k_\perp}H(\mathbf{k})
   \right],
   \nonumber
\end{align}
where $\mathbf{k}_\parallel$ and $k_\perp$ are wave vectors parallel and 
perpendicular to a certain surface, respectively.
Although $\Delta_\mathrm{3d}^\nu$ ($\nu\geq1$) does not 
have time reversal symmetry,
the BdG Hamiltonian hosts a momentum-dependent pseudo time reversal symmetry 
\cite{Kobayashi2015}:
\begin{align}
   U_{\varphi_\mathbf{k}}^\dag T U_{\varphi_\mathbf{k}}
   H(\mathbf{k})
   U_{\varphi_\mathbf{k}}^\dag T^\dag U_{\varphi_\mathbf{k}}
   =
   H(-\mathbf{k}),
   \nonumber
\end{align}
with $U_{\varphi_\mathbf{k}}=\exp(-i\nu\varphi_\mathbf{k}\sigma_0\tau_3/2)$ 
and 
$\varphi_\mathbf{k}=\tan^{-1}(k_y/k_x)$.
Replacing $\Gamma$ with $\Gamma_{\varphi_\mathbf{k}}$, 
we can define the winding number where $\Gamma_{\varphi_\mathbf{k}}$ is given by
\begin{align}
   \Gamma_{\varphi_\mathbf{k}}
   =
   \left\{
      \begin{aligned}
         &U_{\varphi_\mathbf{k}}^\dag\Gamma U_{\varphi_\mathbf{k}} 
         &\:(\nu:\:\mathrm{odd}),
         \\
         &U_{\varphi_\mathbf{k}}^\dag S_zCTU_{\varphi_\mathbf{k}}
         &\:(\nu:\:\mathrm{even}),
      \end{aligned}
   \right.
   \nonumber
\end{align}
with $S_z=\sigma_3\tau_3$.

The Chern number \cite{TKNN} at a fixed $k_{\parallel,1}$ is defined by
\begin{align}
   N(k_{\parallel,1})
   =
   \frac{i}{2\pi}\sum_{n\in\mathrm{occ}}\int_\mathrm{BZ} 
   dk_\perp dk_{\parallel,2}\epsilon^{ab}\partial_{k_a}
   \langle u_n(\mathbf{k})|\partial_{k_b}|u_n(\mathbf{k})\rangle,
   \nonumber
\end{align}
where $|u_n(\mathbf{k})\rangle$ is an eigenstate of $H(\mathbf{k})$ and the 
summation is taken over all of the occupied states.

\begin{figure*}[htbp]
   \centering
   \includegraphics[width=17.5cm,bb = 0 0 1500 1050]{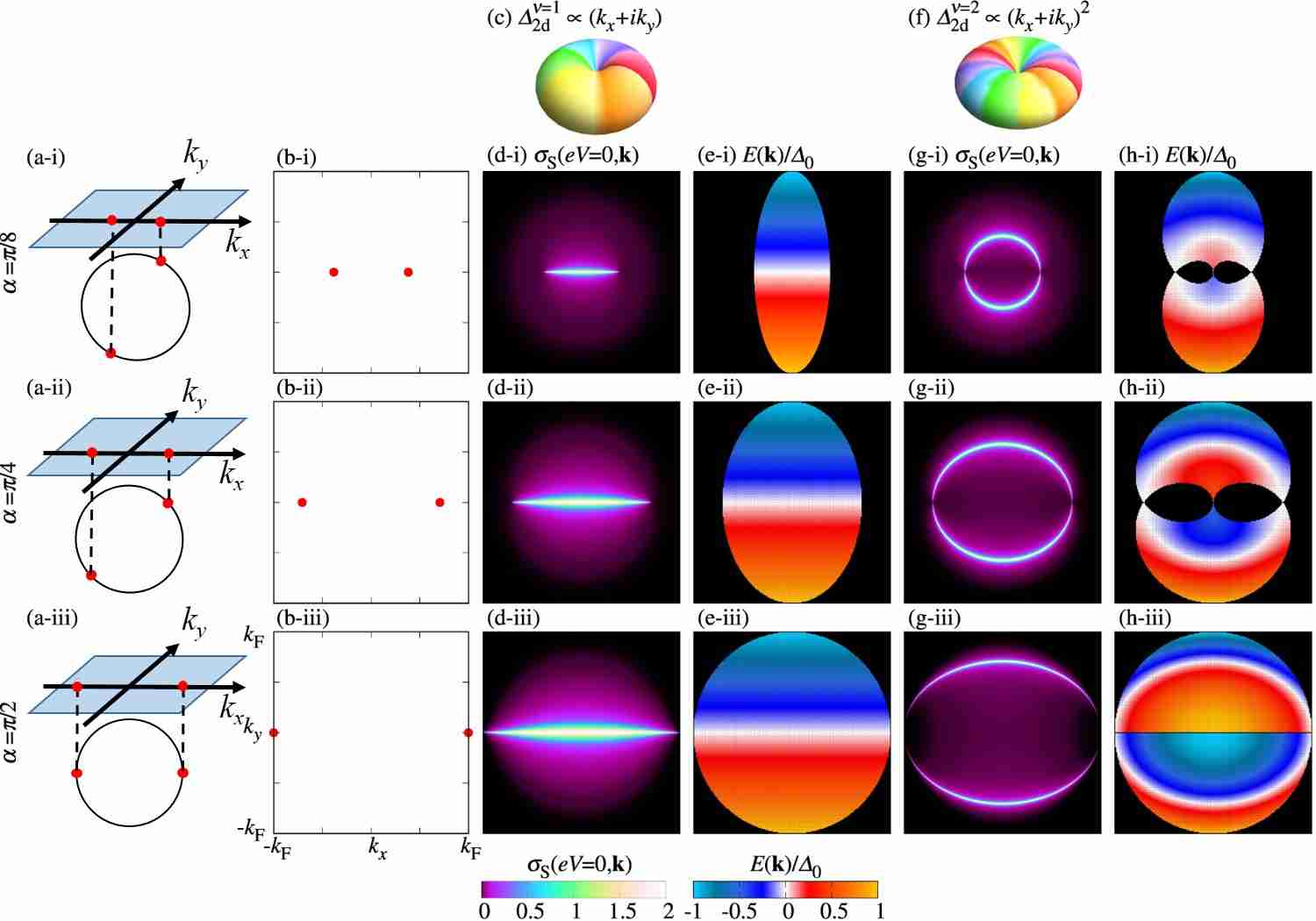}
   \caption{%
      Schematic illustration of point nodes (red dots) for (a-i) $\alpha=\pi/8$,
      (a-ii) $\alpha=\pi/4$, and (a-iii) $\alpha=\pi/2$.
      Point nodes projected on the $k_x-k_y$ plane corresponding to (a-i)--(a-iii) 
      are shown in (b-i)--(b-iii).
      Schematic picture of pair potential for (c) $\Delta_{2\mathrm{d}}^{\nu=1}$ 
      and (f) $\Delta_{2\mathrm{d}}^{\nu=2}$. 
      The angle-resolved zero bias conductance 
      $\sigma_\mathrm{S}(eV=0,\mathbf{k}_\parallel)$ 
      at $Z=6$ are plotted as functions of $k_x$ and $k_y$ for 
      $\Delta_{2\mathrm{d}}^{\nu=1}$ 
      (d-i) $\alpha=\pi/8$, (d-ii) $\alpha=\pi/4$, and (d-iii) $\alpha=\pi/2$. 
      Similar plots for $\Delta_{2\mathrm{d}}^{\nu=2}$  
      for (g-i) $\alpha=\pi/8$, (g-ii) $\alpha=\pi/4$, and (g-iii) $\alpha=\pi/2$. 
      The energy dispersions of the SABS $E(\mathbf{k}_\parallel)$ are plotted as functions 
      of $k_x$ and $k_y$ for $\Delta_{2\mathrm{d}}^{\nu=1}$
      [(e-i)--(e-iii)] and for $\Delta_{2\mathrm{d}}^{\nu=2}$
      [(h-i)--(h-iii)]. 
   }
   \label{fig:2D_l}
\end{figure*}

\begin{figure}[htbp]
   \centering
   \includegraphics[width=8.5cm,bb=0 0 562 1125]{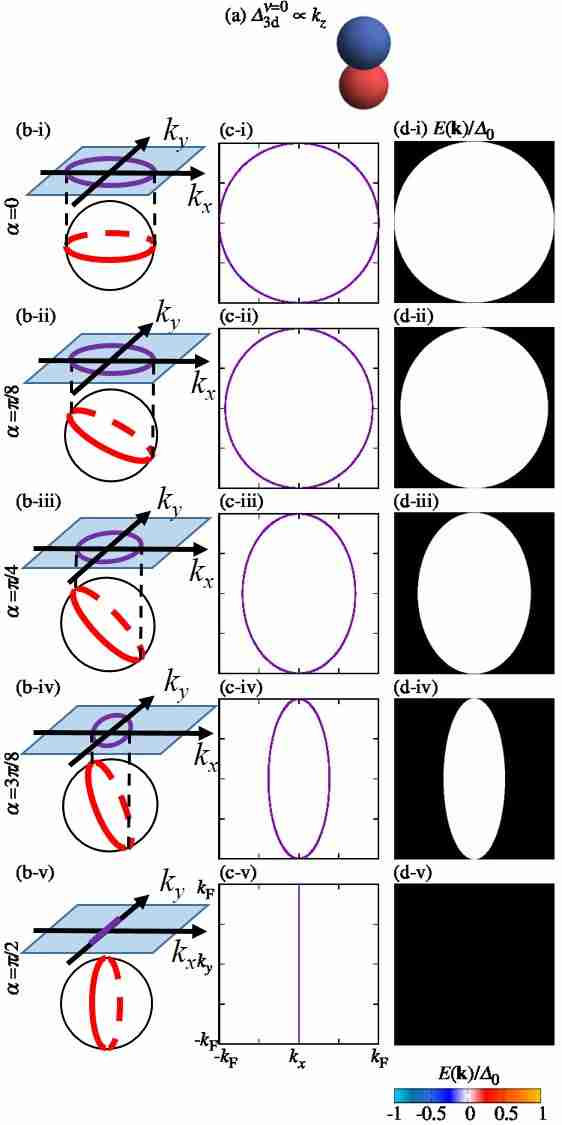}
   \caption{%
      Schematic illustration of pair potential 
      for (a) $\Delta_{3\mathrm{d}}^{\nu=0}$.
      Schematic illustration of a line node (red line) 
      for (b-i) $\alpha=0$, (b-ii) $\alpha=\pi/8$, (b-iii) $\alpha=\pi/4$, 
      (b-iv) $\alpha=3\pi/8$, and (b-v) $\alpha=\pi/2$.
      The line node projected on the $k_x-k_y$ plane corresponding to 
      (b-i)--(b-v) are shown in (c-i)--(c-v).
      The energy dispersion of SABS $E(\mathbf{k}_\parallel)$ for given  $\alpha$ are 
      plotted as functions of $k_x$ and $k_y$ 
      [(d-i)--(d-v)].
   }
   \label{fig:3D_l9}
\end{figure}

\begin{figure*}[htbp]
   \centering
   \includegraphics[width=17.5cm,bb=0 0 1125 1125]{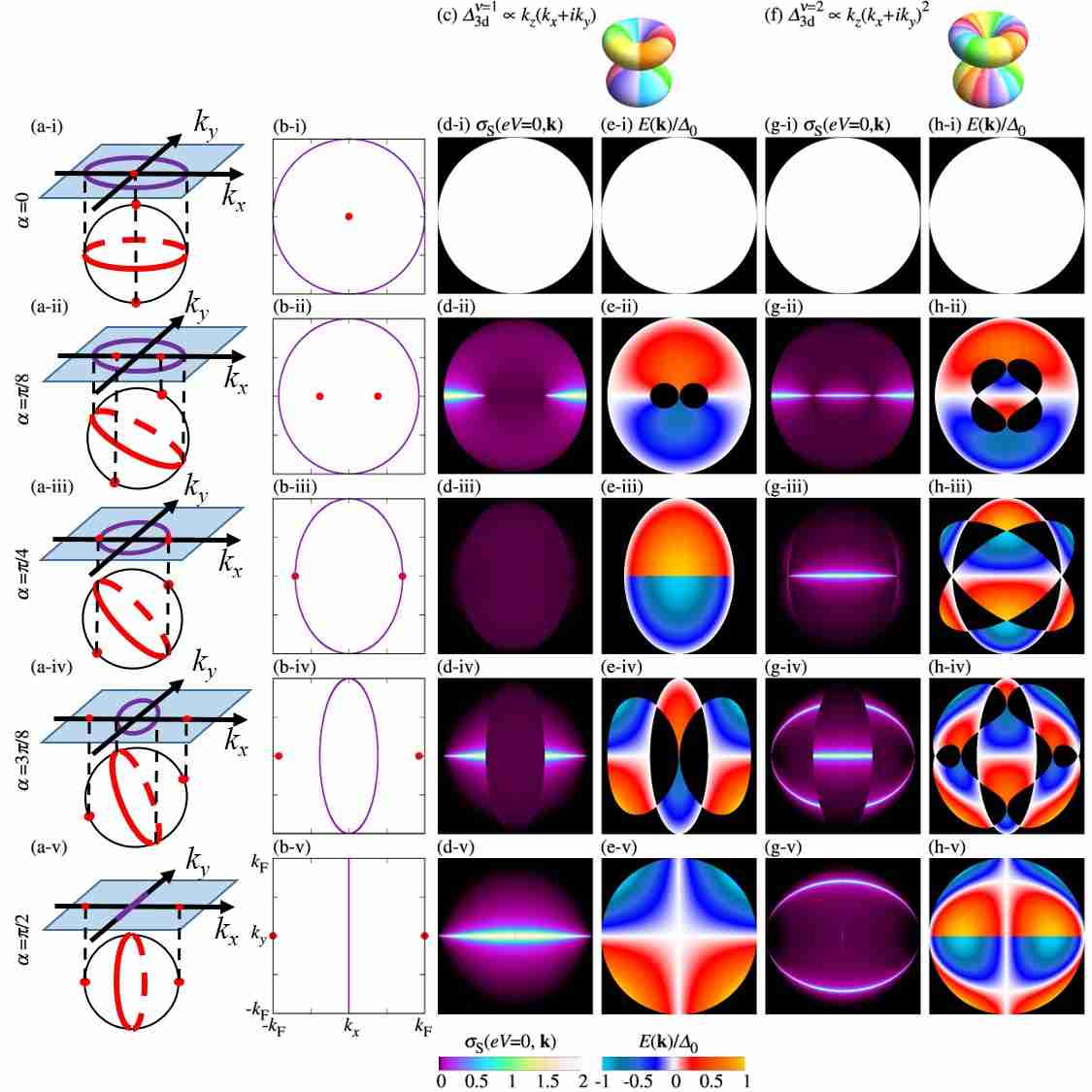}
   \caption{%
      Schematic illustration of point nodes (red dots) and a line node (red line) 
      for (a-i) $\alpha=0$, (a-ii) $\alpha=\pi/8$, (a-iii) $\alpha=\pi/4$, 
      (a-iv) $\alpha=3\pi/8$, and (a-v) $\alpha=\pi/2$.
      Point nodes and a line node projected on the $k_x-k_y$ plane corresponding to 
      (a-i)--(a-v) are shown in (b-i)--(b-v).
      Schematic illustration of pair potential 
      for 
      (c) $\Delta_{3\mathrm{d}}^{\nu=1}$ 
      and (f) $\Delta_{3\mathrm{d}}^{\nu=2}$.
      The angle-resolved zero bias conductance 
      $\sigma_\mathrm{S}(eV=0,\mathbf{k}_\parallel)$ with $Z=6$ is plotted as functions
      of $k_x$ and $k_y$ for $\Delta_{3\mathrm{d}}^{\nu=1}$ [(d-i)--(d-v)] 
      and for $\Delta_{3\mathrm{d}}^{\nu=2}$ [(g-i)--(g-v)].
      The energy dispersion of the SABS $E(\mathbf{k}_\parallel)$ for given  $\alpha$ are 
      plotted as functions of 
      $k_x$ and $k_y$ for 
      $\Delta_{3\mathrm{d}}^{\nu=1}$ [(e-i)--(e-v)] and 
     for $\Delta_{3\mathrm{d}}^{\nu=2}$ [(h-i)--(h-v)].
   }
   \label{fig:3D_l5}
\end{figure*}

These topological numbers connect the number of the ZESABSs by the 
bulk-boundary correspondence.
We show the angle-resolved zero voltage conductance 
calculated by using Eq.~(\ref{eq:conductance})
in
Fig.~\ref{fig:2D_l} ($\Delta_\mathrm{2d}^\nu$),
Fig.~\ref{fig:3D_l9} ($\Delta_\mathrm{3d}^{\nu=0}$), and
Fig.~\ref{fig:3D_l5} ($\Delta_\mathrm{3d}^{\nu=1,2}$).
The corresponding energy dispersion of SABS calculated 
by using Eq.~(\ref{eq:SABS1})--(\ref{eq:det_SABS2}) 
are also shown in the same figures and we discuss them in Sec.~\ref{sec:H0}.
\begin{table}[htbp]
   \caption{\label{tab:topo_num}%
      Origin of ZESABSs for each pair potential.\
      $w$ and $c$ indicate winding number and Chern number, respectively.
   }
   \begin{ruledtabular}
      \begin{tabular}{cccccc}
         $\Delta$               & $\alpha=0$   & $0<\alpha<\frac{\pi}{4}$ &
         $\alpha=\frac{\pi}{4}$ & $\frac{\pi}{4}<\alpha<\frac{\pi}{2}$ & 
         $\alpha=\frac{\pi}{2}$
         \\
         \hline
         $\Delta_\mathrm{2d}^{\nu}$   & $-$ & $c$ & $c$ & $c$         & $c$ \\
         $\Delta_\mathrm{3d}^{\nu=0}$ & $w$ & $w$ & $w$ & $w$         & $-$ \\
         $\Delta_\mathrm{3d}^{\nu=1}$ & $w$ & $w$ & $-$ & $c$         & $c$ \\
         $\Delta_\mathrm{3d}^{\nu=2}$ & $w$ & $w$ & $w$ & $w$ and $c$ & $c$
      \end{tabular}
   \end{ruledtabular}
\end{table}

The position of a line node or point nodes or both on the Fermi surface 
for each pair potential are shown in 
Fig.~\ref{fig:2D_l} (a-i)--(a-iii),
Fig.~\ref{fig:3D_l9} (b-i)--(b-v), and Fig.~\ref{fig:3D_l5} (a-i)--(a-v),
and those projected on the $k_{x}-k_{y}$ (001) plane are shown in
Fig.~\ref{fig:2D_l} (b-i)--(b-iii),
Fig.~\ref{fig:3D_l9} (c-i)--(c-v), and Fig.~\ref{fig:3D_l5} (b-i)--(b-v).
For $\Delta_{3d}^\nu$, the position of a projected line node is given by 
\begin{align}
   \frac{k_{x}^{2}}{\cos^{2}\alpha} + k_{y}^{2} =k_\mathrm{F}^{2}. 
   \label{eq:p_node}
\end{align}

The ZESABSs, including the spin degrees of freedom for 
$\Delta_\mathrm{3d}^{\nu=0}$, are shown in 
Fig.~\ref{fig:3D_l9} (d-i)--(d-v).
The angle-resolved conductance at zero bias voltage reflects on the ZESABSs\@. 
They are shown in 
Fig.~\ref{fig:2D_l} (d-i)--(d-iii) for $\Delta_\mathrm{2d}^{\nu=1}$ and
Fig.~\ref{fig:2D_l} (g-i)--(g-iii) for $\Delta_\mathrm{2d}^{\nu=2}$.
They are also shown in Fig.~\ref{fig:3D_l5} (d-i)--(d-v) for 
$\Delta_\mathrm{3d}^{\nu=1}$ and
Fig.~\ref{fig:3D_l5} (g-i)--(g-v) for $\Delta_\mathrm{3d}^{\nu=2}$.

In the case of $\Delta_\mathrm{3d}^{\nu=0}$, there are flat band SABSs within
the ellipse of Eq.~(\ref{eq:p_node}) originated from the winding number 
[Fig.~\ref{fig:3D_l9} (d-i)--(d-v)].
There are two bands due to spin degrees of freedom.
For $\Delta_\mathrm{3d}^{\nu\geq1}$ with $\alpha=0$, there are flat band 
SABSs within $k_x^2+k_y^2\leq k_\mathrm{F}^2$  
[See Fig.~\ref{fig:3D_l5} (d-i) and (g-i)]. 
The origin of these flat bands is explained from the winding number. 
On the other hand, there is no ZESABS for $\Delta_\mathrm{2d}^\nu$ 
with $\alpha=0$ (not shown).
In the case of $\alpha=\pi/4$ for $\Delta_\mathrm{3d}^{\nu=1}$ 
there is no ZESABS 
[Fig.~\ref{fig:3D_l5} (d-iii)], while  
they exist on $k_y=0$
for $\alpha=\pi/4$ with $\Delta_\mathrm{3d}^{\nu=2}$  
[Fig.~\ref{fig:3D_l5} (g-iii)].
In other cases, 
(except for $\alpha=0,\pi/4$ for $\Delta_\mathrm{3d}^{\nu\geq1}$ and $\alpha=0$ for 
$\Delta_\mathrm{2d}^\nu$), the number of 
the arc shaped ZESABSs which terminate at projected point nodes on the 
$k_x-k_y$ plane
is $2\nu$, where two comes from the spin degeneracy 
[see Fig.~\ref{fig:2D_l} (d-i)--(d-iii), (g-i)--(g-iii),
and Fig.~\ref{fig:3D_l5} (d-ii), (d-iv), (d-v), (g-ii), (g-iv), and (g-v)].
For $\Delta_\mathrm{2d}^\nu$, 
the number of ZESABSs connecting 
two projected point nodes is $2\nu$. 

In addition to the ZESABSs those terminate at the projected point nodes, 
the ZESABSs located on $k_{y}=0$ and $\cos \alpha >|k_{x}|$ appear for 
$\pi/4<\alpha\leq\pi/2$ with $\Delta_\mathrm{3d}^{\nu=2}$. 

For $0<\alpha<\pi/4$ with $\Delta_\mathrm{3d}^{\nu=1,2}$, the ZESABSs are 
originated from the winding number.
For $\pi/4<\alpha<\pi/2$ with $\Delta_\mathrm{3d}^{\nu=2}$, the ZESABSs 
are originated from both the winding number and the Chern number.
The ZESABSs for $\Delta_\mathrm{2d}^\nu$ originate from the Chern number 
for arbitrary $\alpha$
(Table~\ref{tab:topo_num}).

\section{\label{sec:results}Results}

\subsection{\label{sec:H0}%
   \texorpdfstring{%
Andreev bound state with $H=0$}{Andreev bound state with H=0}}
In this subsection, we calculate the energy dispersion of the SABS with $H=0$ for 
two cases of chiral superconductors   
where pair potentials are given by $(k'_{x}+ik'_{y})^{\nu}$ ($\nu=1,2$) and
$k'_{z}(k'_{x}+ik'_{y})^{\nu}$ ($\nu=1,2$).
Although the ZESABS has been discussed 
in a previous paper \cite{Kobayashi2015}, 
the energy dispersion of the  
SABS with nonzero energy has not been clarified at all. 
To resolve this problem, we calculate the energy spectrum of the SABSs 
from Eqs.~(\ref{eq:SABS1}) to (\ref{eq:det_SABS2}).

Here, we apply this formula for normal metal / 3D chiral 
superconductor junctions. 
First, we calculate the energy dispersion
and tunneling conductance 
of the 2D-like chiral superconductor, where the pair potential is given by 
$\Delta_\mathrm{2d}^\nu\propto(k'_{x} + i k'_{y})^{\nu}$.  
The angle-resolved zero bias conductance and the energy dispersion of  
SABS are shown in Fig.~\ref{fig:2D_l}. 
For $\nu=1$, the angle-resolved zero voltage conductance is plotted from 
Fig.~\ref{fig:2D_l} (d-i) to (d-iii). 
As explained in Sec.~\ref{sec:topological_n}, we can see that 
the ZESABS appears on the line connecting 
two point nodes at $k_{y}=0$. 
The region of the ZESABS spreads with the increase of $\alpha$. 
The corresponding 
energy dispersion of the SABS $E(\mathbf{k}_\parallel)$ is shown in 
Fig.~\ref{fig:2D_l} (e-i)--(e-iii).  
In this case, we can obtain an analytical formula for the SABS, 
given by 
\begin{align}
   &
   E(\mathbf{k}_\parallel)=-\Delta_0k_y/k_\mathrm{F}
   \label{eq:2dnu1_1}
   \\
   \mathrm{with}\:\:
   &
   k_x^2+k_y^2\sin^2\alpha\leq k_\mathrm{F}^2\sin^2\alpha. 
   \label{eq:2dnu1_2}
\end{align}
The number of SABSs including the zero energy state is two (including the spin 
degeneracy). 
For $\nu=2$, two branches of the ZESABS appear 
as arcs on the 
$k_{x}-k_{y}$ plane connecting two point nodes as shown from 
Fig.~\ref{fig:2D_l} (g-i) to (g-iii). 
The length of the arcs increases with the increase of $\alpha$.
The corresponding $E(\mathbf{k}_\parallel)$ is shown from 
Fig.~\ref{fig:2D_l} (h-i) to (h-iii). 
The number of SABSs including zero energy state is four. 
Then, we can summarize that the number of SABSs stemming from topological origins, including the ZESABS, is $2\nu$. 

Next, we focus on the case of 3D chiral superconductors 
[$\Delta_\mathrm{3d}^\nu\propto k'_z(k'_x+ik'_y)^\nu$, including the $p$-wave case].
In the $p$-wave ($\nu=0$) case, (see Fig.~\ref{fig:3D_l9}),
the energy dispersion of the SABS is shown from  
Fig.~\ref{fig:3D_l9} (d-i) to (d-v). 
It is located inside the ellipse 
given by Eq.~(\ref{eq:p_node})
and there only appears the ZESABS.

For the $\nu=1$ and $2$ cases, (see Fig.~\ref{fig:3D_l5}),
the angle-resolved zero voltage conductance is plotted in 
Fig.~\ref{fig:3D_l5} (d-i)--(d-v) for $\nu=1$ and
the corresponding 
energy dispersion of the SABS $E(\mathbf{k}_\parallel)$ is shown in 
Fig.~\ref{fig:3D_l5} (e-i)--(e-v).  
We can derive an analytical formula of $E(\mathbf{k}_\parallel)$ for 
$\alpha=\pi/4$ and $\alpha=\pi/2$ 
given by 
\begin{align}
   E(\mathbf{k}_\parallel)
   =&
   -\frac{\Delta_0}{2\sqrt{2}r_{\mathrm{3d},\nu=1}}
   \left(
      \hat{k}_x+\sqrt{1-\hat{k}_x^2-\hat{k}_y^2}
   \right) ^2
   \nonumber\\
   &
   \times
   \left(
      \hat{k}_x-\sqrt{1-\hat{k}_x^2-\hat{k}_y^2}
   \right)\mathrm{sgn}(\hat{k}_y),
   \nonumber
   \\
   \mathrm{with}\:\:&2k_x^2+k_y^2\leq k_\mathrm{F}^2,
   \nonumber
\end{align}
where $\hat{k}_i=k_i/k_\mathrm{F}$ ($i=x,y,z$)
and 
\begin{align}
   E(\mathbf{k}_\parallel)=-\frac{\Delta_0}{r_{\mathrm{3d},\nu=1}}
   \hat{k}_y \mid \hat{k}_{x} \mid , 
   \nonumber
\end{align}
respectively. 
The number of SABSs $n_\mathrm{ABS}$,
which includes the ZESABS,
is classified by whether $(k_{x},k_{y})$ is 
inside the ellipse [Eq.~(\ref{eq:p_node})] or not. 
Inside the ellipse, $n_\mathrm{ABS}$ becomes two 
(including the spin degeneracy) for $\alpha < \pi/4$ and 
zero for $\alpha \geq \pi/4$. 
On the other hand, outside the ellipse, $n_\mathrm{ABS}$ becomes zero for 
$\alpha \leq \pi/4$ and four for $\alpha > \pi/4$.  
Besides this SABS with topological origin, inside the ellipse,
nonzero non-topologically SABS which does not include the zero-energy state 
exists.

Next, we focus on the $\nu=2$ case. 
Angle-resolved zero voltage conductance is plotted in 
Fig.~\ref{fig:3D_l5} (g-i)--(g-v) and 
the corresponding energy dispersion of the SABS $E(\mathbf{k}_\parallel)$ is shown in 
Fig.~\ref{fig:3D_l5} (h-i)--(h-v).  
We can derive an analytical formula of $E(\mathbf{k}_\parallel)$ for 
$\alpha=\pi/2$ 
given by 
\begin{align}
   E(\mathbf{k}_\parallel)=\frac{\Delta_0}{r_{\mathrm{3d},\nu=2}}
   \mathrm{sgn}(\hat{k}_y)|\hat{k}_x|(1-\hat{k}_x^2-2\hat{k}_y^2). 
   \nonumber
\end{align}
$n_\mathrm{ABS}$ is also classified whether $(k_{x},k_{y})$ 
is inside the ellipse or not. 
Inside the ellipse, $n_\mathrm{ABS}$ becomes two for $\alpha=0$. 
For $0<\alpha < \pi/4$, $n_\mathrm{ABS}$ is four and 
$n_\mathrm{ABS}=2$ for $\alpha \geq \pi/4$. 
On the other hand, outside the ellipse, $n_\mathrm{ABS}$ becomes zero for 
$\alpha \leq \pi/4$  
and eight for $\alpha > \pi/4$.  
Beside this SABS with topological origin, nonzero 
non-topologically SABSs also exist. \par

We further calculate $n_\mathrm{ABS}$ up to $\nu=4$ 
(The ZESABSs for $0<\alpha<\pi/4$ with $\nu=3,4$ are discussed in
Appendix~\ref{sec:App_top2}). 
A summary of $n_\mathrm{ABS}$ as a 
function of $\nu$ is shown in Table.~\ref{tab:num_of_SABS}. 
We can also obtain the analytical formula of $E(\mathbf{k}_\parallel)$ 
both for the 3D and 2D-like chiral superconductors for $\alpha=\pi/2$. 
The results are summarized in Table~\ref{tab:Andreev_b}.

\begin{table}[htbp]
   \caption{\label{tab:num_of_SABS}%
      The number of SABS ($n_\mathrm{ABS}$) which cross zero-energy for 
      $\Delta_\mathrm{3d}^{\nu}$.
      It is classified by inside or outside of the ellipse [Eq.~(\ref{eq:p_node})]
      corresponding to projected line node 
      [Fig.~\ref{fig:3D_l9} (c-i)--(c-v) and 
      Fig.~\ref{fig:3D_l5} (b-i)--(b-v)].
      $n_\mathrm{ABS}$ is counted including the spin degeneracy.
      The dash "-" means there is no area inside or outside the ellipse
      [Fig.~\ref{fig:3D_l9} (c-i) and (c-v)].
   }
   \begin{ruledtabular}
      \begin{tabular}{ccc}
          & \multicolumn{2}{c}{$n_\mathrm{ABS}$}\\
         \cline{2-3}
         $\alpha$ & Inside ellipse & Outside ellipse \\
         \hline
         0                                    & 2          & $-$    \\
         $0<\alpha<\frac{\pi}{4}$             & $2\nu$     & $0$    \\
         $\alpha=\pi/4$                       & $2(\nu-1)$ & $0$    \\
         $\frac{\pi}{4}<\alpha<\frac{\pi}{2}$ & $2(\nu-1)$ & $4\nu$ \\
         $\alpha=\pi/2$                       & $-$        & $4\nu$
      \end{tabular}
   \end{ruledtabular}
\end{table}

\begin{table}[htbp]
   \caption{\label{tab:Andreev_b}%
      The energy dispersion of SABS $E(\mathbf{k}_\parallel)/\Delta_0$
      times normalization 
      factor $r$  ($r=r_{\mathrm{3d},\nu}$ for $\Delta_\mathrm{3d}^\nu$ 
      and $r=1$ for $\Delta_\mathrm{2d}^\nu$)
      at $\alpha=\pi/2$ 
      is shown for $\Delta_{3\mathrm{d}}^\nu$ and $\Delta_{2\mathrm{d}}^\nu$.
      $\hat{k}_i=k_i/k_\mathrm{F}(i=x,y,z)$.
      For the chiral 3D superconductor, the line $k_{x}$=0 does not express 
      the energy dispersion of SABS
      since $\Delta_{3\mathrm{d}}^\nu(k_x=0)=0$ is satisfied for $\nu=1,2,3$.
   }
   \begin{ruledtabular}
      \begin{tabular}{cc}
         $\Delta(\mathbf{k})$ & $E(\mathbf{k}_\parallel)r/\Delta_0$ \\
         \hline
         ${k}_x(-{k}_z+i{k}_y)$ & $-\hat{k}_y|\hat{k}_x|$\\
         $(-{k}_z+i{k}_y)$ & $-\hat{k}_y$  \\
         ${k}_x(-{k}_z+i{k}_y)^2$   
         & 
         $\mathrm{sgn}(\hat{k}_y)|\hat{k}_x|(1-\hat{k}_x^2-2\hat{k}_y^2)$\cite{Goswami2015} 
         \\
         $(-{k}_z+i{k}_y)^2$    
         & $\mathrm{sgn}(\hat{k}_y)(1-\hat{k}_x^2-2\hat{k}_y^2)$ 
         \\ 
         ${k}_x(-{k}_z+i{k}_y)^3$ 
         & 
         $-|\hat{k}_x|\hat{k}_y(3-3\hat{k}_x^2-4\hat{k}_y^2)\mathrm{sgn}(1-\hat{k}_x^2-2\hat{k}_y^2)$ 
         \\
         $(-{k}_z+i{k}_y)^3$    
         & $-\hat{k}_y(3-3\hat{k}_x^2-4\hat{k}_y^2)\mathrm{sgn}(1-\hat{k}_x^2-2\hat{k}_y^2)$  
         \\
      \end{tabular}
   \end{ruledtabular}
\end{table}

\subsection{\label{sec:H}Conductance with magnetic field}
\begin{figure}[htbp]
   \centering
   \includegraphics[width=8.5cm,bb=0 0 1642 740]{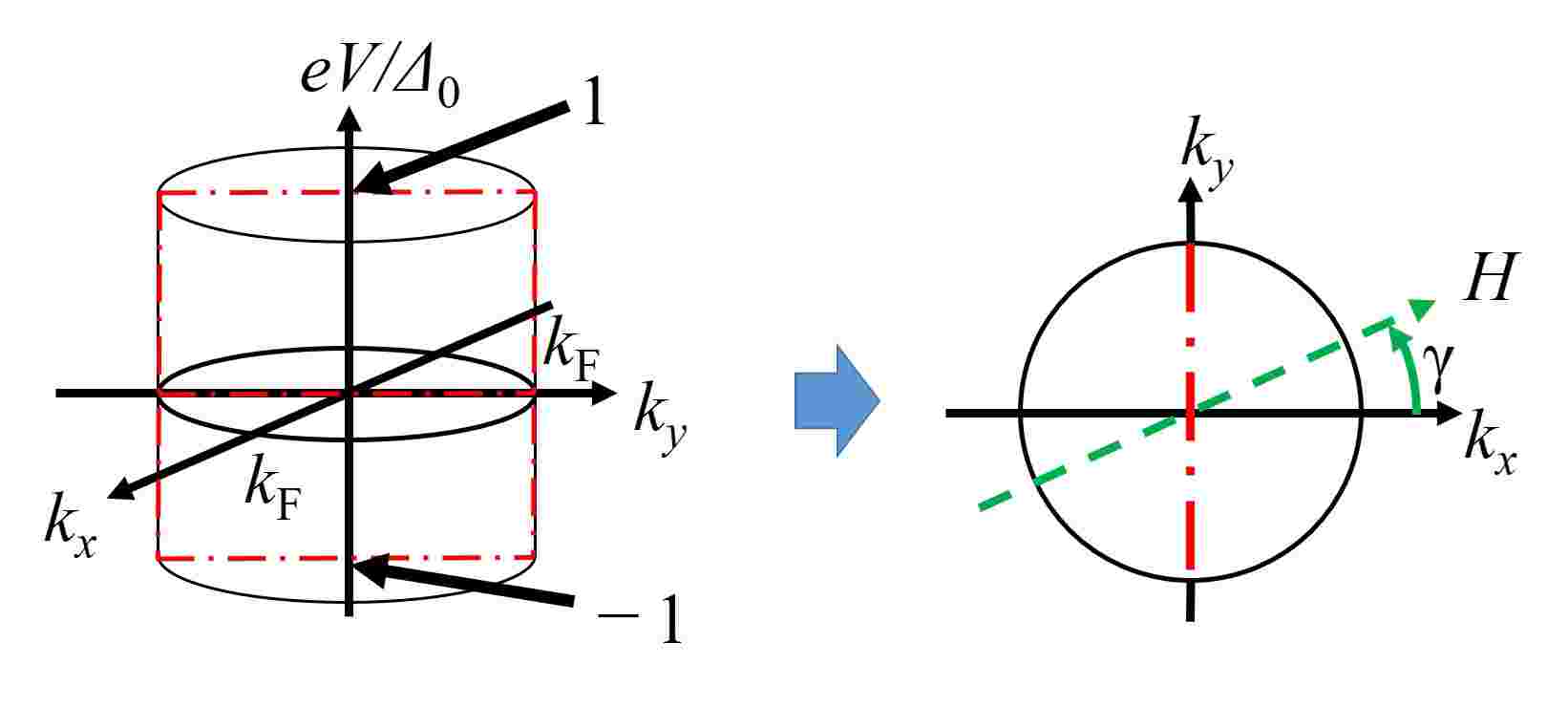}
   \caption{%
      Dash dotted line (red line) represents a rectangular
      located at $k_y/k_\mathrm{F}=[-1,1]$ and 
      $eV/\Delta_0=[-1, 1]$ on $k_{y}-eV$ plane 
      which are used for (b), (e) and (h) in Fig.~\ref{fig:nu1a1250g0}, 
      Fig.~\ref{fig:nu2a1250g0} and Fig.~\ref{fig:nu1a2500g0}.
      Dashed line (green line) represents the direction of magnetic field
      which corresponds to the arrow in Fig.~\ref{fig:pic_junction}.
   }
   \label{fig:pic_cut}
\end{figure}
\begin{figure}[htbp]
   \centering
   \includegraphics[width=8.5cm, bb=0 0 257 249]{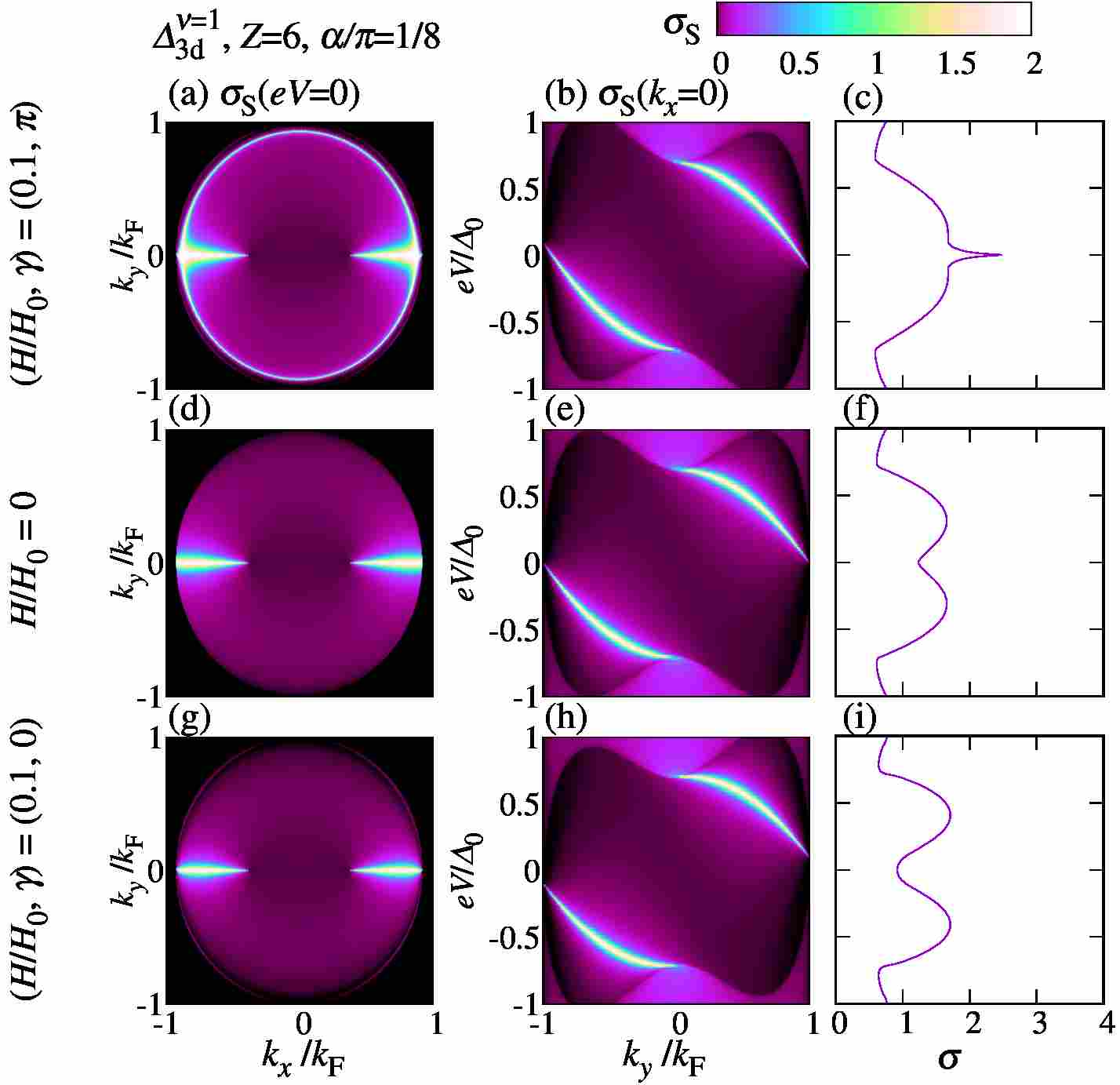}
   \caption{%
      Angle-resolved conductance $\sigma_\mathrm{S}(eV,\mathbf{k}_\parallel)$ for 
      $\Delta_\mathrm{3d}^{\nu=1}$ with $\alpha=\pi/8$ are 
      plotted as functions of ($k_x,k_y$) with $eV=0$ [(a), (d), (g)]
      and ($k_y,eV$) with $k_x=0$
      [(b), (e), (h)] (see Fig.~\ref{fig:pic_cut}) and normalized conductance plotted 
      as a function of $eV$ [(c), (f), (i)].
      Normalized magnetic field $H/H_{0}$ is chosen to be 
      0.1 with $\gamma=\pi$ [(a), (b), (c)], 
      0 [(d), (e), (f)], 
      0.1 with $\gamma=0$ [(g), (h), (i)]. 
   }
   \label{fig:nu1a1250g0}
\end{figure}
\begin{figure}[htbp]
   \centering
   \includegraphics[width=8.5cm,bb=0 0 257 249]{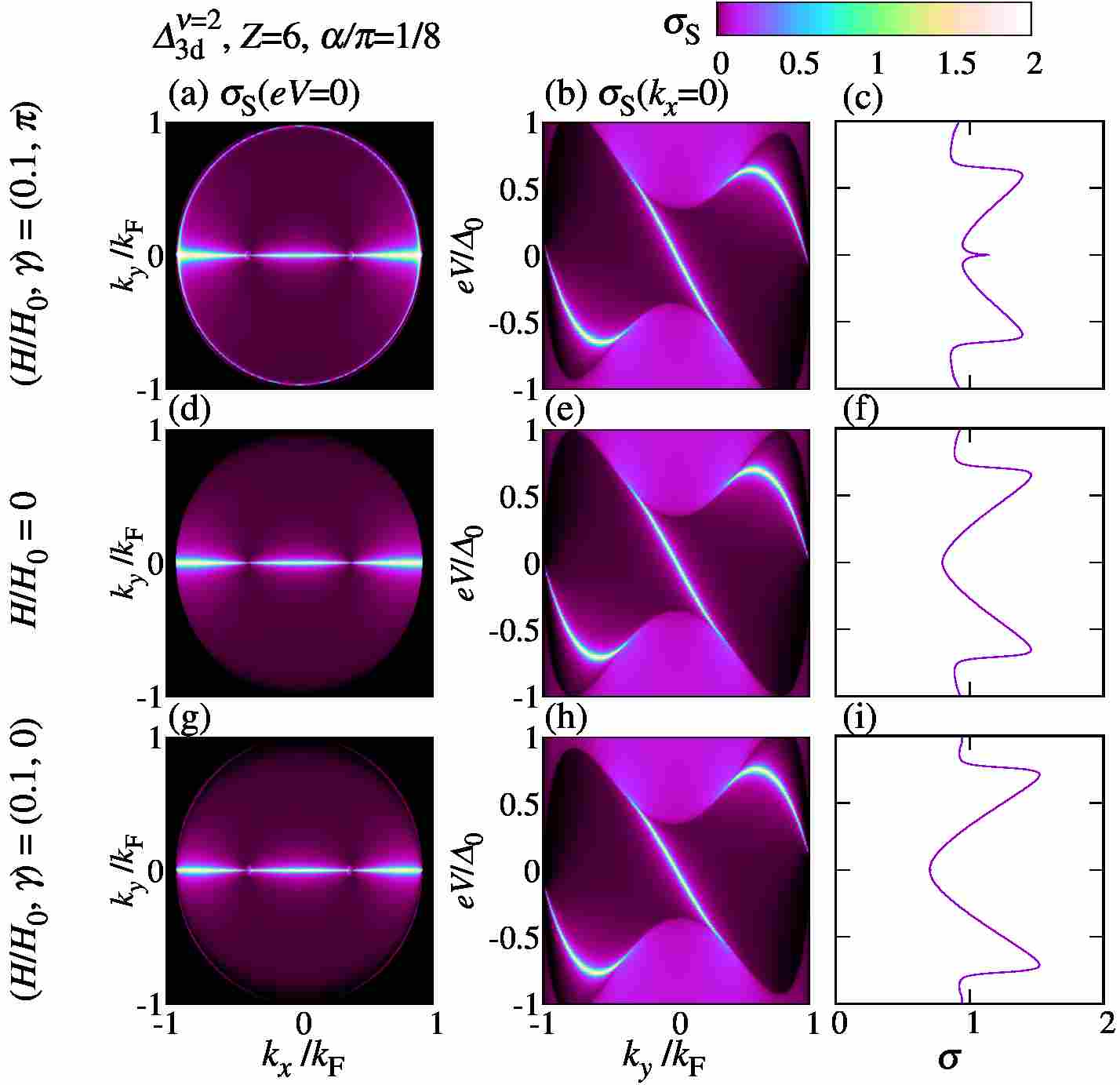}
   \caption{%
      Angle-resolved conductance of $\Delta_\mathrm{3d}^{\nu=2}$ with $\alpha=\pi/8$. 
      The model parameters used in this calculation is the same as in 
      Fig.~\ref{fig:nu1a1250g0} except for pair potential $\Delta$.
   }
   \label{fig:nu2a1250g0}
\end{figure}

In this subsection, we discuss the magnetic field dependence of conductance.
We consider the situation in which magnetic field 
is applied in the $x-y$ plane [Eq.~(\ref{eq:vector_potential})] 
and is rotated along the $z$ axis by $\gamma$ (Fig.~\ref{fig:pic_cut}).
It is known that the applied magnetic field shifts the energy of quasiparticle 
as a Doppler effect \cite{Fogel}. 
In the usual case, ZBCP without magnetic field 
is split into two \cite{Fogel,Tanuma2002a} 
or the height of ZBCP is suppressed \cite{Tanaka2002,Tanaka02a} by the Doppler effect. 
For chiral $p$-wave superconductors, the height of ZBCP is controlled by 
the direction of the applied magnetic field \cite{Tanaka02a}. 
In contrast to this standard knowledge, 
we show a unique behavior whereby the Doppler effect can change 
the line shape of conductance 
from zero bias dip to zero bias peak, as shown 
in Fig.~\ref{fig:nu1a1250g0} ($\Delta_\mathrm{3d}^{\nu=1}$)
and Fig.~\ref{fig:nu2a1250g0} ($\Delta_\mathrm{3d}^{\nu=2}$).

\begin{figure}[htbp]
   \centering
   \includegraphics[width=8.5cm,bb=0 0 360 180]{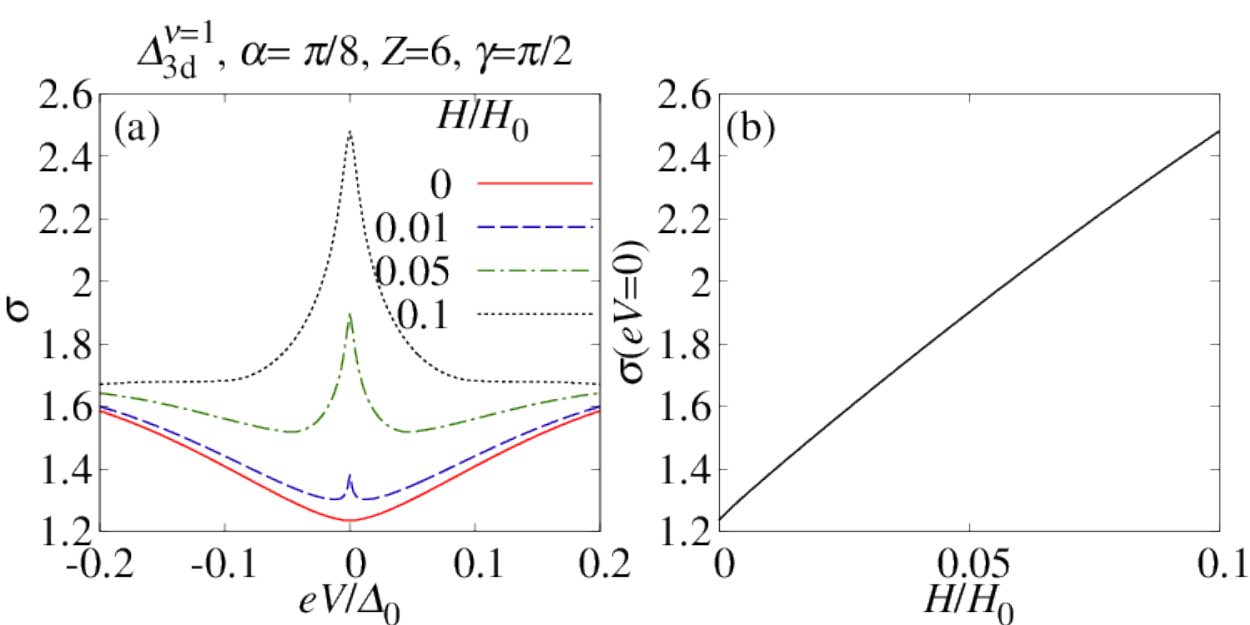}
   \caption{%
      (a) Conductance is plotted as a function of $eV$ for $H/H_0=0$, 0.01, 
      0.05 and 0.1 with $\gamma=\pi/2$.
      (b) Conductance at $eV=0$ is plotted as a function of $H/H_0$.
      This plot is fitted by the linear function $f(H/H_0)=aH/H_0+b$ with
      $(a,b)=(12.4,1.27)$.
   }
   \label{fig:E0_H_nu1}
\end{figure}

For $\alpha=\pi/8$ with $H=0$ [Fig.~\ref{fig:nu1a1250g0} (f), 
Fig.~\ref{fig:nu2a1250g0} (f)], 
$\sigma$ near $eV=0$ has a concave shape,
and it changes into ZBCP for $H/H_0=0.1$ with $\gamma=\pi$ 
[Fig.~\ref{fig:nu1a1250g0} (c), Fig.~\ref{fig:nu2a1250g0} (c)].
The Doppler effect shifts 
the energy dispersion of the SABS along the vector $(\cos \gamma, \sin \gamma, 0)$,
as shown in Fig.~\ref{fig:pic_cut}. 
The SABS that is slightly above or below zero energy for $H=0$ 
can contribute to zero bias conductance in the presence of the magnetic field. 
For $\nu=1$, in Fig.~\ref{fig:nu1a1250g0} (e), there is no ZESABS\@.
However, in the presence of the magnetic field for $\gamma=\pi$, 
an SABS exists around zero energy near $(k_x,k_y)=(0,\pm k_\mathrm{F})$
[Fig.~\ref{fig:nu1a1250g0} (b)]. 
We can also see the circle near $k_x^2/\cos^2(\pi/8)+k_y^2=k_\mathrm{F}^2$ of  
ZESABS in Fig.~\ref{fig:nu1a1250g0} (a), which does not exist 
in Fig.~\ref{fig:nu1a1250g0} (d).
On the other hand, in the case that the direction of the magnetic field is opposite, 
$i.e$., $\gamma=0$, SABS around zero energy remains absent. 
[Fig.~\ref{fig:nu1a1250g0} (g) and Fig.~\ref{fig:nu1a1250g0} (h)].
As seen from Fig.~\ref{fig:nu1a1250g0} (a), the angle-resolved conductance  
near $(k_{x},k_{y})= (\pm k_\mathrm{F},0)$ is enhanced.  
In Fig.~\ref{fig:E0_H_nu1} (a), we can see how the ZBCP 
develops with increasing magnetic field. 
We can see the generation of ZBCP even for small magnitudes of $H$ with 
$H/H_0=0.01$.
The magnitude of zero bias conductance as a function of $H$ is shown in 
Fig.~\ref{fig:E0_H_nu1} (b) and 
it is approximately a linear function of $H$.


\begin{figure}[htbp]
   \centering
   \includegraphics[width=8.5cm,bb=0 0 360 180]{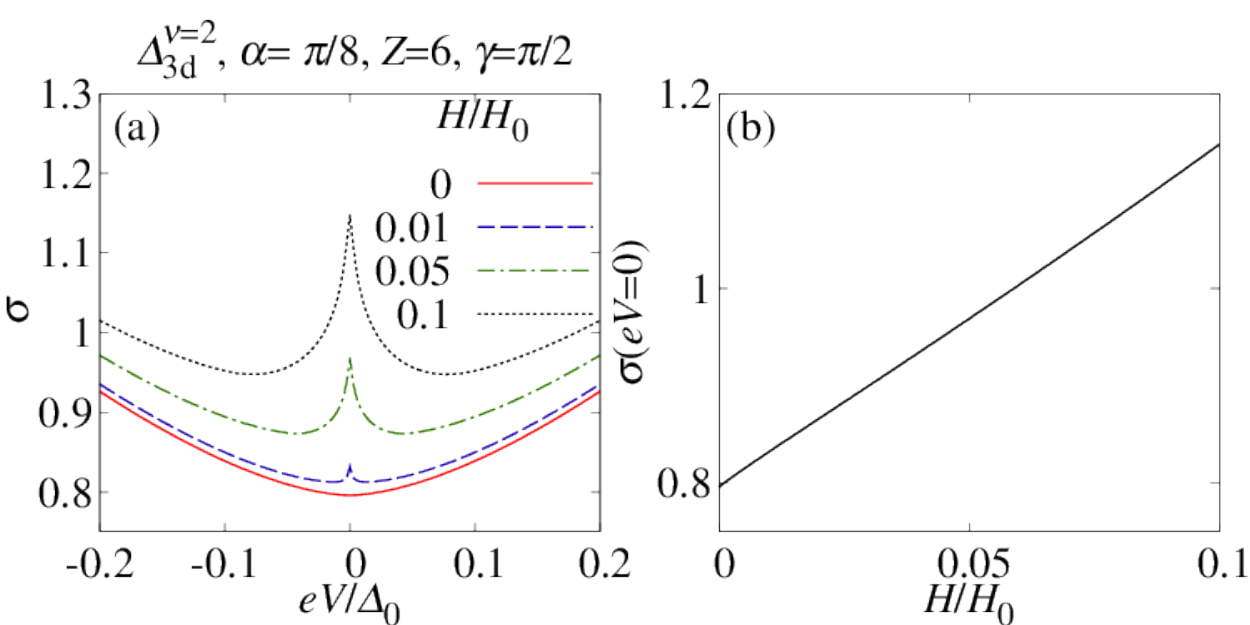}
   \caption{%
      (a) Conductance is plotted as a function of $eV$ for $H/H_0=0$, 0.01, 
      0.05, and 0.1 with $\gamma=\pi/2$.
      (b) Conductance at $eV=0$ is plotted as a function of $H/H_0$.
      This plot is fitted by the linear function $f(H/H_0)=aH/H_0+b$ with
      $(a,b)=(3.49,0.796)$.
   }
   \label{fig:E0_H_nu2}
\end{figure}

A similar plot for $\nu=2$ with the same $\alpha$ is shown in Fig.~\ref{fig:nu2a1250g0}. 
In this case, in contrast to Fig.~\ref{fig:nu1a1250g0}, chiral edge mode crossing 
$k_{y}=0$ exists [see Fig.~\ref{fig:3D_l5} (h-ii) and Fig.~\ref{fig:nu2a1250g0} (e)]. 
There are two kinds of branches of SABS: 
(1) chiral edge mode crossing $(k_{x},k_{y})=(0,0)$ and (2) SABS touching 
the ellipse given by $k_{x}^{2}/\cos^{2}(\pi/8) + k_{y}^{2}=k_\mathrm{F}^{2}$.
The slope of the chiral edge mode becomes gradual (steep) 
for $\gamma=0$ ($\gamma=\pi$) in the presence of the magnetic field. 
The contribution to zero bias conductance becomes 
suppressed (enhanced) for $\gamma=\pi$ ($\gamma=0$). 
On the other hand, qualitative feature of SABS touching the ellipse 
is similar to that for $\nu=1$ shown in Fig.~\ref{fig:nu1a1250g0}. 
In the presence of the magnetic field for $\gamma=\pi$, 
ZESABS exists near $(k_x,k_y)=(\pm k_\mathrm{F},0)$ and 
around the ellipse [Fig.~\ref{fig:nu2a1250g0} (a)]. 
Since the contribution to zero bias conductance from 
SABS touching the ellipse is dominant as compared to that of the chiral edge mode, 
the resulting $\sigma$ has a ZBCP\@. 
On the other hand, if the direction of the magnetic field is opposite, 
ZBCP is absent [Fig.~\ref{fig:nu2a1250g0} (i)]. 
In Fig.~\ref{fig:E0_H_nu2} (a), we show how ZBCP  
develops with increasing magnetic field.  
Even for small magnitude of $H$ ($H/H_0=0.01$), 
ZBCP appears similar to 
Fig.~\ref{fig:E0_H_nu1} (a).
The magnitude of conductance at zero bias as a function of $H$ is shown in 
Fig.~\ref{fig:E0_H_nu2} (b) and it is an approximately linear function of $H$.
The slope $a$ of the fitting function $f(H/H_0)=aH/H_0+b$ is smaller than that for
$\Delta_\mathrm{3d}^{\nu=1}$.

\begin{figure}[htbp]
   \centering
   \includegraphics[width=8.5cm,bb=0 0 257 249]{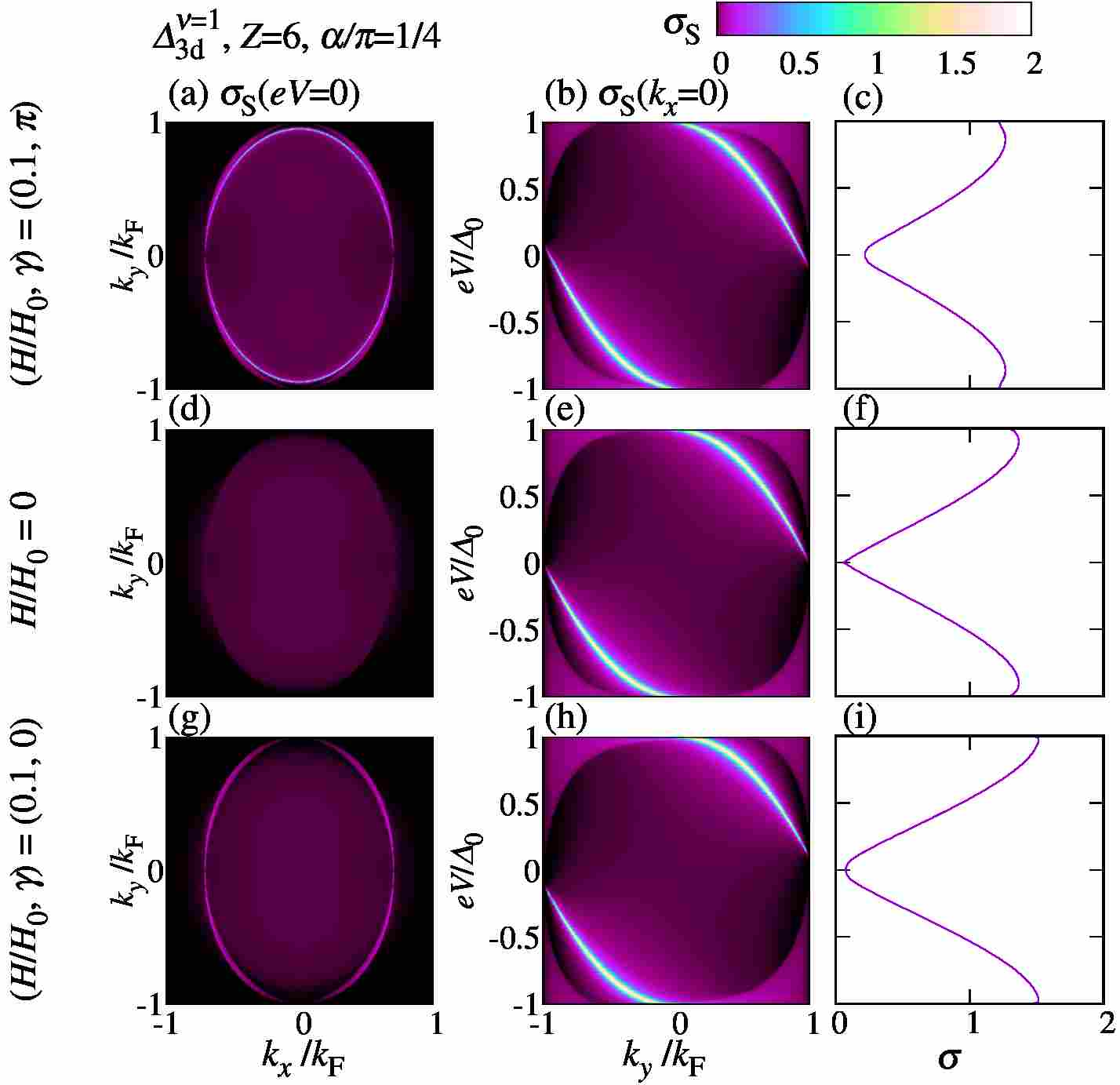}
   \caption{%
      Angle-resolved conductance of $\Delta_\mathrm{3d}^{\nu=1}$ with $\alpha=\pi/4$.
      The parameters are the same as Fig.~\ref{fig:nu1a1250g0} except for $\alpha$.
   }
   \label{fig:nu1a2500g0}
\end{figure}

Next, let us discuss special case of $\Delta_\mathrm{3d}^{\nu=1}$ for 
$\alpha=\pi/4$, at which there is no ZESABS without magnetic field
[Fig.~\ref{fig:3D_l5} (d-iii)].
The magnitude of the zero bias conductance for $H=0$ 
is very small [Fig.~\ref{fig:nu1a2500g0} (f)] and 
it becomes larger for $\gamma=\pi$ [Fig.~\ref{fig:nu1a2500g0} (c)] 
due to the similar mechanism explained 
in Fig.~\ref{fig:nu1a1250g0}.  
For $\gamma=0$ [Fig.~\ref{fig:nu1a2500g0} (i)], 
although there is no SABS at zero-energy, the conductance becomes 
slightly larger than that for $H=0$.

Whether the magnitude of ZBCP becomes larger or smaller by 
applying an infinitesimally small magnitude of magnetic field is 
summarized in Table~\ref{tab:table1}.
For 2D-like chiral superconductors with $\nu=1$, 
since the energy dispersion of the SABS is given by  $E(\mathbf{k}_\parallel)=-k_{y}/k_\mathrm{F}$ 
[Eq.~(\ref{eq:2dnu1_1}) and Eq.~(\ref{eq:2dnu1_2})], 
the magnitude of $\sigma(eV=0)$ becomes larger for $\gamma=0$ 
and smaller for $\gamma=\pi$.
In other cases, there is no simple law owing to the complicated energy 
dispersion of the SABS\@.
Using this table, we can classify five cases. 
If we make a junction for $\alpha=\pi/2$, we can distinguish between 
$\nu=0$ ($k'_{z}$), $\nu=1$ [$k'_{z}(k'_{x}+ ik'_{y})$, $(k'_{x}+ ik'_{y})$],   
and $\nu=2$ [$k'_{z}(k'_{x}+ ik'_{y})^{2}$, $(k'_{x}+ ik'_{y})^{2}$]. 
Further, if we make a junction for $\alpha=0$, 
we can distinguish between 
$k'_{z}(k'_{x}+ ik'_{y})^{\nu}$ and $(k'_{x}+ ik'_{y})^{\nu}$ 
with $(\nu=1,2)$. 

\begin{table}[htbp]
   \caption{\label{tab:table1}%
      Line shape of $\sigma(eV)$ near $eV=0$ for $H=0$.
      p: peak, d: dip. Whether the magnitude of $\sigma(eV=0)$ 
      is enhanced (suppressed) by magnetic field indicated by        
      $\uparrow$ ($\downarrow$). 
      We choose the infinitesimal applied magnetic field as $H/H_0=10^{-4}$.
      In the case $\Delta\propto k'_z$,
      the value of $\sigma(eV=0)$ at $H>0$ with $\gamma=0$ and 
      $\gamma=\pi$ are equivalent because $\sigma$ is a 
      $\pi$ periodic function of $\gamma$.
   }
   \begin{ruledtabular}
      \begin{tabular}{cccccccc}
         & & &\multicolumn{5}{c}{$\alpha$}\\
         \cline{4-8}
         $\Delta(\mathbf{k})$ & $H/H_0$ & $\gamma$ & $0$ & $\pi/8$ 
                              & $\pi/4$ & $3\pi/8$ & $\pi/2$\\
         \hline
         $k^\prime_z$ & 0 & -  & p &p & p & p & d
         \\
         & $10^{-4}$ &$0,\pi$& $\downarrow$ & $\downarrow$ & $\downarrow$ & $\downarrow$ &  $\uparrow$ 
         \\
         \hline
         $k^\prime_z(k'_x+ik'_y)$ & 0 & - & p & d & d & d & p
         \\ 
         & $10^{-4}$ & $0$   & $\downarrow$ & $\downarrow$ & $\uparrow$ & $\uparrow$ & $\uparrow$ \\ 
         & $10^{-4}$ & $\pi$ & $\downarrow$ & $\uparrow$ & $\uparrow$ & $\downarrow$ & $\downarrow$ \\
         $(k'_x+ik'_y)$ & 0 & - & d & d & p & p & p
         \\ 
         & $10^{-4}$ &$0$   & $\uparrow$ &  $\uparrow$ & $\uparrow$ &  $\uparrow$ &  $\uparrow$\\ 
         & $10^{-4}$ &$\pi$ & $\downarrow$ &  $\downarrow$ &  $\downarrow$ &  $\downarrow$ &  $\downarrow$\\ 
         \hline
         $k'_z(k'_x+ik'_y)^2$ & 0 & - & p & d & d & p & d
         \\ 
         & $10^{-4}$ &$0$  &  $\downarrow$ &  $\downarrow$ &  $\uparrow$ &  $\uparrow$ &  $\downarrow$\\ 
         & $10^{-4}$ &$\pi$ &  $\downarrow$ &  $\uparrow$ & $\downarrow$ &  $\downarrow$ &  $\uparrow$\\ 
         $(k'_x+ik'_y)^2$& 0 & - & d & p & p & d & d
         \\ 
         & $10^{-4}$ &$0$   &  $\uparrow$ &  $\uparrow$ &  $\uparrow$ &  $\downarrow$ &  $\downarrow$\\ 
         & $10^{-4}$ &$\pi$ &  $\uparrow$ &  $\downarrow$ &  $\downarrow$ &  $\uparrow$ &  $\uparrow$\\ 
      \end{tabular}
   \end{ruledtabular}
\end{table}

\begin{figure}[htbp]
   \centering
   \includegraphics[width=8.5cm,bb=0 0 703 1125]{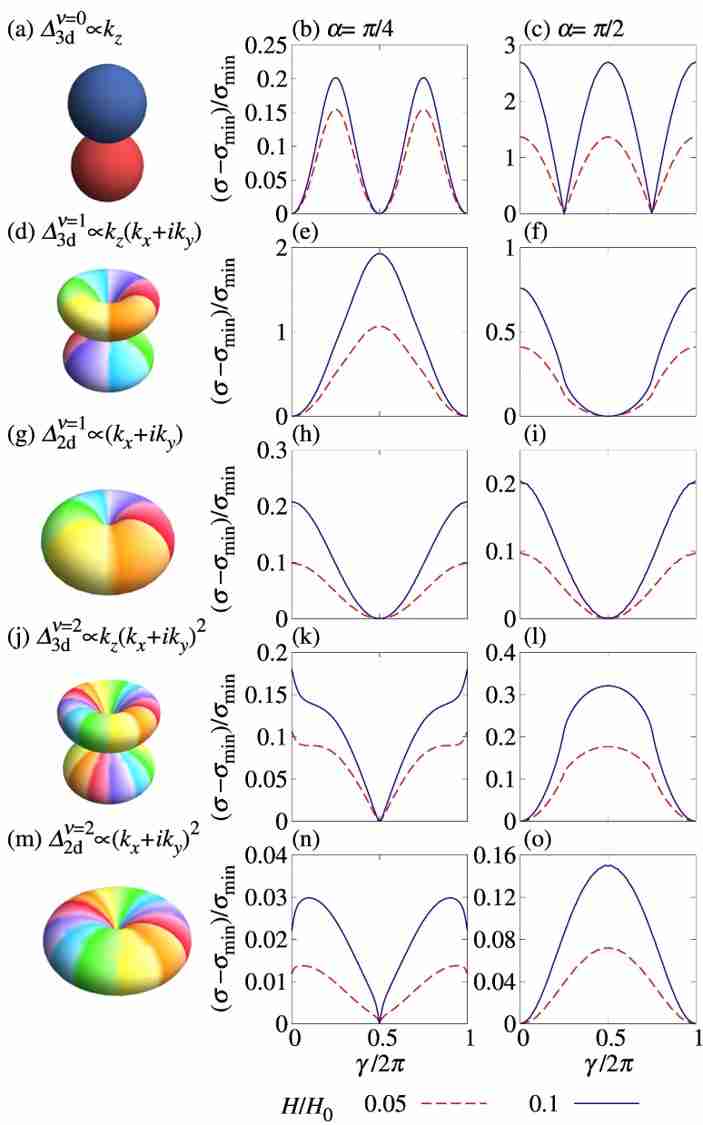}
   \caption{%
      Normalized conductance at $eV=0$ 
      is plotted as a function of 
      $\gamma$ for $\nu=0,1,2$ with $Z=6$. 
      Schematic pictures of the pair potential are shown in (a), (d), (g), (j), and (m). 
      $\alpha=\pi/4$ for (b), (e), (h), (k), and (n). 
      $\alpha=\pi/2$ for (c), (f), (i), (l), and (o). 
      $\sigma_\mathrm{min}$ is a minimum value of $\sigma(eV=0)$ as a function of 
      $\gamma$ for $H/H_0=0.05$ and $H/H_0=0.1$.    
   }
   \label{fig:int7_1}
\end{figure}

To clarify $\gamma$ dependence of conductance, we 
plot $\sigma(eV=0)$ as a function of $\gamma$  for 
$\alpha=\pi/4$, and $\pi/2$ in Fig.~\ref{fig:int7_1}.
For $\alpha=0$, $\sigma(eV=0)$ is constant as a function of $\gamma$ 
owing to the rotational symmetry of the pair potentials (not shown).
Since $\Delta_\mathrm{3d}^{\nu=0}$ conserves time reversal symmetry, 
$i.e$., the energy dispersion of the SABS has two-fold rotational symmetry
[Fig.~\ref{fig:3D_l9} (d-i)--(d-v)],
$\sigma$ has $\pi$ periodicity [Fig.~\ref{fig:int7_1} (b), (c)].
In other cases, $\sigma(eV=0)$ has $2\pi$ periodicity due to time reversal symmetry breaking.

$\gamma$, which makes $\sigma(eV=0)$ largest 
is summarized in Table~\ref{tab:table2}.
In the case of $\Delta_\mathrm{3d}^{\nu=0}$ with $\alpha=\pi/2$, $\sigma(eV=0)$ becomes
the smallest when the direction of the magnetic field is parallel to the 
direction of the projected line node.
This result is consistent with that for the 2D $d$-wave case 
\cite{Vekhter,Tanuma2002a}.
However, in the case of 3D chiral superconductors with 
$\Delta_\mathrm{3d}^{\nu=1,2}$, this does not hold.
\begin{table}[htbp]
   \caption{\label{tab:table2}%
      $\gamma$ at which $\sigma(eV=0)$ becomes maximum as a function of $\gamma$
      with $H/H_0=0.1$.
   }
   \begin{ruledtabular}
      \begin{tabular}{ccccc}
         &\multicolumn{4}{c}{$\alpha$}\\
         \cline{2-5}
         $\Delta$ & $\pi/8$ & $\pi/4$ & $3\pi/8$ & $\pi/2$\\
         \hline
         $k'_z$ &  $\pi/2$,$3\pi/2$ &$\pi/2$, $3\pi/2$ &$\pi/2$, $3\pi/2$ &$0$, $\pi$ \\
         $k'_z(k'_x+ik'_y)$ &$\pi$ &$\pi$ &$0$ &$0$ \\
         $(k'_x+ik'_y)$&$\pi$ &0 &$\pi$ &0 \\
         $k'_z(k'_x+ik'_y)^2$&$\pi$ &$0$ &$0$ &$\pi$ \\
         $(k'_x+ik'_y)^2$&$\sim0$ &$\sim0$ &$\pi$ &$\pi$ \\
      \end{tabular}
   \end{ruledtabular}
\end{table}

\subsection{\label{sec:UPt3}%
   \texorpdfstring{%
Symmetry of pairing potential of UPt$_3$}{symmetry of pairing potential for UPt3}}
Recently, 
the guide to determine the pairing symmetry of UPt$_3$ by using 
quasiparticle interference in a slab model was 
theoretically proposed \cite{Eremin2016}.
However, the role of the SABS in determining the charge transport in junctions 
has not yet been revealed. 
We also propose a way to determine the pairing symmetry by using Doppler shift.
In this subsection, we consider $\Delta_{3\mathrm{d}}^{\nu=2}$, 
$\Delta_{E_{1u}}^\mathrm{chiral}$, and 
$\Delta_{E_{1u}}^\mathrm{planar}$ as candidates of the pairing symmetry.
   Crystal symmetry
      of UPt$_3$ is $P6_3/mmc$, so the pair potential around the 
      $\Gamma$ point respects $D_{6h}$. 
      In addition, various experiments \cite{Joynt} indicate a spin-triplet paring 
      and coexistence of point and line nodes.
      $\Delta_{3\mathrm{d}}^{\nu=2}$ and $\Delta_{E_{1u}}^{\rm planar}$ 
      satisfy these properties and are possible candidates for the pairing 
      symmetry of UPt$_3$ \cite{Mizushima2016}.
$\Delta_{E_{1u}}^\mathrm{chiral}$ does not have time reversal symmetry and 
$\Delta_{E_{1u}}^\mathrm{planar}$ has time reversal symmetry.
For $\Delta_{E_{1u}}^\mathrm{planar}$, conductance is given by
Eq.~(\ref{eq:sigma_planar}).

\begin{figure*}[htbp]
   \centering
   \includegraphics[width=17cm,bb=0 0 1500 1125]{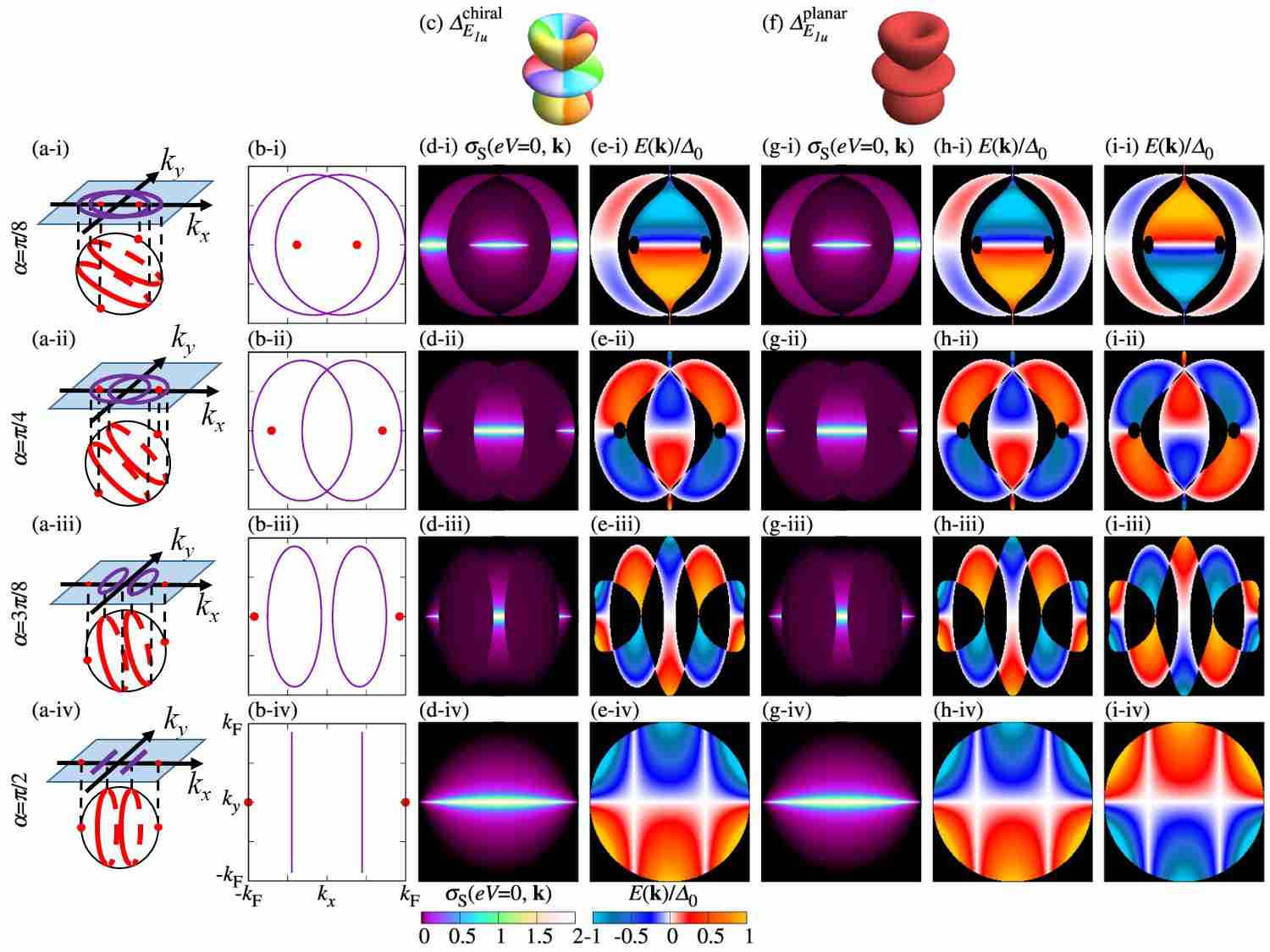}
   \caption{%
      Schematic illustration of point nodes (red dots) and line nodes 
      (red lines)
      for (a-i) $\alpha=\pi/8$,
      (a-ii) $\alpha=\pi/4$, (a-iii) $\alpha=3\pi/8$ and (a-iv) $\alpha=\pi/2$.
      Point nodes and line nodes projected on $k_x-k_y$ plane corresponding to 
      (a-i)--(a-iv) are shown in (b-i)--(b-iv).
      Schematic picture of the pair potential for 
      (c) $\Delta_{E_{1u}}^\mathrm{chiral}$ 
      and (f) $\Delta_{E_{1u}}^\mathrm{planar}$. 
      Angle-resolved zero bias conductance $\sigma_\mathrm{S}(eV=0,\mathbf{k}_\parallel)$
      with $Z=6$ is plotted as functions of $k_x$ and $k_y$ 
      for $\Delta_{E_{1u}}^\mathrm{chiral}$ [(d-i)--(d-iv)] 
      and for $\Delta_{E_{1u}}^\mathrm{planar}$ [(g-i)--(g-iv)].
      The energy dispersion of SABS $E(\mathbf{k}_\parallel)$ for given 
      $\alpha$ are plotted as functions 
      of $k_x$ and $k_y$
      for $\Delta_{E_{1u}}^\mathrm{chiral}$ [(e-i)--(e-iv)]
      and for $\Delta_{E_{1u}}^\mathrm{planar}$ 
      [(h-i) and (i-i)]--[(h-iv) and (i-iv)].
   }
   \label{fig:E1u_both}
\end{figure*}

Firstly, let us discuss the SABS of $\Delta_{E_{1u}}^\mathrm{chiral}$ and 
$\Delta_{E_{1u}}^\mathrm{planar}$ (Fig.~\ref{fig:E1u_both}).
The position of point and line nodes are shown on the Fermi surface 
for 
$\alpha=\pi/8$ [Fig.~\ref{fig:E1u_both} (a-i)], $\alpha=\pi/4$ (a-ii), 
$\alpha=3\pi/8$ (a-iii), and $\alpha=\pi/2$ (a-iv). 
Corresponding projected point and line nodes on the $k_{x}-k_{y}$ plane are 
replotted from Fig.~\ref{fig:E1u_both} (b-i) to (b-iv). 
In the case of $\Delta_{E_{1u}}^\mathrm{planar}$, 
as explained in Sec.~\ref{sec:44},
the energy dispersion of the SABS
for $\Delta_{E_{1u}}^\mathrm{planar}$ 
is 
$\pm E^\mathrm{chiral}(\mathbf{k})$.
For example, Fig.~\ref{fig:E1u_both} (e-i) is the same as (h-i) and 
Fig.~\ref{fig:E1u_both} (h-i) and (i-i) are the energy dispersion of the SABS for 
$\Delta_{E_{1u}}^\mathrm{planar}$. 
Owing to the presence of two line nodes, there is no SABS at $\alpha=0$ for either 
$\Delta_{E_{1u}}^\mathrm{chiral}$ or $\Delta_{E_{1u}}^\mathrm{planar}$ 
(not shown).
For $\alpha>0$, ZESABS of 
$\Delta_{E_{1u}}^\mathrm{chiral}$ [Fig.~\ref{fig:E1u_both} (d-i)--(d-iv)] 
and that of $\Delta_{E_{1u}}^\mathrm{planar}$ [Fig.~\ref{fig:E1u_both} (g-i)--(g-iv)] 
are the same because $S_1=S_2$ [Eq.~(\ref{eq:A}) and Eq.~(\ref{eq:B})].
The number of ZESABSs is discussed in Appendix~\ref{sec:App_top1}.

The energy dispersion of the SABS becomes very complicated as seen from 
Fig.~\ref{fig:E1u_both} (e-i) to (e-iv) for $\Delta_{E_{1u}}^\mathrm{chiral}$ 
and from [(h-i) and (i-i)] to [(h-iv) and (i-iv)] for 
$\Delta_{E_{1u}}^\mathrm{planar}$. 
This is due to the presence of two line nodes and two point nodes. 
The energy dispersion of the SABS at $\alpha=\pi/2$ is summarized in 
Table~\ref{tab:Andreev_E1u}.
\begin{table}[htbp]
   \caption{\label{tab:Andreev_E1u}%
      The energy dispersion of SABS $E(\mathbf{k}_\parallel)r_{E_{1u}}/\Delta_0$ at $\alpha=\pi/2$ 
      is shown for $\Delta_{E_{1u}}^\mathrm{chiral}(\mathbf{k})$ and
      $\Delta_{E_{1u}}^\mathrm{planar}(\mathbf{k})$.
      $\hat{k}_i=k_i/k_\mathrm{F}(i=x,y,z)$.
   }
   \begin{ruledtabular}
      \begin{tabular}{cc}
         $\Delta(\mathbf{k})$ & $E(\mathbf{k}_\parallel)r_{E_{1u}}/\Delta_0$ \\
         \hline
         $\Delta_{E_{1u}}^\mathrm{chiral}(\mathbf{k})$ &
         $-\hat{k}_y|5\hat{k}_x^2-1|$
         \\
         $\Delta_{E_{1u}}^\mathrm{planar}(\mathbf{k})$ &
         $\pm\hat{k}_y|5\hat{k}_x^2-1|$
      \end{tabular}
   \end{ruledtabular}
\end{table}

\begin{figure}[htbp]
   \centering
   \includegraphics[width=8.5cm,bb=0 0 360 237]{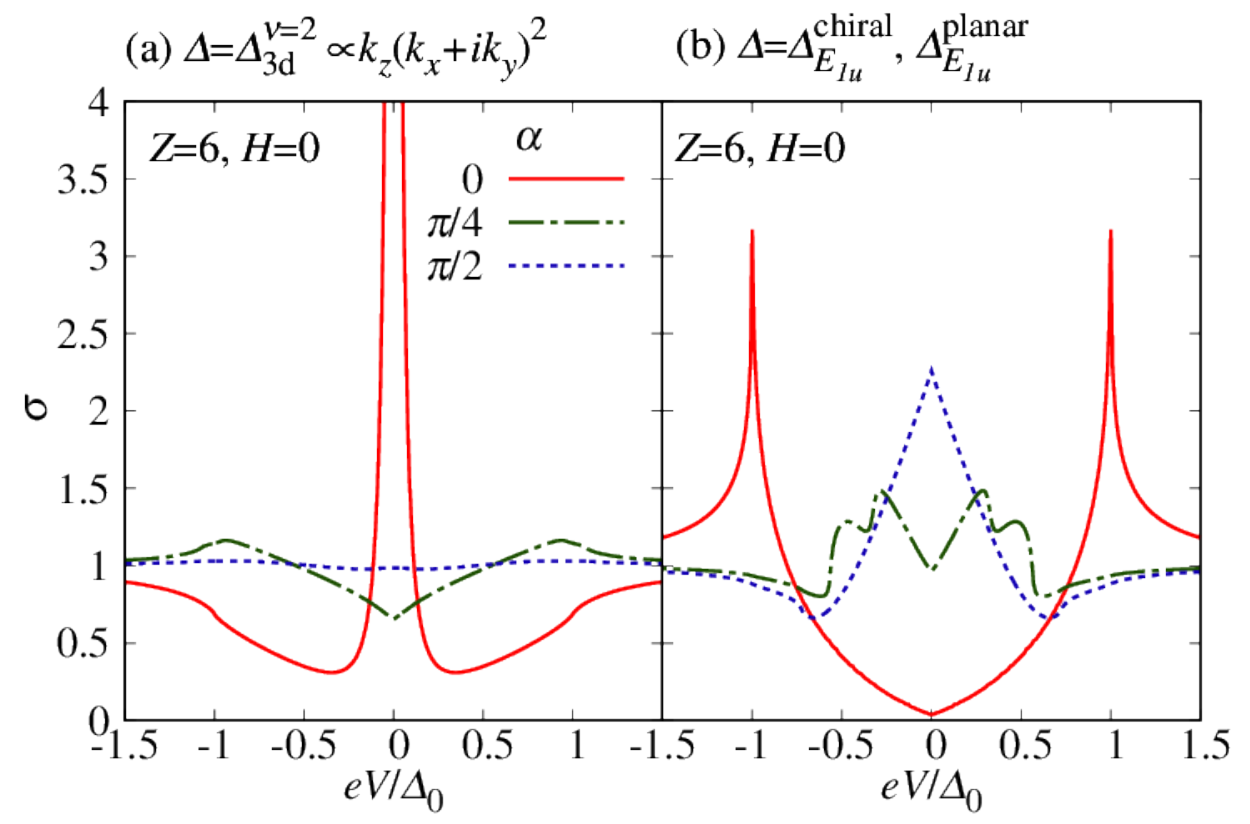}
   \caption{%
      Normalized conductance for spin-triplet $f$-wave symmetry as a function of $eV$ for 
      $\alpha=0$ (solid line), $\alpha=\pi/4$ (dash-dotted line), and 
      $\alpha=\pi/2$ (dotted line).
      (a) $\Delta_{3\mathrm{d}}^{\nu=2}$ and 
      (b) $\Delta_{E_{1u}}^\mathrm{chiral}$,
      $\sigma$ of $\Delta_{E_{1u}}^\mathrm{planar}$ coincides with that of
      $\Delta_{E_{1u}}^\mathrm{chiral}$.
   }
   \label{fig:int_f}
\end{figure}

The conductance for $\Delta_{3\mathrm{d}}^{\nu=2}$, 
$\Delta_{E_{1u}}^\mathrm{chiral}$, and 
$\Delta_{E_{1u}}^\mathrm{planar}$ is shown in Fig.~\ref{fig:int_f}.
Although the angle-resolved conductance ($\sigma_\mathrm{S}$) of 
$\Delta_{E_{1u}}^\mathrm{planar}$ has a counter-propagating mode in addition to 
the same chiral edge mode shown in $\Delta_{E_{1u}}^\mathrm{chiral}$, 
the angular averaged conductance ($\sigma$) is the same [Fig.~\ref{fig:int_f} (b)].
Therefore, we can distinguish $\Delta_{3\mathrm{d}}^{\nu=2}$ from 
($\Delta_{E_{1u}}^\mathrm{chiral}$,
$\Delta_{E_{1u}}^\mathrm{planar}$) by conductance but 
we cannot distinguish between $\Delta_{E_{1u}}^\mathrm{chiral}$ and
$\Delta_{E_{1u}}^\mathrm{planar}$.

\begin{figure}[htbp]
   \centering
   \includegraphics[width=8.5cm,bb=0 0 360 180]{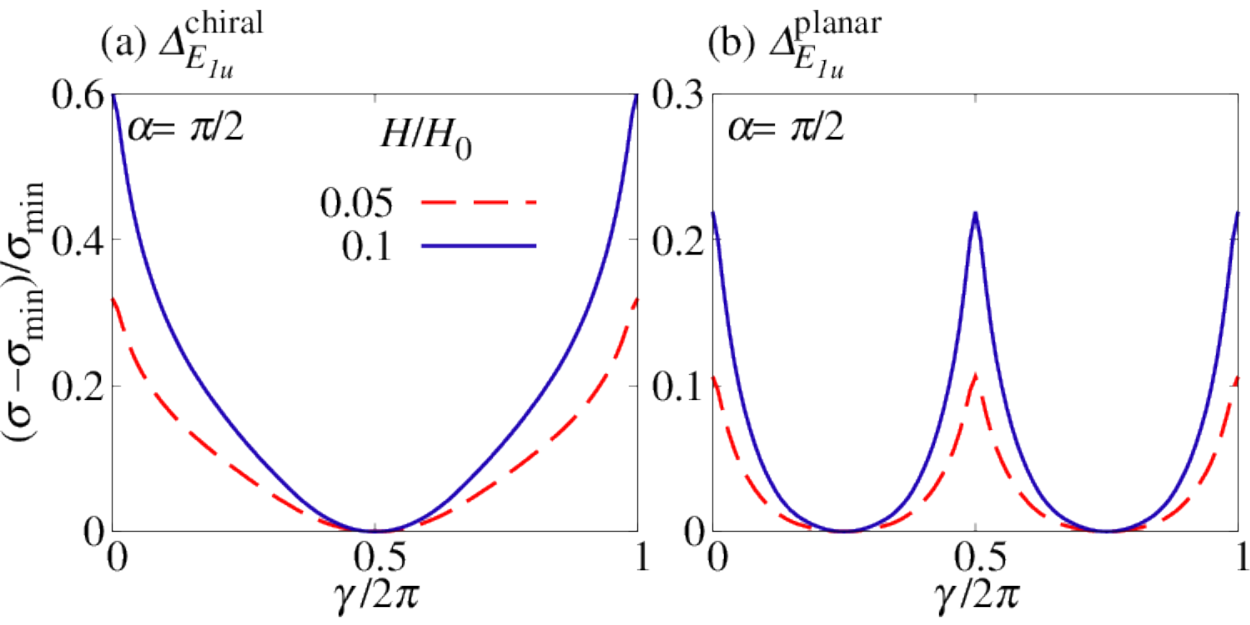}
   \caption{%
      Normalized conductance at $Z=6$ as a function of $\gamma$ 
      with $\alpha=\pi/2$.
      $H/H_0=0.1$ (solid line) and 0.05 (dashed line). 
      (a) $\Delta_{E_{1u}}^\mathrm{chiral}$ 
      and (b) $\Delta_{E_{1u}}^\mathrm{planar}$. 
   }
   \label{fig:int_E12}
\end{figure}

To distinguish $\Delta_{E_{1u}}^\mathrm{chiral}$ from 
$\Delta_{E_{1u}}^\mathrm{planar}$, we discuss the directional dependence 
of the magnetic field on conductance (Fig.~\ref{fig:int_E12}).
Because $\Delta_{E_{1u}}^\mathrm{chiral}$ ($\Delta_{E_{1u}}^\mathrm{planar}$) 
does not have (has) time reversal symmetry, $\sigma(eV=0)$ has $2\pi$ ($\pi$) 
periodicity.
The difference in conductance as a function of $eV$ with $H=0$ and that of 
$\gamma$ with the magnetic field between $\Delta_\mathrm{3d}^{\nu=2}$,
$\Delta_{E_{1u}}^\mathrm{chiral}$ and $\Delta_{E_{1u}}^\mathrm{planar}$ 
are summarized in Table~\ref{tab:UPt_3_dist}.
\begin{table}[htbp]
   \caption{\label{tab:UPt_3_dist}%
      Position of conductance peak with $(\alpha,H)=(0,0)$ 
      (Fig.~\ref{fig:int_f}) and period of $\sigma(eV=0)$ as a function of 
      $\gamma$ with $\alpha=\pi/2$ in the presence of the magnetic field
      [Fig.~\ref{fig:int7_1} (l) and Fig.~\ref{fig:int_E12}].
   }
   \begin{ruledtabular}
      \begin{tabular}{cccc}
         & $\Delta_\mathrm{3d}^{\nu=2}$ 
         & $\Delta_{E_{1u}}^\mathrm{chiral}$ 
         & $\Delta_{E_{1u}}^\mathrm{planar}$
         \\
         \hline
         Conductance peak & $eV=0$ & $eV=\Delta_0$ & $eV=\Delta_0$
         \\
         Period of $\sigma(\gamma)$ & $2\pi$ & $2\pi$ & $\pi$
         \\
      \end{tabular}
   \end{ruledtabular}
\end{table}

Whether ZBCP becomes larger or smaller when an infinitesimal magnitude of 
the magnetic field is 
applied is summarized in Table~\ref{tab:table_E1u} and
$\gamma$ which makes conductance largest is summarized in 
Table~\ref{tab:table_E1u_2}.
\begin{table}[htbp]
   \caption{\label{tab:table_E1u}%
      Line shape of $\sigma(eV)$ near $eV=0$ with $H=0$.
      p: peak, d: dip. Whether the magnitude of $\sigma(eV=0)$ 
      is enhanced (suppressed) by the magnetic field indicated by        
      $\uparrow$ ($\downarrow$). 
      We choose the infinitesimal magnitude of the magnetic field as $H/H_0=10^{-4}$.
      The value of $\sigma(eV)$ for $\Delta_{E_{1u}}^\mathrm{planar}$
      at $\gamma=0$ and $\gamma=\pi$ are equivalent 
      because $\sigma$ is a $\pi$ periodic function of $\gamma$.
   }
   \begin{ruledtabular}
      \begin{tabular}{cccccccc}
         &&&\multicolumn{5}{c}{$\alpha$}\\
         \cline{4-8}
         $\Delta(\mathbf{k})$ & $H/H_0$ & $\gamma$ & $0$ 
                              & $\pi/8$ & $\pi/4$ & $3\pi/8$ & $\pi/2$
         \\
         \hline
         $\Delta_{E_{1u}}^\mathrm{chiral}$ &0 & - & d & d & d & d & p
         \\
         & $10^{-4}$ & $0$   & $\uparrow$ 
         &  $\downarrow$ &  $\uparrow$ &  $\uparrow$   &  $\uparrow$
         \\
         & $10^{-4}$ &$\pi$ & $\uparrow$ &  $\uparrow$   
         &  $\uparrow$ &  $\downarrow$ &  $\downarrow$
         \\
         $\Delta_{E_{1u}}^\mathrm{planar}$ & 0 & - & d & d & d & d & p
         \\
         & $10^{-4}$ &$0, \pi$   & $\uparrow$ &  $\uparrow$ 
         &  $\uparrow$ &  $\uparrow$ &  $\uparrow$
         \\
      \end{tabular}
   \end{ruledtabular}
\end{table}
\begin{table}[htbp]
   \caption{\label{tab:table_E1u_2}%
      The position of $\gamma$ at which $\sigma(eV=0)$ 
      becomes maximum as a function of $\gamma$.
   }
   \begin{ruledtabular}
      \begin{tabular}{ccccc}
         &\multicolumn{4}{c}{$\alpha$}\\
         \cline{2-5}
         $\Delta$ & $\pi/8$ & $\pi/4$ & $3\pi/8$ & $\pi/2$
         \\
         \hline
         $\Delta_{E_{1u}}^\mathrm{chiral}$ & $\pi$ & 0 & 0 & 0
         \\
         $\Delta_{E_{1u}}^\mathrm{planar}$ & 0,$\pi$ 
                                           & 0,$\pi$ & 0,$\pi$ & 0,$\pi$
         \\
      \end{tabular}
   \end{ruledtabular}
\end{table}

\begin{figure}[htbp]
   \centering
   \includegraphics[width=8.5cm,bb=0 0 360 273]{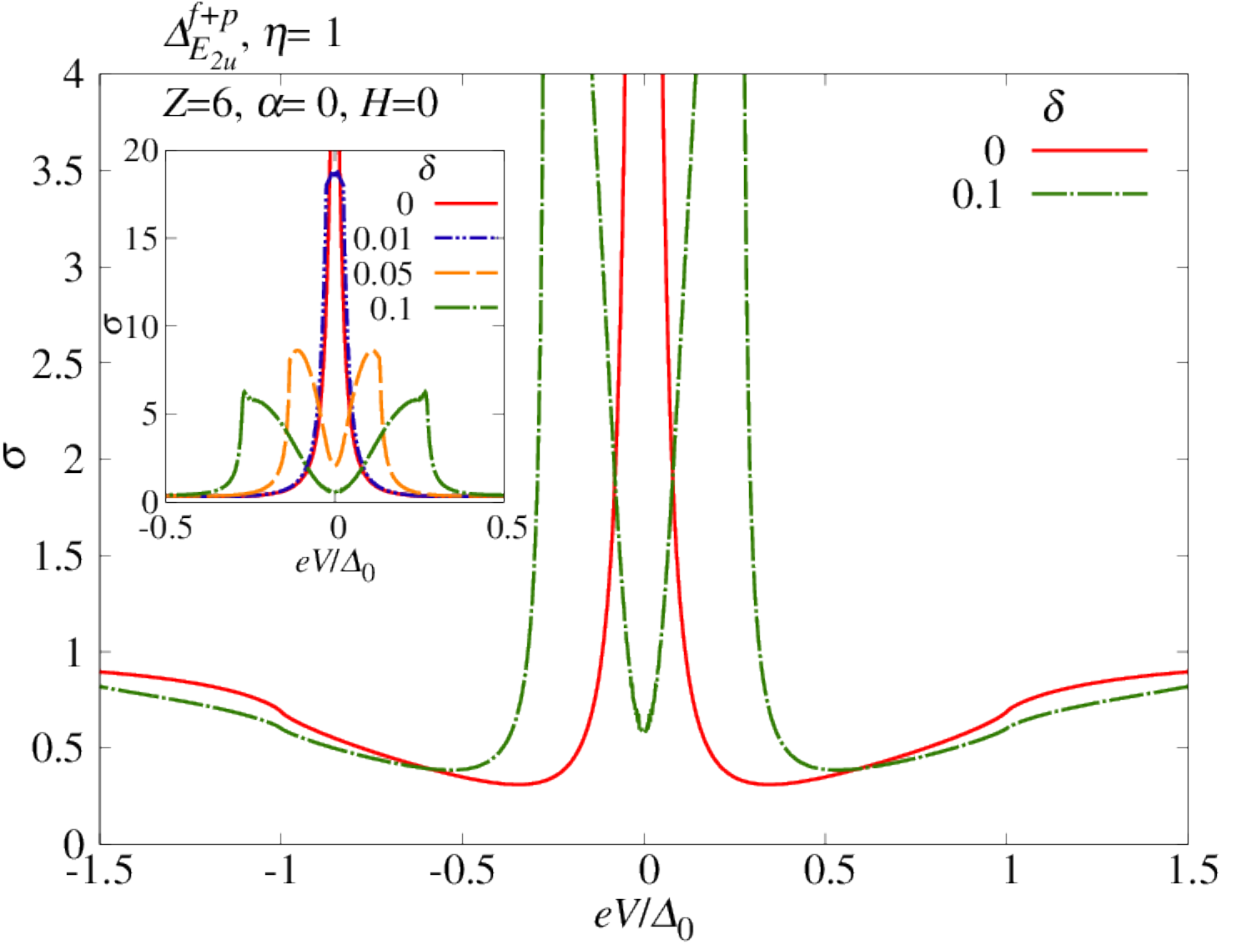}
   \caption{%
      Normalized conductance $\sigma$ by its value in normal state 
      for $\Delta_{E_{2u}}^{f+p}$ is plotted as a function of $eV$ 
      for several $\delta$ with $\eta=1$ and $Z=6$.
      Inset is the magnification near $eV=0$.
   }
   \label{fig:int_f_p}
\end{figure}

\begin{figure}[htbp]
   \centering
   \includegraphics[width=8.5cm,bb=0 0 360 180]{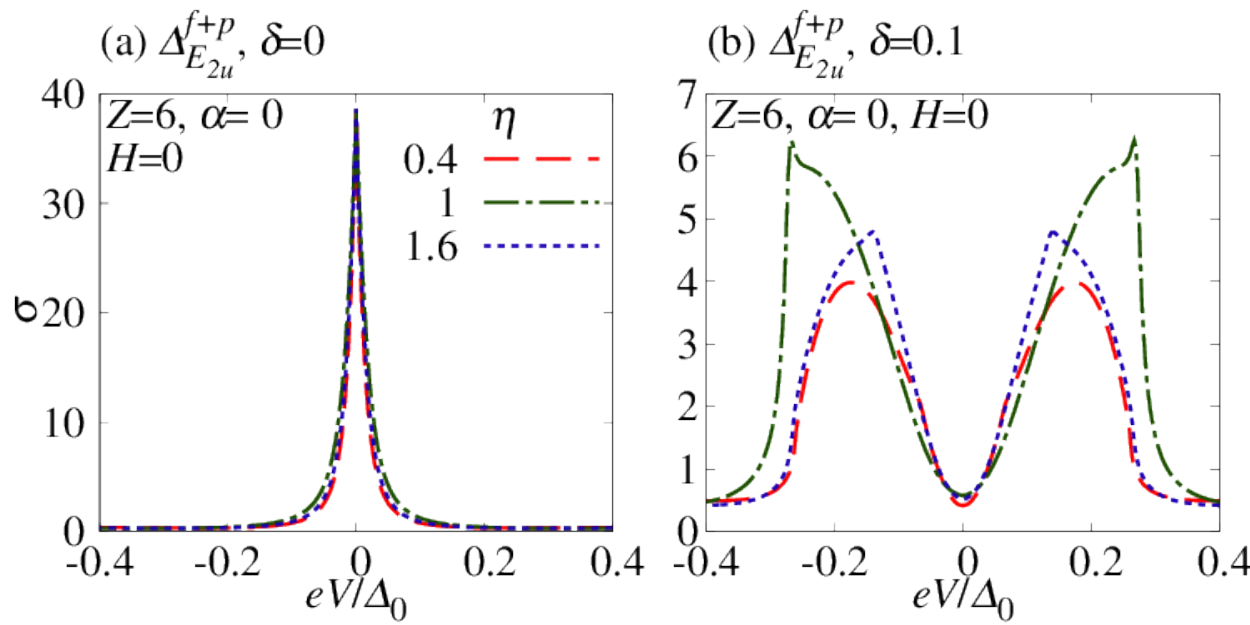}
   \caption{%
      Normalized conductance $\sigma$ by its value in normal state 
      for $\Delta_{E_{2u}}^{f+p}$ is plotted as a function of $eV$ for several $\eta$ with $Z=6$.
      (a) $\delta=0$ and (b) $\delta=0.1$.
   }
   \label{fig:int_f_p_de}
\end{figure}

Recently, 
Y.~Yanase has proposed an extended version of $E_{2u}$ symmetry\cite{Yanase2016} 
as a model of the pairing symmetry 
of UPt$_3$. 
It is a linear combination of the spin-triplet $p$ and $f$-wave pairings.
It is interesting to clarify whether the obtained results in Table~\ref{tab:UPt_3_dist}
are changed by the disappearance of a line node by the mixing of 
a small magnitude of chiral $p$-wave pair potential and
$\Delta_{3\mathrm{d}}^{\nu=2}$.  
We consider the non-unitary pair potential given by Eq.~(\ref{eq:d_nonu}).
For this purpose, we have used a general formula for 
tunneling conductance Eq.~(\ref{eq:cond_gen1}) derived in 
Appendix~\ref{sec:App_cond}. 
For $\eta=1$, the obtained conductance is shown in Fig.~\ref{fig:int_f_p}. 
In this case, there remain two point nodes and a line node
[Fig.~\ref{fig:nodes_f_p_d0} (b-i)--(b-iii) and 
Fig.~\ref{fig:nodes_f_p} (b-i)--(b-iii)]. 
When nonzero $\delta$ is introduced, ZBCP splits, but the shape of conductance 
is still distinct from that of $E_{1u}$ symmetry 
[Fig.~\ref{fig:int_f} (b)].
For fixed $\delta$, the line shape of conductance $\sigma$ does not change 
qualitatively with the change of $\eta$ [Fig.~\ref{fig:int_f_p_de} (a) and (b)]. 
Hence, it is expected that if the spin-triplet $p$-wave pair potential 
is additionally introduced into 
$\Delta_{3\mathrm{d}}^{\nu=2}$, $E_{1u}$ and the extended version of 
$E_{2u}$ can be
distinguished by tunneling conductance.

\section{\label{sec:conclusion}Discussion and conclusion}
In this paper, we have studied the surface Andreev bound state (SABS) 
and quasiparticle tunneling spectroscopy of 
three-dimensional (3D) chiral superconductors 
by changing the misorientation angle of 
superconductors. 
We have analytically derived a  
formula of the energy dispersion of SABS available for general 
pair potentials when an original $4\times 4$ matrix of 
BdG Hamiltonian can be decomposed into two blocks of 
$2 \times 2$ matrices. 
We apply this formula to calculate the SABS 
for 3D chiral superconductors, where the pair potential is given by 
$\Delta_{0}k_{z}(k_{x} + i k_{y})^{\nu}/k_\mathrm{F}^{\nu+1}$ 
($\nu=1,2$). 
The SABS has a complex momentum dependence, 
owing to the coexistence of point and line nodes. 
The number of branches of the energy dispersion of 
SABS with topological origin can be 
understood based on the winding and Chern numbers. 
We have calculated the tunneling conductance of 
normal metal / insulator / chiral superconductor junctions 
in the presence of the applied magnetic field, which induces 
a Doppler shift. 
In contrast to previous studies of Doppler effect on 
tunneling conductance, zero bias conductance dip can change into 
zero bias conductance peak by applying the magnetic field. 
This unique feature originates from 
the complicated energy dispersion of the SABS\@. 
We have also studied the SABS and tunneling conductance of 
UPt$_{3}$ focusing on four possible candidates of the 
pairing symmetry:
$E_{2u}$, $E_{1u}$-planar, $E_{1u}$-chiral,  
and extended version of $E_{2u}$ pairings. 
Since the last pairing is non-unitary, we have developed a 
conductance formula, which is available for a general pair potential. 
By using this formula, we have shown that 
these four parings can be identified by 
tunneling spectroscopy both with and without magnetic field. 
Thus, our theory serves as a guide to determine the 
pairing symmetry of UPt$_{3}$. \par
In this paper, we are focusing on SABS and quasiparticle tunneling 
spectroscopy. As a next step, it will be interesting to calculate the Josephson effect 
in chiral superconductors, including spin-triplet 
non-unitary pairings \cite{Asano2001},  
where both point and line nodes exist. 
Especially, it is known from the studies of $d$-wave 
superconductors, the role of $\alpha$ induces non-monotonic temperature 
dependence of maximum Josephson current 
\cite{Josephsond,Josephson3,tanaka97,Barash}. 
The study of such a kind of exotic temperature dependence of the Josephson 
current will be really interesting.

In this paper, we have studied the case where a normal metal is ballistic. 
It is a challenging issue to extend our theory to 
diffusive normal metal (DN) / superconductor junctions 
where penetration of the Cooper pair owing to the proximity effect 
modifies the total resistance of the junctions \cite{Proximityd,Proximityd2}. 
In particular, it is known that the anomalous proximity effect with 
zero energy peak of the local density of states in DN 
via odd-frequency pairing occurs in spin-triplet superconductor junctions 
\cite{Proximityp,Proximityp2,Proximityp3,Meissner3,odd1,odd3,odd3b}. 
An extension of previous two-dimensional studies to three dimensions 
is also promising.\

   In the remainder, we discuss points related to the conductance formula
   [Eq.~(\ref{eq:cond_gen1}) or Eq.~(\ref{eq:cond_gen})].

   (i) Doppler shift:
   Although the Doppler shift approximation is not quantitatively perfect, 
   it is useful for the classification of pairing symmetry \cite{Tanuma2002b}.
   When the penetration depth is much larger than 
   the coherence length of superconductor, this approximation does work as far as 
   we are discussing surface Andreev bound states and 
   innergap tunneling conductance.
   This approximation has been actually done in the previous work 
   in Eilenberger equation \cite{Tanaka02a}
   and qualitatively reasonable results are obtained.
   Also, these results are qualitatively same as the results obtained by 
   extended version of BTK theory\cite{Tanaka2002}.

   (ii) Isotropic Fermi surface:
      In the point of view of the relation between topological invariant and surface Andreev
      bound state (SABS), the essential feature of SABS is determined 
      by the node structure of pair potential in momentum space and 
      symmetry of Hamiltonian
      when the magnitude of the pair potential is much smaller than Fermi energy.
      Thus, the qualitative nature of SABS is not so sensitive to the band structure. 
      In the case that an actual Fermi surface is not topologically equivalent 
      to the single isotropic Fermi surface, we must calculate SABS with 
      the Fermi surface.
      For UPt$_3$, as far as superconducting pairing is formed on the Fermi surface
      nearest to $\Gamma$ point, it is expected that the qualitative shape of SABS 
      in our paper can be compared to experimental results.\

      On the other hand, line shape of tunneling conductance is more or less 
      influenced by band structures. 
      When the obtained SABS has a flat band dispersion, 
      the tunneling conductance has a  zero bias conductance peak (ZBCP) shown 
      by the previous studies of tight binding model \cite{Tanuma1997}. 
      On the other hand, when the SABS has a linear dispersion like chiral 
      $p$-wave superconductor, the resulting line shape of tunneling conductance 
      depends on band structures \cite{Yada2014}. 

      As regards UPt$_{3}$, the energy band structures are complex. 
      However, we think that one can design the experiment to greatly reduce such effect. 
      For example, one can choose a kind of material belonging to UPt$_3$ family or with 
      similar band structures as a normal side. 
      As a result, the tunneling characteristic would depend mainly on the nodal 
      structure of superconducting gap instead of the anisotropy of the continuous energy 
      band.\
      the superconducting gap structure will play a dominant role in the 
      tunneling spectroscopy which this paper is focused on. 

   (iii) Applicability: 
   If we focus on the low energy physics and a effective Hamiltonian
   of a material has a single Fermi surface, the conductance
   formula can be useful to discuss qualitative nature 
   of superconductors.\
   Especially to three dimensional superconductors which have complex nodes,
   for example, heavy Fermion compounds\cite{Schnyder2015}.

\begin{acknowledgments}
The authors are grateful to M. Sato for discussions.
This work was supported
by a Grant-in-Aid for Scientific Research on Innovative
Areas Topological Material Science JPSJ KAKENHI (Grants No. JP15H05851,
and No. JP15H05853), a Grant-in-Aid for
Scientific Research B (Grant No. JP15H03686), a Grant-in-Aid
for Challenging Exploratory Research (Grant No. JP15K13498)
from the Ministry of Education, Culture, Sports, Science, and
Technology, Japan (MEXT).
The work of S.K. was supported by Building of Consortia for the Development 
of Human Resources in Science and Technology and Grant-in-Aid for Research 
Activity Start-up (Grand No. JP16H06861).
\end{acknowledgments}

\bibliography{Manuscript1.bbl}

\begin{thebibliography}{87}%
\makeatletter
\providecommand \@ifxundefined [1]{%
 \@ifx{#1\undefined}
}%
\providecommand \@ifnum [1]{%
 \ifnum #1\expandafter \@firstoftwo
 \else \expandafter \@secondoftwo
 \fi
}%
\providecommand \@ifx [1]{%
 \ifx #1\expandafter \@firstoftwo
 \else \expandafter \@secondoftwo
 \fi
}%
\providecommand \natexlab [1]{#1}%
\providecommand \enquote  [1]{``#1''}%
\providecommand \bibnamefont  [1]{#1}%
\providecommand \bibfnamefont [1]{#1}%
\providecommand \citenamefont [1]{#1}%
\providecommand \href@noop [0]{\@secondoftwo}%
\providecommand \href [0]{\begingroup \@sanitize@url \@href}%
\providecommand \@href[1]{\@@startlink{#1}\@@href}%
\providecommand \@@href[1]{\endgroup#1\@@endlink}%
\providecommand \@sanitize@url [0]{\catcode `\\12\catcode `\$12\catcode
  `\&12\catcode `\#12\catcode `\^12\catcode `\_12\catcode `\%12\relax}%
\providecommand \@@startlink[1]{}%
\providecommand \@@endlink[0]{}%
\providecommand \url  [0]{\begingroup\@sanitize@url \@url }%
\providecommand \@url [1]{\endgroup\@href {#1}{\urlprefix }}%
\providecommand \urlprefix  [0]{URL }%
\providecommand \Eprint [0]{\href }%
\providecommand \doibase [0]{http://dx.doi.org/}%
\providecommand \selectlanguage [0]{\@gobble}%
\providecommand \bibinfo  [0]{\@secondoftwo}%
\providecommand \bibfield  [0]{\@secondoftwo}%
\providecommand \translation [1]{[#1]}%
\providecommand \BibitemOpen [0]{}%
\providecommand \bibitemStop [0]{}%
\providecommand \bibitemNoStop [0]{.\EOS\space}%
\providecommand \EOS [0]{\spacefactor3000\relax}%
\providecommand \BibitemShut  [1]{\csname bibitem#1\endcsname}%
\let\auto@bib@innerbib\@empty
\bibitem [{\citenamefont {Kashiwaya}\ and\ \citenamefont
  {Tanaka}(2000)}]{kashiwaya00}%
  \BibitemOpen
  \bibfield  {author} {\bibinfo {author} {\bibfnamefont {S.}~\bibnamefont
  {Kashiwaya}}\ and\ \bibinfo {author} {\bibfnamefont {Y.}~\bibnamefont
  {Tanaka}},\ }\href {http://stacks.iop.org/0034-4885/63/i=10/a=202} {\bibfield
   {journal} {\bibinfo  {journal} {Reports on Progress in Physics}\ }\textbf
  {\bibinfo {volume} {63}},\ \bibinfo {pages} {1641} (\bibinfo {year}
  {2000})}\BibitemShut {NoStop}%
\bibitem [{\citenamefont {L\"ofwander}\ \emph {et~al.}(2001)\citenamefont
  {L\"ofwander}, \citenamefont {Shumeiko},\ and\ \citenamefont
  {Wendin}}]{Lofwander}%
  \BibitemOpen
  \bibfield  {author} {\bibinfo {author} {\bibfnamefont {T.}~\bibnamefont
  {L\"ofwander}}, \bibinfo {author} {\bibfnamefont {V.~S.}\ \bibnamefont
  {Shumeiko}}, \ and\ \bibinfo {author} {\bibfnamefont {G.}~\bibnamefont
  {Wendin}},\ }\href {http://stacks.iop.org/0953-2048/14/i=5/a=201} {\bibfield
  {journal} {\bibinfo  {journal} {Supercond. Sci. Tech.}\ }\textbf {\bibinfo
  {volume} {14}},\ \bibinfo {pages} {R53} (\bibinfo {year} {2001})}\BibitemShut
  {NoStop}%
\bibitem [{\citenamefont {Golubov}\ \emph {et~al.}(2004)\citenamefont
  {Golubov}, \citenamefont {Kupriyanov},\ and\ \citenamefont
  {Il'ichev}}]{Golubov_RMP}%
  \BibitemOpen
  \bibfield  {author} {\bibinfo {author} {\bibfnamefont {A.~A.}\ \bibnamefont
  {Golubov}}, \bibinfo {author} {\bibfnamefont {M.~Y.}\ \bibnamefont
  {Kupriyanov}}, \ and\ \bibinfo {author} {\bibfnamefont {E.}~\bibnamefont
  {Il'ichev}},\ }\href {\doibase 10.1103/RevModPhys.76.411} {\bibfield
  {journal} {\bibinfo  {journal} {Rev. Mod. Phys.}\ }\textbf {\bibinfo {volume}
  {76}},\ \bibinfo {pages} {411} (\bibinfo {year} {2004})}\BibitemShut
  {NoStop}%
\bibitem [{\citenamefont {Tanaka}\ \emph {et~al.}(2012)\citenamefont {Tanaka},
  \citenamefont {Sato},\ and\ \citenamefont {Nagaosa}}]{tanaka12}%
  \BibitemOpen
  \bibfield  {author} {\bibinfo {author} {\bibfnamefont {Y.}~\bibnamefont
  {Tanaka}}, \bibinfo {author} {\bibfnamefont {M.}~\bibnamefont {Sato}}, \ and\
  \bibinfo {author} {\bibfnamefont {N.}~\bibnamefont {Nagaosa}},\ }\href
  {\doibase 10.1143/JPSJ.81.011013} {\bibfield  {journal} {\bibinfo  {journal}
  {J. Phys. Soc. Jpn.}\ }\textbf {\bibinfo {volume} {81}},\ \bibinfo {pages}
  {011013} (\bibinfo {year} {2012})}\BibitemShut {NoStop}%
\bibitem [{\citenamefont {Qi}\ and\ \citenamefont {Zhang}(2011)}]{qi11}%
  \BibitemOpen
  \bibfield  {author} {\bibinfo {author} {\bibfnamefont {X.-L.}\ \bibnamefont
  {Qi}}\ and\ \bibinfo {author} {\bibfnamefont {S.-C.}\ \bibnamefont {Zhang}},\
  }\href {\doibase 10.1103/RevModPhys.83.1057} {\bibfield  {journal} {\bibinfo
  {journal} {Rev. Mod. Phys.}\ }\textbf {\bibinfo {volume} {83}},\ \bibinfo
  {pages} {1057} (\bibinfo {year} {2011})}\BibitemShut {NoStop}%
\bibitem [{\citenamefont {Hu}(1994)}]{Hu}%
  \BibitemOpen
  \bibfield  {author} {\bibinfo {author} {\bibfnamefont {C.-R.}\ \bibnamefont
  {Hu}},\ }\href {\doibase 10.1103/PhysRevLett.72.1526} {\bibfield  {journal}
  {\bibinfo  {journal} {Phys. Rev. Lett.}\ }\textbf {\bibinfo {volume} {72}},\
  \bibinfo {pages} {1526} (\bibinfo {year} {1994})}\BibitemShut {NoStop}%
\bibitem [{\citenamefont {Ryu}\ and\ \citenamefont {Hatsugai}(2002)}]{RH02}%
  \BibitemOpen
  \bibfield  {author} {\bibinfo {author} {\bibfnamefont {S.}~\bibnamefont
  {Ryu}}\ and\ \bibinfo {author} {\bibfnamefont {Y.}~\bibnamefont {Hatsugai}},\
  }\href {\doibase 10.1103/PhysRevLett.89.077002} {\bibfield  {journal}
  {\bibinfo  {journal} {Phys. Rev. Lett.}\ }\textbf {\bibinfo {volume} {89}},\
  \bibinfo {pages} {077002} (\bibinfo {year} {2002})}\BibitemShut {NoStop}%
\bibitem [{\citenamefont {Sato}\ \emph
  {et~al.}(2011{\natexlab{a}})\citenamefont {Sato}, \citenamefont {Tanaka},
  \citenamefont {Yada},\ and\ \citenamefont {Yokoyama}}]{index}%
  \BibitemOpen
  \bibfield  {author} {\bibinfo {author} {\bibfnamefont {M.}~\bibnamefont
  {Sato}}, \bibinfo {author} {\bibfnamefont {Y.}~\bibnamefont {Tanaka}},
  \bibinfo {author} {\bibfnamefont {K.}~\bibnamefont {Yada}}, \ and\ \bibinfo
  {author} {\bibfnamefont {T.}~\bibnamefont {Yokoyama}},\ }\href {\doibase
  10.1103/PhysRevB.83.224511} {\bibfield  {journal} {\bibinfo  {journal} {Phys.
  Rev. B}\ }\textbf {\bibinfo {volume} {83}},\ \bibinfo {pages} {224511}
  (\bibinfo {year} {2011}{\natexlab{a}})}\BibitemShut {NoStop}%
\bibitem [{\citenamefont {Sato}\ and\ \citenamefont
  {Fujimoto}(2016)}]{Sato2016}%
  \BibitemOpen
  \bibfield  {author} {\bibinfo {author} {\bibfnamefont {M.}~\bibnamefont
  {Sato}}\ and\ \bibinfo {author} {\bibfnamefont {S.}~\bibnamefont
  {Fujimoto}},\ }\href {\doibase 10.7566/JPSJ.85.072001} {\bibfield  {journal}
  {\bibinfo  {journal} {Journal of the Physical Society of Japan}\ }\textbf
  {\bibinfo {volume} {85}},\ \bibinfo {pages} {072001} (\bibinfo {year}
  {2016})},\ \Eprint
  {http://arxiv.org/abs/http://dx.doi.org/10.7566/JPSJ.85.072001}
  {http://dx.doi.org/10.7566/JPSJ.85.072001} \BibitemShut {NoStop}%
\bibitem [{\citenamefont {Tanaka}\ and\ \citenamefont
  {Kashiwaya}(1995)}]{TK95}%
  \BibitemOpen
  \bibfield  {author} {\bibinfo {author} {\bibfnamefont {Y.}~\bibnamefont
  {Tanaka}}\ and\ \bibinfo {author} {\bibfnamefont {S.}~\bibnamefont
  {Kashiwaya}},\ }\href {\doibase 10.1103/PhysRevLett.74.3451} {\bibfield
  {journal} {\bibinfo  {journal} {Phys. Rev. Lett.}\ }\textbf {\bibinfo
  {volume} {74}},\ \bibinfo {pages} {3451} (\bibinfo {year}
  {1995})}\BibitemShut {NoStop}%
\bibitem [{\citenamefont {Kashiwaya}\ \emph
  {et~al.}(1995{\natexlab{a}})\citenamefont {Kashiwaya}, \citenamefont
  {Tanaka}, \citenamefont {Koyanagi}, \citenamefont {Takashima},\ and\
  \citenamefont {Kajimura}}]{Experiment1}%
  \BibitemOpen
  \bibfield  {author} {\bibinfo {author} {\bibfnamefont {S.}~\bibnamefont
  {Kashiwaya}}, \bibinfo {author} {\bibfnamefont {Y.}~\bibnamefont {Tanaka}},
  \bibinfo {author} {\bibfnamefont {M.}~\bibnamefont {Koyanagi}}, \bibinfo
  {author} {\bibfnamefont {H.}~\bibnamefont {Takashima}}, \ and\ \bibinfo
  {author} {\bibfnamefont {K.}~\bibnamefont {Kajimura}},\ }\href {\doibase
  10.1103/PhysRevB.51.1350} {\bibfield  {journal} {\bibinfo  {journal} {Phys.
  Rev. B}\ }\textbf {\bibinfo {volume} {51}},\ \bibinfo {pages} {1350}
  (\bibinfo {year} {1995}{\natexlab{a}})}\BibitemShut {NoStop}%
\bibitem [{\citenamefont {Kashiwaya}\ \emph {et~al.}(1998)\citenamefont
  {Kashiwaya}, \citenamefont {Tanaka}, \citenamefont {Terada}, \citenamefont
  {Koyanagi}, \citenamefont {Ueno}, \citenamefont {Alff}, \citenamefont
  {Takashima}, \citenamefont {Tanuma},\ and\ \citenamefont
  {Kajimura}}]{Experiment2}%
  \BibitemOpen
  \bibfield  {author} {\bibinfo {author} {\bibfnamefont {S.}~\bibnamefont
  {Kashiwaya}}, \bibinfo {author} {\bibfnamefont {Y.}~\bibnamefont {Tanaka}},
  \bibinfo {author} {\bibfnamefont {N.}~\bibnamefont {Terada}}, \bibinfo
  {author} {\bibfnamefont {M.}~\bibnamefont {Koyanagi}}, \bibinfo {author}
  {\bibfnamefont {S.}~\bibnamefont {Ueno}}, \bibinfo {author} {\bibfnamefont
  {L.}~\bibnamefont {Alff}}, \bibinfo {author} {\bibfnamefont {H.}~\bibnamefont
  {Takashima}}, \bibinfo {author} {\bibfnamefont {Y.}~\bibnamefont {Tanuma}}, \
  and\ \bibinfo {author} {\bibfnamefont {K.}~\bibnamefont {Kajimura}},\ }\href
  {\doibase http://dx.doi.org/10.1016/S0022-3697(98)00174-7} {\bibfield
  {journal} {\bibinfo  {journal} {Journal of Physics and Chemistry of Solids}\
  }\textbf {\bibinfo {volume} {59}},\ \bibinfo {pages} {2034 } (\bibinfo {year}
  {1998})}\BibitemShut {NoStop}%
\bibitem [{\citenamefont {Covington}\ \emph {et~al.}(1997)\citenamefont
  {Covington}, \citenamefont {Aprili}, \citenamefont {Paraoanu}, \citenamefont
  {Greene}, \citenamefont {Xu}, \citenamefont {Zhu},\ and\ \citenamefont
  {Mirkin}}]{Experiment3}%
  \BibitemOpen
  \bibfield  {author} {\bibinfo {author} {\bibfnamefont {M.}~\bibnamefont
  {Covington}}, \bibinfo {author} {\bibfnamefont {M.}~\bibnamefont {Aprili}},
  \bibinfo {author} {\bibfnamefont {E.}~\bibnamefont {Paraoanu}}, \bibinfo
  {author} {\bibfnamefont {L.~H.}\ \bibnamefont {Greene}}, \bibinfo {author}
  {\bibfnamefont {F.}~\bibnamefont {Xu}}, \bibinfo {author} {\bibfnamefont
  {J.}~\bibnamefont {Zhu}}, \ and\ \bibinfo {author} {\bibfnamefont {C.~A.}\
  \bibnamefont {Mirkin}},\ }\href {\doibase 10.1103/PhysRevLett.79.277}
  {\bibfield  {journal} {\bibinfo  {journal} {Phys. Rev. Lett.}\ }\textbf
  {\bibinfo {volume} {79}},\ \bibinfo {pages} {277} (\bibinfo {year}
  {1997})}\BibitemShut {NoStop}%
\bibitem [{\citenamefont {Alff}\ \emph {et~al.}(1997)\citenamefont {Alff},
  \citenamefont {Takashima}, \citenamefont {Kashiwaya}, \citenamefont {Terada},
  \citenamefont {Ihara}, \citenamefont {Tanaka}, \citenamefont {Koyanagi},\
  and\ \citenamefont {Kajimura}}]{Experiment4}%
  \BibitemOpen
  \bibfield  {author} {\bibinfo {author} {\bibfnamefont {L.}~\bibnamefont
  {Alff}}, \bibinfo {author} {\bibfnamefont {H.}~\bibnamefont {Takashima}},
  \bibinfo {author} {\bibfnamefont {S.}~\bibnamefont {Kashiwaya}}, \bibinfo
  {author} {\bibfnamefont {N.}~\bibnamefont {Terada}}, \bibinfo {author}
  {\bibfnamefont {H.}~\bibnamefont {Ihara}}, \bibinfo {author} {\bibfnamefont
  {Y.}~\bibnamefont {Tanaka}}, \bibinfo {author} {\bibfnamefont
  {M.}~\bibnamefont {Koyanagi}}, \ and\ \bibinfo {author} {\bibfnamefont
  {K.}~\bibnamefont {Kajimura}},\ }\href {\doibase 10.1103/PhysRevB.55.R14757}
  {\bibfield  {journal} {\bibinfo  {journal} {Phys. Rev. B}\ }\textbf {\bibinfo
  {volume} {55}},\ \bibinfo {pages} {R14757} (\bibinfo {year}
  {1997})}\BibitemShut {NoStop}%
\bibitem [{\citenamefont {Wei}\ \emph {et~al.}(1998)\citenamefont {Wei},
  \citenamefont {Yeh}, \citenamefont {Garrigus},\ and\ \citenamefont
  {Strasik}}]{Experiment5}%
  \BibitemOpen
  \bibfield  {author} {\bibinfo {author} {\bibfnamefont {J.~Y.~T.}\
  \bibnamefont {Wei}}, \bibinfo {author} {\bibfnamefont {N.-C.}\ \bibnamefont
  {Yeh}}, \bibinfo {author} {\bibfnamefont {D.~F.}\ \bibnamefont {Garrigus}}, \
  and\ \bibinfo {author} {\bibfnamefont {M.}~\bibnamefont {Strasik}},\ }\href
  {\doibase 10.1103/PhysRevLett.81.2542} {\bibfield  {journal} {\bibinfo
  {journal} {Phys. Rev. Lett.}\ }\textbf {\bibinfo {volume} {81}},\ \bibinfo
  {pages} {2542} (\bibinfo {year} {1998})}\BibitemShut {NoStop}%
\bibitem [{\citenamefont {Biswas}\ \emph {et~al.}(2002)\citenamefont {Biswas},
  \citenamefont {Fournier}, \citenamefont {Qazilbash}, \citenamefont
  {Smolyaninova}, \citenamefont {Balci},\ and\ \citenamefont
  {Greene}}]{Experiment6}%
  \BibitemOpen
  \bibfield  {author} {\bibinfo {author} {\bibfnamefont {A.}~\bibnamefont
  {Biswas}}, \bibinfo {author} {\bibfnamefont {P.}~\bibnamefont {Fournier}},
  \bibinfo {author} {\bibfnamefont {M.~M.}\ \bibnamefont {Qazilbash}}, \bibinfo
  {author} {\bibfnamefont {V.~N.}\ \bibnamefont {Smolyaninova}}, \bibinfo
  {author} {\bibfnamefont {H.}~\bibnamefont {Balci}}, \ and\ \bibinfo {author}
  {\bibfnamefont {R.~L.}\ \bibnamefont {Greene}},\ }\href {\doibase
  10.1103/PhysRevLett.88.207004} {\bibfield  {journal} {\bibinfo  {journal}
  {Phys. Rev. Lett.}\ }\textbf {\bibinfo {volume} {88}},\ \bibinfo {pages}
  {207004} (\bibinfo {year} {2002})}\BibitemShut {NoStop}%
\bibitem [{\citenamefont {Buchholtz}\ and\ \citenamefont
  {Zwicknagl}(1981)}]{ABS}%
  \BibitemOpen
  \bibfield  {author} {\bibinfo {author} {\bibfnamefont {L.~J.}\ \bibnamefont
  {Buchholtz}}\ and\ \bibinfo {author} {\bibfnamefont {G.}~\bibnamefont
  {Zwicknagl}},\ }\href {\doibase 10.1103/PhysRevB.23.5788} {\bibfield
  {journal} {\bibinfo  {journal} {Phys. Rev. B}\ }\textbf {\bibinfo {volume}
  {23}},\ \bibinfo {pages} {5788} (\bibinfo {year} {1981})}\BibitemShut
  {NoStop}%
\bibitem [{\citenamefont {Hara}\ and\ \citenamefont {Nagai}(1986)}]{ABSb}%
  \BibitemOpen
  \bibfield  {author} {\bibinfo {author} {\bibfnamefont {J.}~\bibnamefont
  {Hara}}\ and\ \bibinfo {author} {\bibfnamefont {K.}~\bibnamefont {Nagai}},\
  }\href {\doibase 10.1143/PTP.76.1237} {\bibfield  {journal} {\bibinfo
  {journal} {Progress of Theoretical Physics}\ }\textbf {\bibinfo {volume}
  {76}},\ \bibinfo {pages} {1237} (\bibinfo {year} {1986})},\ \Eprint
  {http://arxiv.org/abs/http://ptp.oxfordjournals.org/content/76/6/1237.full.pdf+html}
  {http://ptp.oxfordjournals.org/content/76/6/1237.full.pdf+html} \BibitemShut
  {NoStop}%
\bibitem [{\citenamefont {Yamashiro}\ \emph {et~al.}(1998)\citenamefont
  {Yamashiro}, \citenamefont {Tanaka}, \citenamefont {Tanuma},\ and\
  \citenamefont {Kashiwaya}}]{YTK98}%
  \BibitemOpen
  \bibfield  {author} {\bibinfo {author} {\bibfnamefont {M.}~\bibnamefont
  {Yamashiro}}, \bibinfo {author} {\bibfnamefont {Y.}~\bibnamefont {Tanaka}},
  \bibinfo {author} {\bibfnamefont {Y.}~\bibnamefont {Tanuma}}, \ and\ \bibinfo
  {author} {\bibfnamefont {S.}~\bibnamefont {Kashiwaya}},\ }\href {\doibase
  10.1143/JPSJ.67.3224} {\bibfield  {journal} {\bibinfo  {journal} {Journal of
  the Physical Society of Japan}\ }\textbf {\bibinfo {volume} {67}},\ \bibinfo
  {pages} {3224} (\bibinfo {year} {1998})},\ \Eprint
  {http://arxiv.org/abs/http://dx.doi.org/10.1143/JPSJ.67.3224}
  {http://dx.doi.org/10.1143/JPSJ.67.3224} \BibitemShut {NoStop}%
\bibitem [{\citenamefont {Tanaka}\ \emph
  {et~al.}(2002{\natexlab{a}})\citenamefont {Tanaka}, \citenamefont {Tanuma},
  \citenamefont {Kuroki},\ and\ \citenamefont {Kashiwaya}}]{Tanaka2002}%
  \BibitemOpen
  \bibfield  {author} {\bibinfo {author} {\bibfnamefont {Y.}~\bibnamefont
  {Tanaka}}, \bibinfo {author} {\bibfnamefont {Y.}~\bibnamefont {Tanuma}},
  \bibinfo {author} {\bibfnamefont {K.}~\bibnamefont {Kuroki}}, \ and\ \bibinfo
  {author} {\bibfnamefont {S.}~\bibnamefont {Kashiwaya}},\ }\href {\doibase
  10.1143/JPSJ.71.2102} {\bibfield  {journal} {\bibinfo  {journal} {Journal of
  the Physical Society of Japan}\ }\textbf {\bibinfo {volume} {71}},\ \bibinfo
  {pages} {2102} (\bibinfo {year} {2002}{\natexlab{a}})},\ \Eprint
  {http://arxiv.org/abs/http://dx.doi.org/10.1143/JPSJ.71.2102}
  {http://dx.doi.org/10.1143/JPSJ.71.2102} \BibitemShut {NoStop}%
\bibitem [{\citenamefont {Kwon}\ \emph {et~al.}(2004)\citenamefont {Kwon},
  \citenamefont {Sengupta},\ and\ \citenamefont {Yakovenko}}]{Yakovenko}%
  \BibitemOpen
  \bibfield  {author} {\bibinfo {author} {\bibfnamefont {H.-J.}\ \bibnamefont
  {Kwon}}, \bibinfo {author} {\bibfnamefont {K.}~\bibnamefont {Sengupta}}, \
  and\ \bibinfo {author} {\bibfnamefont {V.}~\bibnamefont {Yakovenko}},\ }\href
  {\doibase 10.1140/epjb/e2004-00066-4} {\bibfield  {journal} {\bibinfo
  {journal} {The European Physical Journal B - Condensed Matter and Complex
  Systems}\ }\textbf {\bibinfo {volume} {37}},\ \bibinfo {pages} {349}
  (\bibinfo {year} {2004})}\BibitemShut {NoStop}%
\bibitem [{\citenamefont {Yamashiro}\ \emph {et~al.}(1997)\citenamefont
  {Yamashiro}, \citenamefont {Tanaka},\ and\ \citenamefont
  {Kashiwaya}}]{YTK97}%
  \BibitemOpen
  \bibfield  {author} {\bibinfo {author} {\bibfnamefont {M.}~\bibnamefont
  {Yamashiro}}, \bibinfo {author} {\bibfnamefont {Y.}~\bibnamefont {Tanaka}}, \
  and\ \bibinfo {author} {\bibfnamefont {S.}~\bibnamefont {Kashiwaya}},\ }\href
  {\doibase 10.1103/PhysRevB.56.7847} {\bibfield  {journal} {\bibinfo
  {journal} {Phys. Rev. B}\ }\textbf {\bibinfo {volume} {56}},\ \bibinfo
  {pages} {7847} (\bibinfo {year} {1997})}\BibitemShut {NoStop}%
\bibitem [{\citenamefont {Honerkamp}\ and\ \citenamefont
  {Sigristt}(1998)}]{Honerkamp}%
  \BibitemOpen
  \bibfield  {author} {\bibinfo {author} {\bibfnamefont {C.}~\bibnamefont
  {Honerkamp}}\ and\ \bibinfo {author} {\bibfnamefont {M.}~\bibnamefont
  {Sigristt}},\ }\href {\doibase 10.1023/A:1022281409397} {\bibfield  {journal}
  {\bibinfo  {journal} {Journal of Low Temperature Physics}\ }\textbf {\bibinfo
  {volume} {111}},\ \bibinfo {pages} {895} (\bibinfo {year}
  {1998})}\BibitemShut {NoStop}%
\bibitem [{\citenamefont {Kashiwaya}\ \emph {et~al.}(2011)\citenamefont
  {Kashiwaya}, \citenamefont {Kashiwaya}, \citenamefont {Kambara},
  \citenamefont {Furuta}, \citenamefont {Yaguchi}, \citenamefont {Tanaka},\
  and\ \citenamefont {Maeno}}]{Kashiwaya11}%
  \BibitemOpen
  \bibfield  {author} {\bibinfo {author} {\bibfnamefont {S.}~\bibnamefont
  {Kashiwaya}}, \bibinfo {author} {\bibfnamefont {H.}~\bibnamefont
  {Kashiwaya}}, \bibinfo {author} {\bibfnamefont {H.}~\bibnamefont {Kambara}},
  \bibinfo {author} {\bibfnamefont {T.}~\bibnamefont {Furuta}}, \bibinfo
  {author} {\bibfnamefont {H.}~\bibnamefont {Yaguchi}}, \bibinfo {author}
  {\bibfnamefont {Y.}~\bibnamefont {Tanaka}}, \ and\ \bibinfo {author}
  {\bibfnamefont {Y.}~\bibnamefont {Maeno}},\ }\href {\doibase
  10.1103/PhysRevLett.107.077003} {\bibfield  {journal} {\bibinfo  {journal}
  {Phys. Rev. Lett.}\ }\textbf {\bibinfo {volume} {107}},\ \bibinfo {pages}
  {077003} (\bibinfo {year} {2011})}\BibitemShut {NoStop}%
\bibitem [{\citenamefont {Wu}\ and\ \citenamefont {Samokhin}(2010)}]{Samokhin}%
  \BibitemOpen
  \bibfield  {author} {\bibinfo {author} {\bibfnamefont {S.}~\bibnamefont
  {Wu}}\ and\ \bibinfo {author} {\bibfnamefont {K.~V.}\ \bibnamefont
  {Samokhin}},\ }\href {\doibase 10.1103/PhysRevB.81.214506} {\bibfield
  {journal} {\bibinfo  {journal} {Phys. Rev. B}\ }\textbf {\bibinfo {volume}
  {81}},\ \bibinfo {pages} {214506} (\bibinfo {year} {2010})}\BibitemShut
  {NoStop}%
\bibitem [{\citenamefont {Fogelstr\"om}\ \emph {et~al.}(1997)\citenamefont
  {Fogelstr\"om}, \citenamefont {Rainer},\ and\ \citenamefont {Sauls}}]{Fogel}%
  \BibitemOpen
  \bibfield  {author} {\bibinfo {author} {\bibfnamefont {M.}~\bibnamefont
  {Fogelstr\"om}}, \bibinfo {author} {\bibfnamefont {D.}~\bibnamefont
  {Rainer}}, \ and\ \bibinfo {author} {\bibfnamefont {J.~A.}\ \bibnamefont
  {Sauls}},\ }\href {\doibase 10.1103/PhysRevLett.79.281} {\bibfield  {journal}
  {\bibinfo  {journal} {Phys. Rev. Lett.}\ }\textbf {\bibinfo {volume} {79}},\
  \bibinfo {pages} {281} (\bibinfo {year} {1997})}\BibitemShut {NoStop}%
\bibitem [{\citenamefont {Tanuma}\ \emph
  {et~al.}(2002{\natexlab{a}})\citenamefont {Tanuma}, \citenamefont {Kuroki},
  \citenamefont {Tanaka}, \citenamefont {Arita}, \citenamefont {Kashiwaya},\
  and\ \citenamefont {Aoki}}]{Tanuma2002b}%
  \BibitemOpen
  \bibfield  {author} {\bibinfo {author} {\bibfnamefont {Y.}~\bibnamefont
  {Tanuma}}, \bibinfo {author} {\bibfnamefont {K.}~\bibnamefont {Kuroki}},
  \bibinfo {author} {\bibfnamefont {Y.}~\bibnamefont {Tanaka}}, \bibinfo
  {author} {\bibfnamefont {R.}~\bibnamefont {Arita}}, \bibinfo {author}
  {\bibfnamefont {S.}~\bibnamefont {Kashiwaya}}, \ and\ \bibinfo {author}
  {\bibfnamefont {H.}~\bibnamefont {Aoki}},\ }\href {\doibase
  10.1103/PhysRevB.66.094507} {\bibfield  {journal} {\bibinfo  {journal} {Phys.
  Rev. B}\ }\textbf {\bibinfo {volume} {66}},\ \bibinfo {pages} {094507}
  (\bibinfo {year} {2002}{\natexlab{a}})}\BibitemShut {NoStop}%
\bibitem [{\citenamefont {Tanaka}\ \emph {et~al.}(2009)\citenamefont {Tanaka},
  \citenamefont {Yokoyama}, \citenamefont {Balatsky},\ and\ \citenamefont
  {Nagaosa}}]{TYBN08}%
  \BibitemOpen
  \bibfield  {author} {\bibinfo {author} {\bibfnamefont {Y.}~\bibnamefont
  {Tanaka}}, \bibinfo {author} {\bibfnamefont {T.}~\bibnamefont {Yokoyama}},
  \bibinfo {author} {\bibfnamefont {A.~V.}\ \bibnamefont {Balatsky}}, \ and\
  \bibinfo {author} {\bibfnamefont {N.}~\bibnamefont {Nagaosa}},\ }\href
  {\doibase 10.1103/PhysRevB.79.060505} {\bibfield  {journal} {\bibinfo
  {journal} {Phys. Rev. B}\ }\textbf {\bibinfo {volume} {79}},\ \bibinfo
  {pages} {060505} (\bibinfo {year} {2009})}\BibitemShut {NoStop}%
\bibitem [{\citenamefont {Qi}\ \emph {et~al.}(2009)\citenamefont {Qi},
  \citenamefont {Hughes}, \citenamefont {Raghu},\ and\ \citenamefont
  {Zhang}}]{Qi09}%
  \BibitemOpen
  \bibfield  {author} {\bibinfo {author} {\bibfnamefont {X.-L.}\ \bibnamefont
  {Qi}}, \bibinfo {author} {\bibfnamefont {T.~L.}\ \bibnamefont {Hughes}},
  \bibinfo {author} {\bibfnamefont {S.}~\bibnamefont {Raghu}}, \ and\ \bibinfo
  {author} {\bibfnamefont {S.-C.}\ \bibnamefont {Zhang}},\ }\href {\doibase
  10.1103/PhysRevLett.102.187001} {\bibfield  {journal} {\bibinfo  {journal}
  {Phys. Rev. Lett.}\ }\textbf {\bibinfo {volume} {102}},\ \bibinfo {pages}
  {187001} (\bibinfo {year} {2009})}\BibitemShut {NoStop}%
\bibitem [{\citenamefont {Chung}\ and\ \citenamefont {Zhang}(2009)}]{Chung}%
  \BibitemOpen
  \bibfield  {author} {\bibinfo {author} {\bibfnamefont {S.~B.}\ \bibnamefont
  {Chung}}\ and\ \bibinfo {author} {\bibfnamefont {S.-C.}\ \bibnamefont
  {Zhang}},\ }\href {\doibase 10.1103/PhysRevLett.103.235301} {\bibfield
  {journal} {\bibinfo  {journal} {Phys. Rev. Lett.}\ }\textbf {\bibinfo
  {volume} {103}},\ \bibinfo {pages} {235301} (\bibinfo {year}
  {2009})}\BibitemShut {NoStop}%
\bibitem [{\citenamefont {Murakawa}\ \emph {et~al.}(2009)\citenamefont
  {Murakawa}, \citenamefont {Tamura}, \citenamefont {Wada}, \citenamefont
  {Wasai}, \citenamefont {Saitoh}, \citenamefont {Aoki}, \citenamefont
  {Nomura}, \citenamefont {Okuda}, \citenamefont {Nagato}, \citenamefont
  {Yamamoto}, \citenamefont {Higashitani},\ and\ \citenamefont
  {Nagai}}]{Murakawa}%
  \BibitemOpen
  \bibfield  {author} {\bibinfo {author} {\bibfnamefont {S.}~\bibnamefont
  {Murakawa}}, \bibinfo {author} {\bibfnamefont {Y.}~\bibnamefont {Tamura}},
  \bibinfo {author} {\bibfnamefont {Y.}~\bibnamefont {Wada}}, \bibinfo {author}
  {\bibfnamefont {M.}~\bibnamefont {Wasai}}, \bibinfo {author} {\bibfnamefont
  {M.}~\bibnamefont {Saitoh}}, \bibinfo {author} {\bibfnamefont
  {Y.}~\bibnamefont {Aoki}}, \bibinfo {author} {\bibfnamefont {R.}~\bibnamefont
  {Nomura}}, \bibinfo {author} {\bibfnamefont {Y.}~\bibnamefont {Okuda}},
  \bibinfo {author} {\bibfnamefont {Y.}~\bibnamefont {Nagato}}, \bibinfo
  {author} {\bibfnamefont {M.}~\bibnamefont {Yamamoto}}, \bibinfo {author}
  {\bibfnamefont {S.}~\bibnamefont {Higashitani}}, \ and\ \bibinfo {author}
  {\bibfnamefont {K.}~\bibnamefont {Nagai}},\ }\href {\doibase
  10.1103/PhysRevLett.103.155301} {\bibfield  {journal} {\bibinfo  {journal}
  {Phys. Rev. Lett.}\ }\textbf {\bibinfo {volume} {103}},\ \bibinfo {pages}
  {155301} (\bibinfo {year} {2009})}\BibitemShut {NoStop}%
\bibitem [{\citenamefont {Asano}\ \emph {et~al.}(2003)\citenamefont {Asano},
  \citenamefont {Tanaka}, \citenamefont {Matsuda},\ and\ \citenamefont
  {Kashiwaya}}]{Asano2003}%
  \BibitemOpen
  \bibfield  {author} {\bibinfo {author} {\bibfnamefont {Y.}~\bibnamefont
  {Asano}}, \bibinfo {author} {\bibfnamefont {Y.}~\bibnamefont {Tanaka}},
  \bibinfo {author} {\bibfnamefont {Y.}~\bibnamefont {Matsuda}}, \ and\
  \bibinfo {author} {\bibfnamefont {S.}~\bibnamefont {Kashiwaya}},\ }\href
  {\doibase 10.1103/PhysRevB.68.184506} {\bibfield  {journal} {\bibinfo
  {journal} {Phys. Rev. B}\ }\textbf {\bibinfo {volume} {68}},\ \bibinfo
  {pages} {184506} (\bibinfo {year} {2003})}\BibitemShut {NoStop}%
\bibitem [{\citenamefont {Fu}\ and\ \citenamefont {Berg}(2010)}]{fu10}%
  \BibitemOpen
  \bibfield  {author} {\bibinfo {author} {\bibfnamefont {L.}~\bibnamefont
  {Fu}}\ and\ \bibinfo {author} {\bibfnamefont {E.}~\bibnamefont {Berg}},\
  }\href {\doibase 10.1103/PhysRevLett.105.097001} {\bibfield  {journal}
  {\bibinfo  {journal} {Phys. Rev. Lett.}\ }\textbf {\bibinfo {volume} {105}},\
  \bibinfo {pages} {097001} (\bibinfo {year} {2010})}\BibitemShut {NoStop}%
\bibitem [{\citenamefont {Sasaki}\ \emph {et~al.}(2011)\citenamefont {Sasaki},
  \citenamefont {Kriener}, \citenamefont {Segawa}, \citenamefont {Yada},
  \citenamefont {Tanaka}, \citenamefont {Sato},\ and\ \citenamefont
  {Ando}}]{sasaki11}%
  \BibitemOpen
  \bibfield  {author} {\bibinfo {author} {\bibfnamefont {S.}~\bibnamefont
  {Sasaki}}, \bibinfo {author} {\bibfnamefont {M.}~\bibnamefont {Kriener}},
  \bibinfo {author} {\bibfnamefont {K.}~\bibnamefont {Segawa}}, \bibinfo
  {author} {\bibfnamefont {K.}~\bibnamefont {Yada}}, \bibinfo {author}
  {\bibfnamefont {Y.}~\bibnamefont {Tanaka}}, \bibinfo {author} {\bibfnamefont
  {M.}~\bibnamefont {Sato}}, \ and\ \bibinfo {author} {\bibfnamefont
  {Y.}~\bibnamefont {Ando}},\ }\href {\doibase 10.1103/PhysRevLett.107.217001}
  {\bibfield  {journal} {\bibinfo  {journal} {Phys. Rev. Lett.}\ }\textbf
  {\bibinfo {volume} {107}},\ \bibinfo {pages} {217001} (\bibinfo {year}
  {2011})}\BibitemShut {NoStop}%
\bibitem [{\citenamefont {Hao}\ and\ \citenamefont {Lee}(2011)}]{hao11}%
  \BibitemOpen
  \bibfield  {author} {\bibinfo {author} {\bibfnamefont {L.}~\bibnamefont
  {Hao}}\ and\ \bibinfo {author} {\bibfnamefont {T.~K.}\ \bibnamefont {Lee}},\
  }\href {\doibase 10.1103/PhysRevB.83.134516} {\bibfield  {journal} {\bibinfo
  {journal} {Phys. Rev. B}\ }\textbf {\bibinfo {volume} {83}},\ \bibinfo
  {pages} {134516} (\bibinfo {year} {2011})}\BibitemShut {NoStop}%
\bibitem [{\citenamefont {Hsieh}\ and\ \citenamefont {Fu}(2012)}]{hsieh12}%
  \BibitemOpen
  \bibfield  {author} {\bibinfo {author} {\bibfnamefont {T.~H.}\ \bibnamefont
  {Hsieh}}\ and\ \bibinfo {author} {\bibfnamefont {L.}~\bibnamefont {Fu}},\
  }\href {\doibase 10.1103/PhysRevLett.108.107005} {\bibfield  {journal}
  {\bibinfo  {journal} {Phys. Rev. Lett.}\ }\textbf {\bibinfo {volume} {108}},\
  \bibinfo {pages} {107005} (\bibinfo {year} {2012})}\BibitemShut {NoStop}%
\bibitem [{\citenamefont {Yamakage}\ \emph {et~al.}(2012)\citenamefont
  {Yamakage}, \citenamefont {Yada}, \citenamefont {Sato},\ and\ \citenamefont
  {Tanaka}}]{yamakage12}%
  \BibitemOpen
  \bibfield  {author} {\bibinfo {author} {\bibfnamefont {A.}~\bibnamefont
  {Yamakage}}, \bibinfo {author} {\bibfnamefont {K.}~\bibnamefont {Yada}},
  \bibinfo {author} {\bibfnamefont {M.}~\bibnamefont {Sato}}, \ and\ \bibinfo
  {author} {\bibfnamefont {Y.}~\bibnamefont {Tanaka}},\ }\href {\doibase
  10.1103/PhysRevB.85.180509} {\bibfield  {journal} {\bibinfo  {journal} {Phys.
  Rev. B}\ }\textbf {\bibinfo {volume} {85}},\ \bibinfo {pages} {180509}
  (\bibinfo {year} {2012})}\BibitemShut {NoStop}%
\bibitem [{\citenamefont {Hashimoto}\ \emph {et~al.}(2015)\citenamefont
  {Hashimoto}, \citenamefont {Yada}, \citenamefont {Sato},\ and\ \citenamefont
  {Tanaka}}]{Hashimoto1}%
  \BibitemOpen
  \bibfield  {author} {\bibinfo {author} {\bibfnamefont {T.}~\bibnamefont
  {Hashimoto}}, \bibinfo {author} {\bibfnamefont {K.}~\bibnamefont {Yada}},
  \bibinfo {author} {\bibfnamefont {M.}~\bibnamefont {Sato}}, \ and\ \bibinfo
  {author} {\bibfnamefont {Y.}~\bibnamefont {Tanaka}},\ }\href {\doibase
  10.1103/PhysRevB.92.174527} {\bibfield  {journal} {\bibinfo  {journal} {Phys.
  Rev. B}\ }\textbf {\bibinfo {volume} {92}},\ \bibinfo {pages} {174527}
  (\bibinfo {year} {2015})}\BibitemShut {NoStop}%
\bibitem [{\citenamefont {Lu}\ \emph {et~al.}(2015)\citenamefont {Lu},
  \citenamefont {Yada}, \citenamefont {Sato},\ and\ \citenamefont
  {Tanaka}}]{LuBo}%
  \BibitemOpen
  \bibfield  {author} {\bibinfo {author} {\bibfnamefont {B.}~\bibnamefont
  {Lu}}, \bibinfo {author} {\bibfnamefont {K.}~\bibnamefont {Yada}}, \bibinfo
  {author} {\bibfnamefont {M.}~\bibnamefont {Sato}}, \ and\ \bibinfo {author}
  {\bibfnamefont {Y.}~\bibnamefont {Tanaka}},\ }\href {\doibase
  10.1103/PhysRevLett.114.096804} {\bibfield  {journal} {\bibinfo  {journal}
  {Phys. Rev. Lett.}\ }\textbf {\bibinfo {volume} {114}},\ \bibinfo {pages}
  {096804} (\bibinfo {year} {2015})}\BibitemShut {NoStop}%
\bibitem [{\citenamefont {Hashimoto}\ \emph {et~al.}(2016)\citenamefont
  {Hashimoto}, \citenamefont {Kobayashi}, \citenamefont {Tanaka},\ and\
  \citenamefont {Sato}}]{Hashimoto2}%
  \BibitemOpen
  \bibfield  {author} {\bibinfo {author} {\bibfnamefont {T.}~\bibnamefont
  {Hashimoto}}, \bibinfo {author} {\bibfnamefont {S.}~\bibnamefont
  {Kobayashi}}, \bibinfo {author} {\bibfnamefont {Y.}~\bibnamefont {Tanaka}}, \
  and\ \bibinfo {author} {\bibfnamefont {M.}~\bibnamefont {Sato}},\ }\href
  {\doibase 10.1103/PhysRevB.94.014510} {\bibfield  {journal} {\bibinfo
  {journal} {Phys. Rev. B}\ }\textbf {\bibinfo {volume} {94}},\ \bibinfo
  {pages} {014510} (\bibinfo {year} {2016})}\BibitemShut {NoStop}%
\bibitem [{\citenamefont {Kobayashi}\ \emph {et~al.}(2015)\citenamefont
  {Kobayashi}, \citenamefont {Tanaka},\ and\ \citenamefont
  {Sato}}]{Kobayashi2015}%
  \BibitemOpen
  \bibfield  {author} {\bibinfo {author} {\bibfnamefont {S.}~\bibnamefont
  {Kobayashi}}, \bibinfo {author} {\bibfnamefont {Y.}~\bibnamefont {Tanaka}}, \
  and\ \bibinfo {author} {\bibfnamefont {M.}~\bibnamefont {Sato}},\ }\href
  {\doibase 10.1103/PhysRevB.92.214514} {\bibfield  {journal} {\bibinfo
  {journal} {Phys. Rev. B}\ }\textbf {\bibinfo {volume} {92}},\ \bibinfo
  {pages} {214514} (\bibinfo {year} {2015})}\BibitemShut {NoStop}%
\bibitem [{\citenamefont {Schemm}\ \emph
  {et~al.}(2015{\natexlab{a}})\citenamefont {Schemm}, \citenamefont {Baumbach},
  \citenamefont {Tobash}, \citenamefont {Ronning}, \citenamefont {Bauer},\ and\
  \citenamefont {Kapitulnik}}]{Schemm2015}%
  \BibitemOpen
  \bibfield  {author} {\bibinfo {author} {\bibfnamefont {E.~R.}\ \bibnamefont
  {Schemm}}, \bibinfo {author} {\bibfnamefont {R.~E.}\ \bibnamefont
  {Baumbach}}, \bibinfo {author} {\bibfnamefont {P.~H.}\ \bibnamefont
  {Tobash}}, \bibinfo {author} {\bibfnamefont {F.}~\bibnamefont {Ronning}},
  \bibinfo {author} {\bibfnamefont {E.~D.}\ \bibnamefont {Bauer}}, \ and\
  \bibinfo {author} {\bibfnamefont {A.}~\bibnamefont {Kapitulnik}},\ }\href
  {\doibase 10.1103/PhysRevB.91.140506} {\bibfield  {journal} {\bibinfo
  {journal} {Phys. Rev. B}\ }\textbf {\bibinfo {volume} {91}},\ \bibinfo
  {pages} {140506} (\bibinfo {year} {2015}{\natexlab{a}})}\BibitemShut
  {NoStop}%
\bibitem [{\citenamefont {Kasahara}\ \emph {et~al.}(2007)\citenamefont
  {Kasahara}, \citenamefont {Iwasawa}, \citenamefont {Shishido}, \citenamefont
  {Shibauchi}, \citenamefont {Behnia}, \citenamefont {Haga}, \citenamefont
  {Matsuda}, \citenamefont {Onuki}, \citenamefont {Sigrist},\ and\
  \citenamefont {Matsuda}}]{Kasahara}%
  \BibitemOpen
  \bibfield  {author} {\bibinfo {author} {\bibfnamefont {Y.}~\bibnamefont
  {Kasahara}}, \bibinfo {author} {\bibfnamefont {T.}~\bibnamefont {Iwasawa}},
  \bibinfo {author} {\bibfnamefont {H.}~\bibnamefont {Shishido}}, \bibinfo
  {author} {\bibfnamefont {T.}~\bibnamefont {Shibauchi}}, \bibinfo {author}
  {\bibfnamefont {K.}~\bibnamefont {Behnia}}, \bibinfo {author} {\bibfnamefont
  {Y.}~\bibnamefont {Haga}}, \bibinfo {author} {\bibfnamefont {T.~D.}\
  \bibnamefont {Matsuda}}, \bibinfo {author} {\bibfnamefont {Y.}~\bibnamefont
  {Onuki}}, \bibinfo {author} {\bibfnamefont {M.}~\bibnamefont {Sigrist}}, \
  and\ \bibinfo {author} {\bibfnamefont {Y.}~\bibnamefont {Matsuda}},\ }\href
  {\doibase 10.1103/PhysRevLett.99.116402} {\bibfield  {journal} {\bibinfo
  {journal} {Phys. Rev. Lett.}\ }\textbf {\bibinfo {volume} {99}},\ \bibinfo
  {pages} {116402} (\bibinfo {year} {2007})}\BibitemShut {NoStop}%
\bibitem [{\citenamefont {Shibauchi}\ \emph {et~al.}(2014)\citenamefont
  {Shibauchi}, \citenamefont {Ikeda},\ and\ \citenamefont
  {Matsuda}}]{Shibauchi}%
  \BibitemOpen
  \bibfield  {author} {\bibinfo {author} {\bibfnamefont {T.}~\bibnamefont
  {Shibauchi}}, \bibinfo {author} {\bibfnamefont {H.}~\bibnamefont {Ikeda}}, \
  and\ \bibinfo {author} {\bibfnamefont {Y.}~\bibnamefont {Matsuda}},\ }\href
  {\doibase 10.1080/14786435.2014.887861} {\bibfield  {journal} {\bibinfo
  {journal} {Philos. Mag.}\ }\textbf {\bibinfo {volume} {94}},\ \bibinfo
  {pages} {3747} (\bibinfo {year} {2014})}\BibitemShut {NoStop}%
\bibitem [{\citenamefont {Schemm}\ \emph
  {et~al.}(2015{\natexlab{b}})\citenamefont {Schemm}, \citenamefont {Baumbach},
  \citenamefont {Tobash}, \citenamefont {Ronning}, \citenamefont {Bauer},\ and\
  \citenamefont {Kapitulnik}}]{Schemm2015PRB}%
  \BibitemOpen
  \bibfield  {author} {\bibinfo {author} {\bibfnamefont {E.~R.}\ \bibnamefont
  {Schemm}}, \bibinfo {author} {\bibfnamefont {R.~E.}\ \bibnamefont
  {Baumbach}}, \bibinfo {author} {\bibfnamefont {P.~H.}\ \bibnamefont
  {Tobash}}, \bibinfo {author} {\bibfnamefont {F.}~\bibnamefont {Ronning}},
  \bibinfo {author} {\bibfnamefont {E.~D.}\ \bibnamefont {Bauer}}, \ and\
  \bibinfo {author} {\bibfnamefont {A.}~\bibnamefont {Kapitulnik}},\ }\href
  {\doibase 10.1103/PhysRevB.91.140506} {\bibfield  {journal} {\bibinfo
  {journal} {Phys. Rev. B}\ }\textbf {\bibinfo {volume} {91}},\ \bibinfo
  {pages} {140506} (\bibinfo {year} {2015}{\natexlab{b}})}\BibitemShut
  {NoStop}%
\bibitem [{\citenamefont {Sauls}(1994)}]{Sauls}%
  \BibitemOpen
  \bibfield  {author} {\bibinfo {author} {\bibfnamefont {J.}~\bibnamefont
  {Sauls}},\ }\href {\doibase 10.1080/00018739400101475} {\bibfield  {journal}
  {\bibinfo  {journal} {Advances in Physics}\ }\textbf {\bibinfo {volume}
  {43}},\ \bibinfo {pages} {113} (\bibinfo {year} {1994})}\BibitemShut
  {NoStop}%
\bibitem [{\citenamefont {Joynt}\ and\ \citenamefont
  {Taillefer}(2002)}]{Joynt}%
  \BibitemOpen
  \bibfield  {author} {\bibinfo {author} {\bibfnamefont {R.}~\bibnamefont
  {Joynt}}\ and\ \bibinfo {author} {\bibfnamefont {L.}~\bibnamefont
  {Taillefer}},\ }\href {\doibase 10.1103/RevModPhys.74.235} {\bibfield
  {journal} {\bibinfo  {journal} {Rev. Mod. Phys.}\ }\textbf {\bibinfo {volume}
  {74}},\ \bibinfo {pages} {235} (\bibinfo {year} {2002})}\BibitemShut
  {NoStop}%
\bibitem [{\citenamefont {Schemm}\ \emph {et~al.}(2014)\citenamefont {Schemm},
  \citenamefont {Gannon}, \citenamefont {Wishne}, \citenamefont {Halperin},\
  and\ \citenamefont {Kapitulnik}}]{Schemm2014}%
  \BibitemOpen
  \bibfield  {author} {\bibinfo {author} {\bibfnamefont {E.~R.}\ \bibnamefont
  {Schemm}}, \bibinfo {author} {\bibfnamefont {W.~J.}\ \bibnamefont {Gannon}},
  \bibinfo {author} {\bibfnamefont {C.~M.}\ \bibnamefont {Wishne}}, \bibinfo
  {author} {\bibfnamefont {W.~P.}\ \bibnamefont {Halperin}}, \ and\ \bibinfo
  {author} {\bibfnamefont {A.}~\bibnamefont {Kapitulnik}},\ }\href {\doibase
  10.1126/science.1248552} {\ \textbf {\bibinfo {volume} {345}},\ \bibinfo
  {pages} {190} (\bibinfo {year} {2014})}\BibitemShut {NoStop}%
\bibitem [{\citenamefont {Goswami}\ and\ \citenamefont
  {Nevidomskyy}(2015)}]{Goswami2015}%
  \BibitemOpen
  \bibfield  {author} {\bibinfo {author} {\bibfnamefont {P.}~\bibnamefont
  {Goswami}}\ and\ \bibinfo {author} {\bibfnamefont {A.~H.}\ \bibnamefont
  {Nevidomskyy}},\ }\href {\doibase 10.1103/PhysRevB.92.214504} {\bibfield
  {journal} {\bibinfo  {journal} {Phys. Rev. B}\ }\textbf {\bibinfo {volume}
  {92}},\ \bibinfo {pages} {214504} (\bibinfo {year} {2015})}\BibitemShut
  {NoStop}%
\bibitem [{\citenamefont {Tsutsumi}\ \emph {et~al.}(2012)\citenamefont
  {Tsutsumi}, \citenamefont {Machida}, \citenamefont {Ohmi},\ and\
  \citenamefont {aki Ozaki}}]{Tsutsumi2012}%
  \BibitemOpen
  \bibfield  {author} {\bibinfo {author} {\bibfnamefont {Y.}~\bibnamefont
  {Tsutsumi}}, \bibinfo {author} {\bibfnamefont {K.}~\bibnamefont {Machida}},
  \bibinfo {author} {\bibfnamefont {T.}~\bibnamefont {Ohmi}}, \ and\ \bibinfo
  {author} {\bibfnamefont {M.}~\bibnamefont {aki Ozaki}},\ }\href {\doibase
  10.1143/JPSJ.81.074717} {\bibfield  {journal} {\bibinfo  {journal} {Journal
  of the Physical Society of Japan}\ }\textbf {\bibinfo {volume} {81}},\
  \bibinfo {pages} {074717} (\bibinfo {year} {2012})},\ \Eprint
  {http://arxiv.org/abs/http://dx.doi.org/10.1143/JPSJ.81.074717}
  {http://dx.doi.org/10.1143/JPSJ.81.074717} \BibitemShut {NoStop}%
\bibitem [{\citenamefont {Tsutsumi}\ \emph {et~al.}(2013)\citenamefont
  {Tsutsumi}, \citenamefont {Ishikawa}, \citenamefont {Kawakami}, \citenamefont
  {Mizushima}, \citenamefont {Sato}, \citenamefont {Ichioka},\ and\
  \citenamefont {Machida}}]{Tsutsumi2013}%
  \BibitemOpen
  \bibfield  {author} {\bibinfo {author} {\bibfnamefont {Y.}~\bibnamefont
  {Tsutsumi}}, \bibinfo {author} {\bibfnamefont {M.}~\bibnamefont {Ishikawa}},
  \bibinfo {author} {\bibfnamefont {T.}~\bibnamefont {Kawakami}}, \bibinfo
  {author} {\bibfnamefont {T.}~\bibnamefont {Mizushima}}, \bibinfo {author}
  {\bibfnamefont {M.}~\bibnamefont {Sato}}, \bibinfo {author} {\bibfnamefont
  {M.}~\bibnamefont {Ichioka}}, \ and\ \bibinfo {author} {\bibfnamefont
  {K.}~\bibnamefont {Machida}},\ }\href {\doibase 10.7566/JPSJ.82.113707}
  {\bibfield  {journal} {\bibinfo  {journal} {J. Phys. Soc. Jpn.}\ }\textbf
  {\bibinfo {volume} {82}},\ \bibinfo {pages} {113707} (\bibinfo {year}
  {2013})}\BibitemShut {NoStop}%
\bibitem [{\citenamefont {Kashiwaya}\ \emph {et~al.}(1996)\citenamefont
  {Kashiwaya}, \citenamefont {Tanaka}, \citenamefont {Koyanagi},\ and\
  \citenamefont {Kajimura}}]{KT96}%
  \BibitemOpen
  \bibfield  {author} {\bibinfo {author} {\bibfnamefont {S.}~\bibnamefont
  {Kashiwaya}}, \bibinfo {author} {\bibfnamefont {Y.}~\bibnamefont {Tanaka}},
  \bibinfo {author} {\bibfnamefont {M.}~\bibnamefont {Koyanagi}}, \ and\
  \bibinfo {author} {\bibfnamefont {K.}~\bibnamefont {Kajimura}},\ }\href
  {\doibase 10.1103/PhysRevB.53.2667} {\bibfield  {journal} {\bibinfo
  {journal} {Phys. Rev. B}\ }\textbf {\bibinfo {volume} {53}},\ \bibinfo
  {pages} {2667} (\bibinfo {year} {1996})}\BibitemShut {NoStop}%
\bibitem [{\citenamefont {Blonder}\ \emph {et~al.}(1982)\citenamefont
  {Blonder}, \citenamefont {Tinkham},\ and\ \citenamefont {Klapwijk}}]{BTK}%
  \BibitemOpen
  \bibfield  {author} {\bibinfo {author} {\bibfnamefont {G.~E.}\ \bibnamefont
  {Blonder}}, \bibinfo {author} {\bibfnamefont {M.}~\bibnamefont {Tinkham}}, \
  and\ \bibinfo {author} {\bibfnamefont {T.~M.}\ \bibnamefont {Klapwijk}},\
  }\href {\doibase 10.1103/PhysRevB.25.4515} {\bibfield  {journal} {\bibinfo
  {journal} {Phys. Rev. B}\ }\textbf {\bibinfo {volume} {25}},\ \bibinfo
  {pages} {4515} (\bibinfo {year} {1982})}\BibitemShut {NoStop}%
\bibitem [{\citenamefont {Bruder}(1990)}]{Bruder90}%
  \BibitemOpen
  \bibfield  {author} {\bibinfo {author} {\bibfnamefont {C.}~\bibnamefont
  {Bruder}},\ }\href {\doibase 10.1103/PhysRevB.41.4017} {\bibfield  {journal}
  {\bibinfo  {journal} {Phys. Rev. B}\ }\textbf {\bibinfo {volume} {41}},\
  \bibinfo {pages} {4017} (\bibinfo {year} {1990})}\BibitemShut {NoStop}%
\bibitem [{\citenamefont {Tanuma}\ \emph
  {et~al.}(2002{\natexlab{b}})\citenamefont {Tanuma}, \citenamefont {Tanaka},
  \citenamefont {Kuroki},\ and\ \citenamefont {Kashiwaya}}]{Tanuma2002a}%
  \BibitemOpen
  \bibfield  {author} {\bibinfo {author} {\bibfnamefont {Y.}~\bibnamefont
  {Tanuma}}, \bibinfo {author} {\bibfnamefont {Y.}~\bibnamefont {Tanaka}},
  \bibinfo {author} {\bibfnamefont {K.}~\bibnamefont {Kuroki}}, \ and\ \bibinfo
  {author} {\bibfnamefont {S.}~\bibnamefont {Kashiwaya}},\ }\href {\doibase
  10.1103/PhysRevB.66.174502} {\bibfield  {journal} {\bibinfo  {journal} {Phys.
  Rev. B}\ }\textbf {\bibinfo {volume} {66}},\ \bibinfo {pages} {174502}
  (\bibinfo {year} {2002}{\natexlab{b}})}\BibitemShut {NoStop}%
\bibitem [{\citenamefont {Broholm}\ \emph {et~al.}(1990)\citenamefont
  {Broholm}, \citenamefont {Aeppli}, \citenamefont {Kleiman}, \citenamefont
  {Harshman}, \citenamefont {Bishop}, \citenamefont {Bucher}, \citenamefont
  {Williams}, \citenamefont {Ansaldo},\ and\ \citenamefont
  {Heffner}}]{Broholm90}%
  \BibitemOpen
  \bibfield  {author} {\bibinfo {author} {\bibfnamefont {C.}~\bibnamefont
  {Broholm}}, \bibinfo {author} {\bibfnamefont {G.}~\bibnamefont {Aeppli}},
  \bibinfo {author} {\bibfnamefont {R.~N.}\ \bibnamefont {Kleiman}}, \bibinfo
  {author} {\bibfnamefont {D.~R.}\ \bibnamefont {Harshman}}, \bibinfo {author}
  {\bibfnamefont {D.~J.}\ \bibnamefont {Bishop}}, \bibinfo {author}
  {\bibfnamefont {E.}~\bibnamefont {Bucher}}, \bibinfo {author} {\bibfnamefont
  {D.~L.}\ \bibnamefont {Williams}}, \bibinfo {author} {\bibfnamefont {E.~J.}\
  \bibnamefont {Ansaldo}}, \ and\ \bibinfo {author} {\bibfnamefont {R.~H.}\
  \bibnamefont {Heffner}},\ }\href {\doibase 10.1103/PhysRevLett.65.2062}
  {\bibfield  {journal} {\bibinfo  {journal} {Phys. Rev. Lett.}\ }\textbf
  {\bibinfo {volume} {65}},\ \bibinfo {pages} {2062} (\bibinfo {year}
  {1990})}\BibitemShut {NoStop}%
\bibitem [{\citenamefont {Marabelli}\ \emph {et~al.}(1986)\citenamefont
  {Marabelli}, \citenamefont {Wachter},\ and\ \citenamefont
  {Franse}}]{Marabelli86}%
  \BibitemOpen
  \bibfield  {author} {\bibinfo {author} {\bibfnamefont {F.}~\bibnamefont
  {Marabelli}}, \bibinfo {author} {\bibfnamefont {P.}~\bibnamefont {Wachter}},
  \ and\ \bibinfo {author} {\bibfnamefont {J.}~\bibnamefont {Franse}},\ }\href
  {\doibase http://dx.doi.org/10.1016/0304-8853(86)90157-5} {\bibfield
  {journal} {\bibinfo  {journal} {Journal of Magnetism and Magnetic Materials}\
  }\textbf {\bibinfo {volume} {62}},\ \bibinfo {pages} {287 } (\bibinfo {year}
  {1986})}\BibitemShut {NoStop}%
\bibitem [{\citenamefont {Matsumoto}\ and\ \citenamefont
  {Shiba}(1995)}]{Matsumoto95}%
  \BibitemOpen
  \bibfield  {author} {\bibinfo {author} {\bibfnamefont {M.}~\bibnamefont
  {Matsumoto}}\ and\ \bibinfo {author} {\bibfnamefont {H.}~\bibnamefont
  {Shiba}},\ }\href {\doibase 10.1143/JPSJ.64.4867} {\bibfield  {journal}
  {\bibinfo  {journal} {Journal of the Physical Society of Japan}\ }\textbf
  {\bibinfo {volume} {64}},\ \bibinfo {pages} {4867} (\bibinfo {year}
  {1995})},\ \Eprint
  {http://arxiv.org/abs/http://dx.doi.org/10.1143/JPSJ.64.4867}
  {http://dx.doi.org/10.1143/JPSJ.64.4867} \BibitemShut {NoStop}%
\bibitem [{\citenamefont {Kashiwaya}\ \emph
  {et~al.}(1995{\natexlab{b}})\citenamefont {Kashiwaya}, \citenamefont
  {Tanaka}, \citenamefont {Koyanagi}, \citenamefont {Takashima},\ and\
  \citenamefont {Kajimura}}]{Kashiwayadis}%
  \BibitemOpen
  \bibfield  {author} {\bibinfo {author} {\bibfnamefont {S.}~\bibnamefont
  {Kashiwaya}}, \bibinfo {author} {\bibfnamefont {Y.}~\bibnamefont {Tanaka}},
  \bibinfo {author} {\bibfnamefont {M.}~\bibnamefont {Koyanagi}}, \bibinfo
  {author} {\bibfnamefont {H.}~\bibnamefont {Takashima}}, \ and\ \bibinfo
  {author} {\bibfnamefont {K.}~\bibnamefont {Kajimura}},\ }\href {\doibase
  http://dx.doi.org/10.1016/0022-3697(95)00135-2} {\bibfield  {journal}
  {\bibinfo  {journal} {Journal of Physics and Chemistry of Solids}\ }\textbf
  {\bibinfo {volume} {56}},\ \bibinfo {pages} {1721 } (\bibinfo {year}
  {1995}{\natexlab{b}})}\BibitemShut {NoStop}%
\bibitem [{\citenamefont {Furusaki}\ \emph {et~al.}(2001)\citenamefont
  {Furusaki}, \citenamefont {Matsumoto},\ and\ \citenamefont
  {Sigrist}}]{FMS01}%
  \BibitemOpen
  \bibfield  {author} {\bibinfo {author} {\bibfnamefont {A.}~\bibnamefont
  {Furusaki}}, \bibinfo {author} {\bibfnamefont {M.}~\bibnamefont {Matsumoto}},
  \ and\ \bibinfo {author} {\bibfnamefont {M.}~\bibnamefont {Sigrist}},\ }\href
  {\doibase 10.1103/PhysRevB.64.054514} {\bibfield  {journal} {\bibinfo
  {journal} {Phys. Rev. B}\ }\textbf {\bibinfo {volume} {64}},\ \bibinfo
  {pages} {054514} (\bibinfo {year} {2001})}\BibitemShut {NoStop}%
\bibitem [{\citenamefont {{Yanase}}(2016)}]{Yanase2016}%
  \BibitemOpen
  \bibfield  {author} {\bibinfo {author} {\bibfnamefont {Y.}~\bibnamefont
  {{Yanase}}},\ }\href
  {http://journals.aps.org/prb/pdf/10.1103/PhysRevB.94.174502} {\bibfield
  {journal} {\bibinfo  {journal} {Phys. Rev. B}\ }\textbf {\bibinfo {volume}
  {94}},\ \bibinfo {pages} {174502} (\bibinfo {year} {2016})}\BibitemShut
  {NoStop}%
\bibitem [{\citenamefont {Takami}\ \emph {et~al.}(2014)\citenamefont {Takami},
  \citenamefont {Yada}, \citenamefont {Yamakage}, \citenamefont {Sato},\ and\
  \citenamefont {Tanaka}}]{Takami2014}%
  \BibitemOpen
  \bibfield  {author} {\bibinfo {author} {\bibfnamefont {S.}~\bibnamefont
  {Takami}}, \bibinfo {author} {\bibfnamefont {K.}~\bibnamefont {Yada}},
  \bibinfo {author} {\bibfnamefont {A.}~\bibnamefont {Yamakage}}, \bibinfo
  {author} {\bibfnamefont {M.}~\bibnamefont {Sato}}, \ and\ \bibinfo {author}
  {\bibfnamefont {Y.}~\bibnamefont {Tanaka}},\ }\href {\doibase
  10.7566/JPSJ.83.064705} {\bibfield  {journal} {\bibinfo  {journal} {J. Phys.
  Soc. Jpn.}\ }\textbf {\bibinfo {volume} {83}},\ \bibinfo {pages} {064705}
  (\bibinfo {year} {2014})}\BibitemShut {NoStop}%
\bibitem [{\citenamefont {Sato}\ \emph
  {et~al.}(2011{\natexlab{b}})\citenamefont {Sato}, \citenamefont {Tanaka},
  \citenamefont {Yada},\ and\ \citenamefont {Yokoyama}}]{STYY11}%
  \BibitemOpen
  \bibfield  {author} {\bibinfo {author} {\bibfnamefont {M.}~\bibnamefont
  {Sato}}, \bibinfo {author} {\bibfnamefont {Y.}~\bibnamefont {Tanaka}},
  \bibinfo {author} {\bibfnamefont {K.}~\bibnamefont {Yada}}, \ and\ \bibinfo
  {author} {\bibfnamefont {T.}~\bibnamefont {Yokoyama}},\ }\href {\doibase
  10.1103/PhysRevB.83.224511} {\bibfield  {journal} {\bibinfo  {journal} {Phys.
  Rev. B}\ }\textbf {\bibinfo {volume} {83}},\ \bibinfo {pages} {224511}
  (\bibinfo {year} {2011}{\natexlab{b}})}\BibitemShut {NoStop}%
\bibitem [{\citenamefont {Schnyder}\ and\ \citenamefont
  {Ryu}(2011)}]{Schnyder2011}%
  \BibitemOpen
  \bibfield  {author} {\bibinfo {author} {\bibfnamefont {A.~P.}\ \bibnamefont
  {Schnyder}}\ and\ \bibinfo {author} {\bibfnamefont {S.}~\bibnamefont {Ryu}},\
  }\href {\doibase 10.1103/PhysRevB.84.060504} {\bibfield  {journal} {\bibinfo
  {journal} {Phys. Rev. B}\ }\textbf {\bibinfo {volume} {84}},\ \bibinfo
  {pages} {060504} (\bibinfo {year} {2011})}\BibitemShut {NoStop}%
\bibitem [{\citenamefont {Thouless}\ \emph {et~al.}(1982)\citenamefont
  {Thouless}, \citenamefont {Kohmoto}, \citenamefont {Nightingale},\ and\
  \citenamefont {den Nijs}}]{TKNN}%
  \BibitemOpen
  \bibfield  {author} {\bibinfo {author} {\bibfnamefont {D.~J.}\ \bibnamefont
  {Thouless}}, \bibinfo {author} {\bibfnamefont {M.}~\bibnamefont {Kohmoto}},
  \bibinfo {author} {\bibfnamefont {M.~P.}\ \bibnamefont {Nightingale}}, \ and\
  \bibinfo {author} {\bibfnamefont {M.}~\bibnamefont {den Nijs}},\ }\href
  {\doibase 10.1103/PhysRevLett.49.405} {\bibfield  {journal} {\bibinfo
  {journal} {Phys. Rev. Lett.}\ }\textbf {\bibinfo {volume} {49}},\ \bibinfo
  {pages} {405} (\bibinfo {year} {1982})}\BibitemShut {NoStop}%
\bibitem [{\citenamefont {Tanaka}\ \emph
  {et~al.}(2002{\natexlab{b}})\citenamefont {Tanaka}, \citenamefont
  {Tsuchiura}, \citenamefont {Tanuma},\ and\ \citenamefont
  {Kashiwaya}}]{Tanaka02a}%
  \BibitemOpen
  \bibfield  {author} {\bibinfo {author} {\bibfnamefont {Y.}~\bibnamefont
  {Tanaka}}, \bibinfo {author} {\bibfnamefont {H.}~\bibnamefont {Tsuchiura}},
  \bibinfo {author} {\bibfnamefont {Y.}~\bibnamefont {Tanuma}}, \ and\ \bibinfo
  {author} {\bibfnamefont {S.}~\bibnamefont {Kashiwaya}},\ }\href {\doibase
  10.1143/JPSJ.71.271} {\bibfield  {journal} {\bibinfo  {journal} {Journal of
  the Physical Society of Japan}\ }\textbf {\bibinfo {volume} {71}},\ \bibinfo
  {pages} {271} (\bibinfo {year} {2002}{\natexlab{b}})},\ \Eprint
  {http://arxiv.org/abs/http://dx.doi.org/10.1143/JPSJ.71.271}
  {http://dx.doi.org/10.1143/JPSJ.71.271} \BibitemShut {NoStop}%
\bibitem [{\citenamefont {Vekhter}\ \emph {et~al.}(1999)\citenamefont
  {Vekhter}, \citenamefont {Hirschfeld}, \citenamefont {Carbotte},\ and\
  \citenamefont {Nicol}}]{Vekhter}%
  \BibitemOpen
  \bibfield  {author} {\bibinfo {author} {\bibfnamefont {I.}~\bibnamefont
  {Vekhter}}, \bibinfo {author} {\bibfnamefont {P.~J.}\ \bibnamefont
  {Hirschfeld}}, \bibinfo {author} {\bibfnamefont {J.~P.}\ \bibnamefont
  {Carbotte}}, \ and\ \bibinfo {author} {\bibfnamefont {E.~J.}\ \bibnamefont
  {Nicol}},\ }\href {\doibase 10.1103/PhysRevB.59.R9023} {\bibfield  {journal}
  {\bibinfo  {journal} {Phys. Rev. B}\ }\textbf {\bibinfo {volume} {59}},\
  \bibinfo {pages} {R9023} (\bibinfo {year} {1999})}\BibitemShut {NoStop}%
\bibitem [{\citenamefont {{Lambert}}\ \emph {et~al.}(2016)\citenamefont
  {{Lambert}}, \citenamefont {{Akbari}}, \citenamefont {{Thalmeier}},\ and\
  \citenamefont {{Eremin}}}]{Eremin2016}%
  \BibitemOpen
  \bibfield  {author} {\bibinfo {author} {\bibfnamefont {F.}~\bibnamefont
  {{Lambert}}}, \bibinfo {author} {\bibfnamefont {A.}~\bibnamefont {{Akbari}}},
  \bibinfo {author} {\bibfnamefont {P.}~\bibnamefont {{Thalmeier}}}, \ and\
  \bibinfo {author} {\bibfnamefont {I.}~\bibnamefont {{Eremin}}},\ }\href@noop
  {} {\bibfield  {journal} {\bibinfo  {journal} {ArXiv e-prints}\ } (\bibinfo
  {year} {2016})},\ \Eprint {http://arxiv.org/abs/1608.07946} {arXiv:1608.07946
  [cond-mat.supr-con]} \BibitemShut {NoStop}%
\bibitem [{\citenamefont {Mizushima}\ \emph {et~al.}(2016)\citenamefont
  {Mizushima}, \citenamefont {Tsutsumi}, \citenamefont {Kawakami},
  \citenamefont {Sato}, \citenamefont {Ichioka},\ and\ \citenamefont
  {Machida}}]{Mizushima2016}%
  \BibitemOpen
  \bibfield  {author} {\bibinfo {author} {\bibfnamefont {T.}~\bibnamefont
  {Mizushima}}, \bibinfo {author} {\bibfnamefont {Y.}~\bibnamefont {Tsutsumi}},
  \bibinfo {author} {\bibfnamefont {T.}~\bibnamefont {Kawakami}}, \bibinfo
  {author} {\bibfnamefont {M.}~\bibnamefont {Sato}}, \bibinfo {author}
  {\bibfnamefont {M.}~\bibnamefont {Ichioka}}, \ and\ \bibinfo {author}
  {\bibfnamefont {K.}~\bibnamefont {Machida}},\ }\href {\doibase
  10.7566/JPSJ.85.022001} {\bibfield  {journal} {\bibinfo  {journal} {Journal
  of the Physical Society of Japan}\ }\textbf {\bibinfo {volume} {85}},\
  \bibinfo {pages} {022001} (\bibinfo {year} {2016})},\ \Eprint
  {http://arxiv.org/abs/http://dx.doi.org/10.7566/JPSJ.85.022001}
  {http://dx.doi.org/10.7566/JPSJ.85.022001} \BibitemShut {NoStop}%
\bibitem [{\citenamefont {Asano}(2001)}]{Asano2001}%
  \BibitemOpen
  \bibfield  {author} {\bibinfo {author} {\bibfnamefont {Y.}~\bibnamefont
  {Asano}},\ }\href {\doibase 10.1103/PhysRevB.64.224515} {\bibfield  {journal}
  {\bibinfo  {journal} {Phys. Rev. B}\ }\textbf {\bibinfo {volume} {64}},\
  \bibinfo {pages} {224515} (\bibinfo {year} {2001})}\BibitemShut {NoStop}%
\bibitem [{\citenamefont {Tanaka}\ and\ \citenamefont
  {Kashiwaya}(1996)}]{Josephsond}%
  \BibitemOpen
  \bibfield  {author} {\bibinfo {author} {\bibfnamefont {Y.}~\bibnamefont
  {Tanaka}}\ and\ \bibinfo {author} {\bibfnamefont {S.}~\bibnamefont
  {Kashiwaya}},\ }\href {\doibase 10.1103/PhysRevB.53.R11957} {\bibfield
  {journal} {\bibinfo  {journal} {Phys. Rev. B}\ }\textbf {\bibinfo {volume}
  {53}},\ \bibinfo {pages} {R11957} (\bibinfo {year} {1996})}\BibitemShut
  {NoStop}%
\bibitem [{\citenamefont {Barash}\ \emph
  {et~al.}(1996{\natexlab{a}})\citenamefont {Barash}, \citenamefont
  {Burkhardt},\ and\ \citenamefont {Rainer}}]{Josephson3}%
  \BibitemOpen
  \bibfield  {author} {\bibinfo {author} {\bibfnamefont {Y.~S.}\ \bibnamefont
  {Barash}}, \bibinfo {author} {\bibfnamefont {H.}~\bibnamefont {Burkhardt}}, \
  and\ \bibinfo {author} {\bibfnamefont {D.}~\bibnamefont {Rainer}},\ }\href
  {\doibase 10.1103/PhysRevLett.77.4070} {\bibfield  {journal} {\bibinfo
  {journal} {Phys. Rev. Lett.}\ }\textbf {\bibinfo {volume} {77}},\ \bibinfo
  {pages} {4070} (\bibinfo {year} {1996}{\natexlab{a}})}\BibitemShut {NoStop}%
\bibitem [{\citenamefont {Tanaka}\ and\ \citenamefont
  {Kashiwaya}(1997)}]{tanaka97}%
  \BibitemOpen
  \bibfield  {author} {\bibinfo {author} {\bibfnamefont {Y.}~\bibnamefont
  {Tanaka}}\ and\ \bibinfo {author} {\bibfnamefont {S.}~\bibnamefont
  {Kashiwaya}},\ }\href {\doibase 10.1103/PhysRevB.56.892} {\bibfield
  {journal} {\bibinfo  {journal} {Phys. Rev. B}\ }\textbf {\bibinfo {volume}
  {56}},\ \bibinfo {pages} {892} (\bibinfo {year} {1997})}\BibitemShut
  {NoStop}%
\bibitem [{\citenamefont {Barash}\ \emph
  {et~al.}(1996{\natexlab{b}})\citenamefont {Barash}, \citenamefont
  {Burkhardt},\ and\ \citenamefont {Rainer}}]{Barash}%
  \BibitemOpen
  \bibfield  {author} {\bibinfo {author} {\bibfnamefont {Y.~S.}\ \bibnamefont
  {Barash}}, \bibinfo {author} {\bibfnamefont {H.}~\bibnamefont {Burkhardt}}, \
  and\ \bibinfo {author} {\bibfnamefont {D.}~\bibnamefont {Rainer}},\ }\href
  {\doibase 10.1103/PhysRevLett.77.4070} {\bibfield  {journal} {\bibinfo
  {journal} {Phys. Rev. Lett.}\ }\textbf {\bibinfo {volume} {77}},\ \bibinfo
  {pages} {4070} (\bibinfo {year} {1996}{\natexlab{b}})}\BibitemShut {NoStop}%
\bibitem [{\citenamefont {Tanaka}\ \emph {et~al.}(2003)\citenamefont {Tanaka},
  \citenamefont {Nazarov},\ and\ \citenamefont {Kashiwaya}}]{Proximityd}%
  \BibitemOpen
  \bibfield  {author} {\bibinfo {author} {\bibfnamefont {Y.}~\bibnamefont
  {Tanaka}}, \bibinfo {author} {\bibfnamefont {Y.~V.}\ \bibnamefont {Nazarov}},
  \ and\ \bibinfo {author} {\bibfnamefont {S.}~\bibnamefont {Kashiwaya}},\
  }\href {\doibase 10.1103/PhysRevLett.90.167003} {\bibfield  {journal}
  {\bibinfo  {journal} {Phys. Rev. Lett.}\ }\textbf {\bibinfo {volume} {90}},\
  \bibinfo {pages} {167003} (\bibinfo {year} {2003})}\BibitemShut {NoStop}%
\bibitem [{\citenamefont {Tanaka}\ \emph {et~al.}(2004)\citenamefont {Tanaka},
  \citenamefont {Nazarov}, \citenamefont {Golubov},\ and\ \citenamefont
  {Kashiwaya}}]{Proximityd2}%
  \BibitemOpen
  \bibfield  {author} {\bibinfo {author} {\bibfnamefont {Y.}~\bibnamefont
  {Tanaka}}, \bibinfo {author} {\bibfnamefont {Y.~V.}\ \bibnamefont {Nazarov}},
  \bibinfo {author} {\bibfnamefont {A.~A.}\ \bibnamefont {Golubov}}, \ and\
  \bibinfo {author} {\bibfnamefont {S.}~\bibnamefont {Kashiwaya}},\ }\href
  {\doibase 10.1103/PhysRevB.69.144519} {\bibfield  {journal} {\bibinfo
  {journal} {Phys. Rev. B}\ }\textbf {\bibinfo {volume} {69}},\ \bibinfo
  {pages} {144519} (\bibinfo {year} {2004})}\BibitemShut {NoStop}%
\bibitem [{\citenamefont {Tanaka}\ and\ \citenamefont
  {Kashiwaya}(2004)}]{Proximityp}%
  \BibitemOpen
  \bibfield  {author} {\bibinfo {author} {\bibfnamefont {Y.}~\bibnamefont
  {Tanaka}}\ and\ \bibinfo {author} {\bibfnamefont {S.}~\bibnamefont
  {Kashiwaya}},\ }\href {\doibase 10.1103/PhysRevB.70.012507} {\bibfield
  {journal} {\bibinfo  {journal} {Phys. Rev. B}\ }\textbf {\bibinfo {volume}
  {70}},\ \bibinfo {pages} {012507} (\bibinfo {year} {2004})}\BibitemShut
  {NoStop}%
\bibitem [{\citenamefont {Tanaka}\ \emph
  {et~al.}(2005{\natexlab{a}})\citenamefont {Tanaka}, \citenamefont
  {Kashiwaya},\ and\ \citenamefont {Yokoyama}}]{Proximityp2}%
  \BibitemOpen
  \bibfield  {author} {\bibinfo {author} {\bibfnamefont {Y.}~\bibnamefont
  {Tanaka}}, \bibinfo {author} {\bibfnamefont {S.}~\bibnamefont {Kashiwaya}}, \
  and\ \bibinfo {author} {\bibfnamefont {T.}~\bibnamefont {Yokoyama}},\ }\href
  {\doibase 10.1103/PhysRevB.71.094513} {\bibfield  {journal} {\bibinfo
  {journal} {Phys. Rev. B}\ }\textbf {\bibinfo {volume} {71}},\ \bibinfo
  {pages} {094513} (\bibinfo {year} {2005}{\natexlab{a}})}\BibitemShut
  {NoStop}%
\bibitem [{\citenamefont {Asano}\ \emph {et~al.}(2006)\citenamefont {Asano},
  \citenamefont {Tanaka},\ and\ \citenamefont {Kashiwaya}}]{Proximityp3}%
  \BibitemOpen
  \bibfield  {author} {\bibinfo {author} {\bibfnamefont {Y.}~\bibnamefont
  {Asano}}, \bibinfo {author} {\bibfnamefont {Y.}~\bibnamefont {Tanaka}}, \
  and\ \bibinfo {author} {\bibfnamefont {S.}~\bibnamefont {Kashiwaya}},\ }\href
  {\doibase 10.1103/PhysRevLett.96.097007} {\bibfield  {journal} {\bibinfo
  {journal} {Phys. Rev. Lett.}\ }\textbf {\bibinfo {volume} {96}},\ \bibinfo
  {pages} {097007} (\bibinfo {year} {2006})}\BibitemShut {NoStop}%
\bibitem [{\citenamefont {Tanaka}\ \emph
  {et~al.}(2005{\natexlab{b}})\citenamefont {Tanaka}, \citenamefont {Asano},
  \citenamefont {Golubov},\ and\ \citenamefont {Kashiwaya}}]{Meissner3}%
  \BibitemOpen
  \bibfield  {author} {\bibinfo {author} {\bibfnamefont {Y.}~\bibnamefont
  {Tanaka}}, \bibinfo {author} {\bibfnamefont {Y.}~\bibnamefont {Asano}},
  \bibinfo {author} {\bibfnamefont {A.~A.}\ \bibnamefont {Golubov}}, \ and\
  \bibinfo {author} {\bibfnamefont {S.}~\bibnamefont {Kashiwaya}},\ }\href
  {\doibase 10.1103/PhysRevB.72.140503} {\bibfield  {journal} {\bibinfo
  {journal} {Phys. Rev. B}\ }\textbf {\bibinfo {volume} {72}},\ \bibinfo
  {pages} {140503} (\bibinfo {year} {2005}{\natexlab{b}})}\BibitemShut
  {NoStop}%
\bibitem [{\citenamefont {Tanaka}\ and\ \citenamefont {Golubov}(2007)}]{odd1}%
  \BibitemOpen
  \bibfield  {author} {\bibinfo {author} {\bibfnamefont {Y.}~\bibnamefont
  {Tanaka}}\ and\ \bibinfo {author} {\bibfnamefont {A.~A.}\ \bibnamefont
  {Golubov}},\ }\href {\doibase 10.1103/PhysRevLett.98.037003} {\bibfield
  {journal} {\bibinfo  {journal} {Phys. Rev. Lett.}\ }\textbf {\bibinfo
  {volume} {98}},\ \bibinfo {pages} {037003} (\bibinfo {year}
  {2007})}\BibitemShut {NoStop}%
\bibitem [{\citenamefont {Tanaka}\ \emph
  {et~al.}(2007{\natexlab{a}})\citenamefont {Tanaka}, \citenamefont {Golubov},
  \citenamefont {Kashiwaya},\ and\ \citenamefont {Ueda}}]{odd3}%
  \BibitemOpen
  \bibfield  {author} {\bibinfo {author} {\bibfnamefont {Y.}~\bibnamefont
  {Tanaka}}, \bibinfo {author} {\bibfnamefont {A.~A.}\ \bibnamefont {Golubov}},
  \bibinfo {author} {\bibfnamefont {S.}~\bibnamefont {Kashiwaya}}, \ and\
  \bibinfo {author} {\bibfnamefont {M.}~\bibnamefont {Ueda}},\ }\href {\doibase
  10.1103/PhysRevLett.99.037005} {\bibfield  {journal} {\bibinfo  {journal}
  {Phys. Rev. Lett.}\ }\textbf {\bibinfo {volume} {99}},\ \bibinfo {pages}
  {037005} (\bibinfo {year} {2007}{\natexlab{a}})}\BibitemShut {NoStop}%
\bibitem [{\citenamefont {Tanaka}\ \emph
  {et~al.}(2007{\natexlab{b}})\citenamefont {Tanaka}, \citenamefont {Tanuma},\
  and\ \citenamefont {Golubov}}]{odd3b}%
  \BibitemOpen
  \bibfield  {author} {\bibinfo {author} {\bibfnamefont {Y.}~\bibnamefont
  {Tanaka}}, \bibinfo {author} {\bibfnamefont {Y.}~\bibnamefont {Tanuma}}, \
  and\ \bibinfo {author} {\bibfnamefont {A.~A.}\ \bibnamefont {Golubov}},\
  }\href {\doibase 10.1103/PhysRevB.76.054522} {\bibfield  {journal} {\bibinfo
  {journal} {Phys. Rev. B}\ }\textbf {\bibinfo {volume} {76}},\ \bibinfo
  {pages} {054522} (\bibinfo {year} {2007}{\natexlab{b}})}\BibitemShut
  {NoStop}%
\bibitem [{\citenamefont {Tanuma}\ \emph {et~al.}(1998)\citenamefont {Tanuma},
  \citenamefont {Tanaka}, \citenamefont {Yamashiro},\ and\ \citenamefont
  {Kashiwaya}}]{Tanuma1997}%
  \BibitemOpen
  \bibfield  {author} {\bibinfo {author} {\bibfnamefont {Y.}~\bibnamefont
  {Tanuma}}, \bibinfo {author} {\bibfnamefont {Y.}~\bibnamefont {Tanaka}},
  \bibinfo {author} {\bibfnamefont {M.}~\bibnamefont {Yamashiro}}, \ and\
  \bibinfo {author} {\bibfnamefont {S.}~\bibnamefont {Kashiwaya}},\ }\href
  {\doibase 10.1103/PhysRevB.57.7997} {\bibfield  {journal} {\bibinfo
  {journal} {Phys. Rev. B}\ }\textbf {\bibinfo {volume} {57}},\ \bibinfo
  {pages} {7997} (\bibinfo {year} {1998})}\BibitemShut {NoStop}%
\bibitem [{\citenamefont {Yada}\ \emph {et~al.}(2014)\citenamefont {Yada},
  \citenamefont {Golubov}, \citenamefont {Tanaka},\ and\ \citenamefont
  {Kashiwaya}}]{Yada2014}%
  \BibitemOpen
  \bibfield  {author} {\bibinfo {author} {\bibfnamefont {K.}~\bibnamefont
  {Yada}}, \bibinfo {author} {\bibfnamefont {A.~A.}\ \bibnamefont {Golubov}},
  \bibinfo {author} {\bibfnamefont {Y.}~\bibnamefont {Tanaka}}, \ and\ \bibinfo
  {author} {\bibfnamefont {S.}~\bibnamefont {Kashiwaya}},\ }\href {\doibase
  10.7566/JPSJ.83.074706} {\bibfield  {journal} {\bibinfo  {journal} {Journal
  of the Physical Society of Japan}\ }\textbf {\bibinfo {volume} {83}},\
  \bibinfo {pages} {074706} (\bibinfo {year} {2014})},\ \Eprint
  {http://arxiv.org/abs/http://dx.doi.org/10.7566/JPSJ.83.074706}
  {http://dx.doi.org/10.7566/JPSJ.83.074706} \BibitemShut {NoStop}%
\bibitem [{\citenamefont {Schnyder}\ and\ \citenamefont
  {Brydon}(2015)}]{Schnyder2015}%
  \BibitemOpen
  \bibfield  {author} {\bibinfo {author} {\bibfnamefont {A.~P.}\ \bibnamefont
  {Schnyder}}\ and\ \bibinfo {author} {\bibfnamefont {P.~M.~R.}\ \bibnamefont
  {Brydon}},\ }\href {http://stacks.iop.org/0953-8984/27/i=24/a=243201}
  {\bibfield  {journal} {\bibinfo  {journal} {Journal of Physics: Condensed
  Matter}\ }\textbf {\bibinfo {volume} {27}},\ \bibinfo {pages} {243201}
  (\bibinfo {year} {2015})}\BibitemShut {NoStop}%
\bibitem [{\citenamefont {Sigrist}\ and\ \citenamefont
  {Ueda}(1991)}]{Sigrist_RMP}%
  \BibitemOpen
  \bibfield  {author} {\bibinfo {author} {\bibfnamefont {M.}~\bibnamefont
  {Sigrist}}\ and\ \bibinfo {author} {\bibfnamefont {K.}~\bibnamefont {Ueda}},\
  }\href {\doibase 10.1103/RevModPhys.63.239} {\bibfield  {journal} {\bibinfo
  {journal} {Rev. Mod. Phys.}\ }\textbf {\bibinfo {volume} {63}},\ \bibinfo
  {pages} {239} (\bibinfo {year} {1991})}\BibitemShut {NoStop}%
\end{thebibliography}%
\appendix
\begin{widetext}
\section{\label{sec:App_top}Topological numbers}
The 3D chiral superconductors host both a winding number and a Chern number relevant to ZESABSs. 
In a previous study, two of the present authors have discussed 
relevant topological numbers in 3D chiral superconductors with 
$\Delta_{\rm 3d}^{\nu=1}$ and $\Delta_{\rm 3d}^{\nu=2}$ by taking into account 
an additional symmetry \cite{Kobayashi2015}. 
At this point, we explain topological numbers in 3D chiral superconductors 
with $\Delta^{\rm chiral}_{E_{1u}}$ and $\Delta_{\rm 3d}^{\nu>2}$.
\subsection{\label{sec:App_top1}Winding number in the case $\Delta^{\rm chiral}_{E_{1u}}$}
Following the discussion in Ref.~[\onlinecite{Kobayashi2015}], a winding number 
is defined in a similar way to the case $\Delta^{\nu=2}_{\rm 3d}$. 
However, $\Delta^{\rm chiral}_{E_{1u}}$ has two line nodes and 
is an even function of $k_z$, leading to a vanishing of zero-energy flat band for $\alpha=0$. 
Thus, in the following, we consider winding number for $\alpha \neq 0$ and $k_y=0$. 
In the $k_x-k_z$ plane, $\Delta^{\rm chiral}_{E_{1u}}$ is real, 
so we can define the winding number using time reversal symmetry and spin-rotation symmetry:
\begin{align}
   W(k_x) 
   \equiv 
   \frac{-1}{4 \pi i} \int^{\infty}_{-\infty} d k_z 
   \text{Tr} [\Gamma H^{-1} (\mathbf{k}) \partial_{k_z} H(\mathbf{k})] \Big|_{k_y=0}, 
   \label{eq:winding}
\end{align}
where $H(\mathbf{k})$ is the BdG Hamiltonian with 
$\epsilon (\mathbf{k})= \frac{\hbar^2}{2m} (\mathbf{k}^2 -k_\mathrm{F}^2) $ 
and $\Delta^{\rm chiral}_{E_{1u}}$. 
The chiral operator, which anti-commutes with $H(\mathbf{k})$, is given by 
$\Gamma = - \sigma_1 \tau_2$, where $\sigma_{\mu}$ and $\tau_{\mu}$ ($\mu=0,1,2,3$) 
are the identity matrix and the Pauli matrices in the spin and Nambu spaces, 
respectively. 
Furthermore, using the weak pairing assumption, Eq.~(\ref{eq:winding}) is reduced into 
\begin{align}
   W(k_x) 
   = 
   \sum_{\epsilon (\mathbf{k})=0} \text{sgn} [ \partial_{k_z} \epsilon (\mathbf{k})] \cdot 
   \text{sgn} [(5{k'}_z^2-k_\mathrm{F}^2) {k'}_x], 
   \label{eq:winding2}
\end{align}
where $k_x' = k_x \cos \alpha -k_z \sin \alpha$ and $k_z' = k_x \sin \alpha + k_z \cos \alpha$ 
and the summation is taken for $k_z$ satisfying $\epsilon (\mathbf{k})=0$.

To evaluate the winding number (\ref{eq:winding2}), 
we define the characteristic angles: 
$\alpha_1 = \tan^{-1} \left(\frac{2}{\sqrt{5}+1} \right)$,  
$\alpha_2 = \tan^{-1} \left(\frac{2}{\sqrt{5}-1} \right)$, 
and $\alpha_3 = \tan^{-1} (2)$ and the positions of point and line nodes 
projected onto the $k_x$ line in the $(001)$ plane: 
$k^{\rm line}_1= \frac{k_\mathrm{F}}{\sqrt{5}}(\sin \alpha + 2 \cos \alpha)$ 
and $k^{\rm line}_2 = \frac{k_\mathrm{F}}{\sqrt{5}}(-\sin \alpha + 2 \cos \alpha)$ 
for the line nodes and $k^{\rm point} = k_\mathrm{F} \sin \alpha$ for the point nodes
(see Fig.~\ref{fig:E1u_App}). 
These angles satisfy 
$0 < \frac{\pi}{8} <\alpha_1 < \frac{\pi}{4} < \alpha_2 < \alpha_3 
<\frac{3 \pi}{8}< \frac{\pi}{2}$. 
Calculating Eq.~(\ref{eq:winding2}), we obtain the winding number as follows:
\begin{itemize}
\item $0 < \alpha \le \alpha_1$ \\
\begin{align}
   W(k_x) 
   = 
   \begin{cases}
      2  & -k^{\rm line}_1 < k_x < -k^{\rm line}_2 \\
      -2 & -k^{\rm point} < k_x < k^{\rm point} \\
      2  & k^{\rm line}_2 < k_x < k^{\rm line}_1 \\
      0  & \text{otherwise}
   \end{cases} 
   \nonumber
\end{align}

\item $\alpha_1 < \alpha \le \alpha_2$ \\
\begin{align}
   W(k_x) 
   = 
   \begin{cases} 
      2  & -k^{\rm line}_1 < k_x < -k^{\rm point} \\
      -2 & -k^{\rm line}_2 < k_x < k^{\rm line}_2 \\
      2  & k^{\rm point} < k_x < k^{\rm line}_1 \\
      0  & \text{otherwise}
   \end{cases} 
   \nonumber
\end{align}

\item $\alpha_2 < \alpha \le \alpha_3$ \\
\begin{align}
   W(k_x) 
   = 
   \begin{cases}
      -2 & -k^{\rm point} < k_x < -k^{\rm line}_1 \\
      -2 & -k^{\rm line}_2 < k_x < k^{\rm line}_2 \\
      -2 & k^{\rm line}_1 < k_x < k^{\rm point} \\
      0  & \text{otherwise}
   \end{cases} 
   \nonumber
\end{align}

\item $\alpha_3 < \alpha \le \frac{\pi}{2}$ \\
   \begin{align}
      W(k_x) 
      = 
      \begin{cases}
         -2 & -k^{\rm point} < k_x < -k^{\rm line}_1 \\
         2  & k^{\rm line}_2 < k_x < -k^{\rm line}_2 \\
         -2 & k^{\rm line}_1 < k_x < k^{\rm point} \\
         0  & \text{otherwise}
      \end{cases} 
      \nonumber
   \end{align}
\end{itemize}
The factor $2$ comes from the spin degrees of freedom. 
The obtained results are consistent with the ZESABSs in Fig.~\ref{fig:E1u_both}.  
\begin{figure}[htbp]
   \centering
   \includegraphics[width=17.5cm,bb=0 0 720 115]{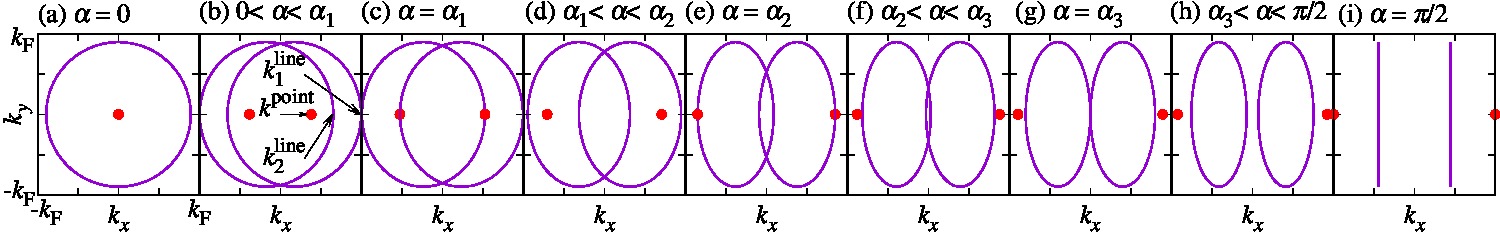}
   \caption{%
      The position of point nodes and line nodes for $\Delta_{E_{1u}}^\mathrm{chiral}$ 
      projected on the $k_x-k_y$ plane 
      are shown for 
      (a) $\alpha=0$, 
      (b) $0<\alpha<\alpha_1$,
      (c) $\alpha=\alpha_1$, 
      (d) $\alpha_1<\alpha<\alpha_2$,
      (e) $\alpha=\alpha_2$, 
      (f) $\alpha_2<\alpha<\alpha_3$,
      (f) $\alpha=\alpha_3$, 
      (h) $\alpha_3<\alpha<\pi/2$ and
      (i) $\alpha=\pi/2$. 
      (b), (d), (h), and (i) are the same as 
      Fig.~\ref{fig:E1u_both} (b-i), (b-ii) (b-iii) and (b-iv), respectively.
      $k_1^\mathrm{line}$, $k_2^\mathrm{line}$ and $k^\mathrm{point}$ for
      $0<\alpha<\alpha_1$ are shown in (b).
   }
   \label{fig:E1u_App}
\end{figure}

\subsection{\label{sec:App_top2}Chern number in the case $\Delta_{\rm 3d}^{\nu>2}$}
In 3D chiral superconductors with $\Delta_{\rm 3d}^{\nu=1}$ and $\Delta_{\rm 3d}^{\nu=2}$, 
ZESABSs are understood from both the winding number and the Chern number. 
On the other hand, 3D chiral superconductors with $\Delta_{\rm 3d}^{\nu>2}$ 
has the redundant ZESABSs for $0 < \alpha \le \frac{\pi}{4}$. 
To understand this type of ZESABSs, we introduce the Chern number defined on a cylinder:
\begin{align}
   N 
   = 
   \frac{i}{2\pi} 
   \sum_{n \in \text{occ}} 
   \int^{\infty}_{-\infty} d k_z \int_0^{2\pi} \rho d \theta \, 
   \epsilon^{ab} \partial_{k_a} 
   \langle 
      u_n (\mathbf{k}) | \partial_{k_b} | u_n (\mathbf{k}) 
   \rangle, 
   \label{eq:Chern}
\end{align}
where the cylindrical coordinate is defined by 
$(\rho \cos \theta \pm k_\mathrm{F} \sin \alpha,\rho \sin \theta, k_z )$, 
$| u_n (\mathbf{k}) \rangle$ is an eigenstate of $H(\mathbf{k})$, 
and the summation is taken over all of occupied stats. 
We choose $\rho$ in such a way that the cylinder includes a single point node 
and does not touch the line node. 
Then, Eq. (\ref{eq:Chern}) gives $\pm 2 \nu$ in analogy with the Chern number 
on the $k_y-k_z$ plane \cite{Kobayashi2015}, where $2$ comes from the spin degrees of freedom. 
The nontrivial Chern number predicts $2\nu$ ZESABSs terminated at the point nodes. 
As shown in Fig.~\ref{fig:3D_l5} (d-ii) and (g-ii), single and double ZESABSs terminated at the point nodes appear on the $k_x$ line, respectively. 
For $\nu>2$, we find $2\nu$ ZESABSs terminated at the point nodes 
[Fig.~\ref{fig:3D_App} (d) and (g)]. 
Note that the winding number also exists and explains ZESABSs in the $k_x$ line.

\begin{figure}[htbp]
   \centering
   \includegraphics[width=17cm,bb=0 0 1500 525]{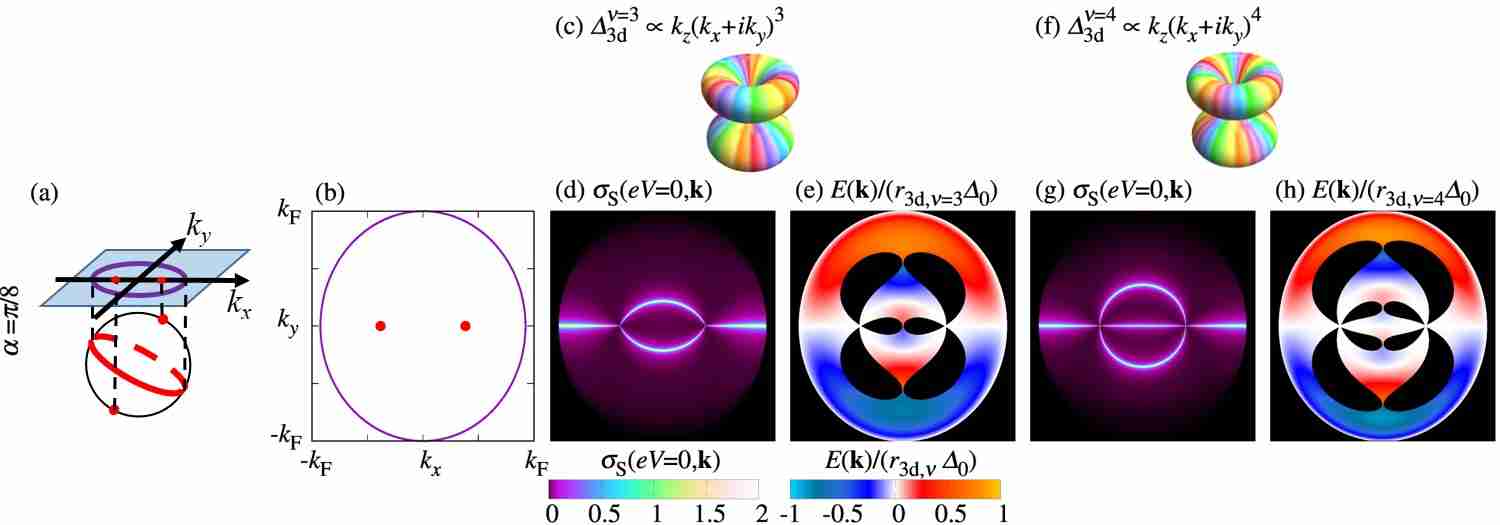}
   \caption{%
      (a) Schematic illustration of point nodes (red dots) and a line node (red line) 
      for $\alpha=\pi/8$.
      Point nodes and a line node projected on the $k_x-k_y$ plane corresponding to (a)
      are shown in (b).
      Schematic picture of pair potential for (c) $\Delta_{3\mathrm{d}}^{\nu=3}$ 
      and (f) $\Delta_{3\mathrm{d}}^{\nu=4}$. 
      (d) The angle-resolved zero bias conductance $\sigma_\mathrm{S}(eV=0,\mathbf{k}_\parallel)$ 
      at $Z=6$ are plotted as functions of $k_x$ and $k_y$ for 
      (d) $\Delta_{3\mathrm{d}}^{\nu=3}$ and (g) $\Delta_{3\mathrm{d}}^{\nu=4}$.
      The energy dispersions of the SABS $E(\mathbf{k}_\parallel)$ are plotted as functions 
      of $k_x$ and $k_y$ for $\Delta_{3\mathrm{d}}^{\nu=3}$
      (e) and for $\Delta_{3\mathrm{d}}^{\nu=4}$
      (h). 
   }
   \label{fig:3D_App}
\end{figure}

\section{\label{sec:App_SABS}SABS}
In Appendix~\ref{sec:App_SABS}, we show how to derive the SABS\@.
Since $|\Gamma_\pm|=1$ [$\Gamma_\pm$ is given by 
Eqs.~(\ref{eq:Gamma_p}) and~(\ref{eq:Gamma_m})] is satisfied in in-gap state, 
we introduce $\theta_\pm$ as follows,
\begin{align}
   E
   &=|\Delta_+|\cos\theta_+,
   \label{eq:plus}
   \\
   &=|\Delta_-|\cos\theta_-.
   \label{eq:minus}
\end{align}
with $\Delta_+=\Delta(\mathbf{k})$, $\Delta_-=\Delta(\tilde{\mathbf{k}})$.
Since $\theta_{+}$ and $\theta_{-}$ are not independent, 
we obtain
\begin{align}
   \kappa\frac{\cos\theta_+}{\cos\theta_-}
   =1,
   \label{eq:theta_ratio}
\end{align}
with 
\begin{align}
   \kappa=\frac{|\Delta_+|}{|\Delta_-|}.
   \nonumber
\end{align}

$\theta_+$ ($\theta_-$) is used for the definition of 
$\Gamma_+$ ($\Gamma_-$) and is confined in the 
domain 
$0\leq\theta_\pm\leq\pi$.
Then, $\Gamma_{\pm}$ is given by 
\begin{align}
   \Gamma_+
   &=\exp{[i(-\phi_+-\theta_+)]},
   \nonumber
   \\
   \Gamma_-&=
   \exp{[i(\phi_--\theta_-)]},
   \nonumber
\end{align}
where $\phi_\pm$ is defined by $\Delta_\pm=|\Delta_\pm|\exp(i\phi_\pm)$.
Then, the SABS satisfies following conditions, 
\begin{align}
   0=\mathrm{Im}\Gamma_+\Gamma_-
      &=\sin
   \left(
      \phi-\theta_+-\theta_-
   \right),
   \label{eq:Im2-1}\\
   1=\mathrm{Re}\Gamma_+\Gamma_-
      &=\cos
   \left(
      \phi-\theta_+-\theta_-
   \right),
   \label{eq:Re2-1} 
\end{align}
with 
\begin{align}
   \phi=-\phi_++\phi_-.
   \nonumber
\end{align}
From Eqs.~(\ref{eq:Im2-1}) and (\ref{eq:Re2-1}),
the relation between $\phi$ and $\theta_{\pm}$ is obtained, 
\begin{align}
   \phi - \theta_+ - \theta_-
   =&
   -2n\pi,\nonumber\\
   \Leftrightarrow
   \theta_+ + \theta_-
   =&
   2n\pi + \phi,
   \label{eq:theta_cond2}
\end{align}
where $n$ is an integer.
The dispersion relation is given by Eq.~(\ref{eq:plus}) and 
Eq.~(\ref{eq:theta_cond2}) or Eq.~(\ref{eq:minus}) 
and Eq.~(\ref{eq:theta_cond2}).

In order to eliminate $\theta_-$, 
substitute Eq.~(\ref{eq:theta_cond2}) for Eq.~(\ref{eq:theta_ratio}). 
Then, we get  
\begin{align}
   &
   \cos\theta_+(\kappa-\cos\phi)=\sin\theta_+\sin\phi,
   \nonumber\\
   \Rightarrow&
   \cos^2\theta_+\left[(\kappa-\cos\phi)^2+\sin^2\phi\right]=\sin^2\phi.
   \label{eq:disp_t}
\end{align}

I)$\phi=2m\pi$ ($m$: integer) with $\kappa=1$ \par
$(\theta_+,\:\theta_-)\neq(\pi/2,\:\pi/2)$ is satisfied 
by Eq.~(\ref{eq:theta_cond2}).
From Eq.~(\ref{eq:theta_ratio}), we obtain 
\begin{align}
   \cos\theta_+=\cos\theta_-.
   \nonumber
\end{align}
This condition is held with
$(\theta_+,\theta_-)=(0,\:0)$ or $(\pi,\:\pi)$.  
However, from Eq.~(\ref{eq:plus}) or Eq.~(\ref{eq:minus}),
$E=\pm|\Delta_+|=\pm|\Delta_-|$ is satisfied. 
This means that the obtained energy dispersion is not inside the energy gap and 
is not the SABS (in-gap state).

II) $\phi\neq 2m\pi$ or $\kappa\neq1$ \par

From Eq.~(\ref{eq:disp_t}), we obtain $\cos \theta_{\pm}$ and 
$\sin \theta_{\pm}$ as follows 
\begin{align}
   \cos\theta_+&=\pm\frac{\sin\phi}{\sqrt{(\kappa-\cos\phi)^2+\sin^2\phi}},
   \label{eq:SABS_cos_p}
   \\
   \sin\theta_+&=\frac{|\kappa-\cos\phi|}{\sqrt{(\kappa-\cos\phi)^2+\sin^2\phi}},\\
   \cos\theta_-&=\pm\frac{\sin\phi}{\sqrt{(\kappa^{-1}-\cos\phi)^2+\sin^2\phi}},\\
   \sin\theta_-&=\frac{|\kappa^{-1}-\cos\phi|}{\sqrt{(\kappa^{-1}-\cos\phi)^2+\sin^2\phi}},
   \label{eq:SABS_sin_m}
\end{align}
where the sign of $\cos\theta_+$ and $\cos\theta_-$ are the same.
We must check 
whether four equations from 
Eq.~(\ref{eq:SABS_cos_p}) to Eq.~(\ref{eq:SABS_sin_m}) 
are consistent with Eq.~(\ref{eq:theta_cond2}) or not.
From Eq.~(\ref{eq:theta_cond2}), we obtain following relations, 
\begin{align}
   &\sin(\theta_++\theta_-)
   =\sin\phi,
   \nonumber\\
   \Leftrightarrow&
   \pm
   \sin\phi
   \left(
      |\kappa-\cos\phi|+|\kappa^{-1}-\cos\phi|
   \right)
   \nonumber\\
   =&\sin\phi\sqrt{(\kappa-\cos\phi)^2+\sin^2\phi}\sqrt{(\kappa^{-1}-\cos\phi)^2+\sin^2\phi},
   \label{eq:sin}\\
   &\cos(\theta_++\theta_-)
   =
   \cos\phi,
   \nonumber\\
   \Leftrightarrow&
   \sin^2\phi-|\kappa-\cos\phi||\kappa^{-1}-\cos\phi|
   \nonumber\\
   =&
   \cos\phi\sqrt{(\kappa-\cos\phi)^2+\sin^2\phi}\sqrt{(\kappa^{-1}-\cos\phi)^2+\sin^2\phi}.
   \label{eq:cos}
\end{align}

From Eq.~(\ref{eq:sin}), we must consider following four cases.

II-1) $\sin\phi=0$

i)\underline{$\phi=2l\pi$ ($l$: integer)}

Substituting $\phi=2l\pi$ for Eq.~(\ref{eq:cos}); the left-hand side of Eq.~(\ref{eq:sin}) 
is negative but the right-hand side of it 
is positive. Therefore, there is no SABS.\

ii)\underline{$\phi=(2l-1)\pi$ ($l$: integer)}

$\phi=(2l-1)\pi$ satisfies Eq.~(\ref{eq:cos}). 
From Eq.~(\ref{eq:theta_cond2}), we obtain 
\begin{align}
   \theta_++\theta_-=\pi.
   \nonumber
\end{align}
This equation contradicts the fact that the sign of $\cos\theta_+$ is equal to 
that of $\cos\theta_-$ except for $\theta_\pm=\pi/2$.
Only for $\theta_+=\theta_-=\pi/2$, 
we obtain
\begin{align}
   \frac{\Delta_+}{|\Delta_+|}
   =
   -\frac{\Delta_-}{|\Delta_-|}. 
   \nonumber
\end{align}
This is the condition known for zero-energy SABS in unconventional 
superconductors \cite{TK95,kashiwaya00}.

II-2) $\sin\phi\neq0$

Hereafter, we suppose $\kappa\leq1$.
In the case of $\kappa>1$, the same discussion can be held with 
replacing $\kappa$ by $\kappa^{-1}$.\

i)\underline{$\kappa-\cos\phi\geq0$}

In this case, 
Eq.~(\ref{eq:sin}) becomes
\begin{align}
   \pm(\kappa+\kappa^{-1}-2\cos\phi)
   =&
   \sqrt{%
		 (\kappa-\cos\phi)^2+\sin^2\phi}
		 \sqrt{(\kappa^{-1}-\cos\phi)^2+\sin^2\phi}.
   \nonumber
\end{align}
The case of negative sign of the lefthand side 
does not satisfy above equation. 
If we choose positive sign, 
it is not difficult to confirm that the above equation is always satisfied.

On the other hand, Eq.~(\ref{eq:cos}) becomes
\begin{align}
   \sin^2\phi-1+\cos\phi(\kappa+\kappa^{-1})-\cos^2\phi
   =&
   \cos\phi\left[(\kappa+\kappa^{-1})-2\cos\phi\right]
   \nonumber\\
   =&
   \cos\phi\sqrt{%
   (\kappa-\cos\phi)^2 +\sin^2\phi}
   \sqrt{(\kappa^{-1}-\cos\phi)^2+\sin^2\phi
   }.
   \label{eq:SABS_App_condition2}
\end{align}
For $\cos\phi=0$, Eq.~(\ref{eq:SABS_App_condition2}) is satisfied and 
and it means that it is a solution of the SABS\@. 
In the case of $\cos\phi\neq0$, 
Eq.~(\ref{eq:cos}) becomes the same as 
Eq.~(\ref{eq:sin}) when we choose plus sign in Eq.~(\ref{eq:cos}).
For $\kappa-\cos\phi\geq0$, there always exists a solution of the SABS for 
arbitrary $\phi$.\

ii)\underline{$\kappa-\cos\phi<0$}

Eq.~(\ref{eq:sin}) becomes
\begin{align}
   \pm(-\kappa+\kappa^{-1})
			&=
			\sqrt{(\kappa-\cos\phi)^2+\sin^2\phi}
			\sqrt{(\kappa^{-1}-\cos\phi)^2+\sin^2\phi}.
   \nonumber
\end{align}
The negative sign of the lefthand side 
does not satisfy this equation. 
By taking the square of this equation, we obtain 
\begin{align}
   (\kappa-\kappa^{-1})^2
   =&
   2+(\kappa^2+\kappa^{-2})+4\cos^2\phi-4\cos\phi(\kappa+\kappa^{-1}),
   \nonumber\\
   \Leftrightarrow
   \cos\phi
   =&\frac{1}{2 }
   \left[
      \kappa+\kappa^{-1}\pm\sqrt{(\kappa-\kappa^{-1})^2}
   \right]
   \nonumber\\
   =&
   \kappa,\:\kappa^{-1}.
   \nonumber
\end{align}
The relation $\cos\phi=\kappa$ contradicts $\kappa-\cos\phi<0$ and
$\cos\phi=\kappa^{-1}$ contradicts $\kappa^{-1}\geq1$.
Then, there is no solution of the SABS. 

To summarize, if $\sin\phi=\sin(-\phi_++\phi_-)\neq0$ is satisfied, 
the energy dispersion of the SABS is given by
\begin{align}
   E(\mathbf{k}_\parallel)
   =&
   \frac{|\Delta_+||\Delta_-|\sin\phi}
	 {\sqrt{|\Delta_+|^2+|\Delta_-|^2-2|\Delta_+||\Delta_-|\cos\phi}}
   \nonumber\\
   =&
   \frac{|\Delta_+||\Delta_-|\sin\phi}{|\Delta_+-\Delta_-|},
   \nonumber
   \\
   \mathrm{with}\:\:
	 &
	 \left(
		 \frac{|\Delta_+|}{|\Delta_-|}-\cos\phi
	 \right)
	 \left(
		 \frac{|\Delta_-|}{|\Delta_+|}-\cos\phi
	 \right)\geq0. 
   \nonumber
\end{align}
For $\sin\phi=0$,
\begin{align}
   E(\mathbf{k}_\parallel)
   =&0,
   \nonumber
   \\
   \mathrm{with}\:\frac{\Delta_+}{|\Delta_+|}
   =&
   -\frac{\Delta_-}{|\Delta_-|}.
   \nonumber
\end{align}
\section{\label{sec:App_cond}Conductance for general pair potential}
In this Appendx, we derive a conductance formula 
for any pair potential with single band superconductor where 
$\hat{\varepsilon}(\mathbf{k})$ in Eq.~\ref{eq:BdG_H} does not have an off-diagonal element.
In Sec.~\ref{sec:App_cond1}, we introduce eigen vectors used for the 
wave function of superconducting side.
In Sec.~\ref{sec:App_cond3}, we solve boundary conditions and derive 
conductance.
In Sec.~\ref{sec:App_cond4}, we explain that conductance is invariant under
the spin rotation.
\subsection{\label{sec:App_cond1}Derivation of eigen vectors}
In this subsection, we derive the eigen vectors of the BdG Hamiltonian.
In Eq.~(\ref{eq:Psi_S}), we define $\psi_{e,\sigma}^\mathrm{S}$ and $\psi_{h,\sigma}^\mathrm{S}$ as
\begin{align}
   \psi_{e,\uparrow}^\mathrm{S}
   =&
   \begin{pmatrix}
      u_{+,p}\\
      v_{+,p}
   \end{pmatrix},
   \:
   \psi_{e,\downarrow}^\mathrm{S}
   =
   \begin{pmatrix}
      u_{+,m}\\
      v_{+,m}
   \end{pmatrix},
   \nonumber\\
   \psi_{h,\uparrow}^\mathrm{S}
   =&
   \begin{pmatrix}
      v_{-,p}\\
      u_{-,p}
   \end{pmatrix},
   \:
   \psi_{h,\downarrow}^\mathrm{S}
   =
   \begin{pmatrix}
      v_{-,m}\\
      u_{-,m}
   \end{pmatrix}.
   \nonumber
\end{align}
$u_{\pm,p(m)}$ and $v_{\pm,p(m)}$ satisfy
\begin{align}
   \begin{pmatrix}
      \tilde{\omega}_{+,p(m)}\sigma_0 & \Delta_+\\
      \Delta_+^\dag & -\tilde{\omega}_{+,p(m)}\sigma_0
   \end{pmatrix}
   \begin{pmatrix}
      u_{+,p(m)}\\
      v_{+,p(m)}
   \end{pmatrix}
   &=
   E
   \begin{pmatrix}
      u_{+,p(m)}\\
      v_{+,p(m)}
   \end{pmatrix},
   \label{eq:Eigen_e}
   \\
   \begin{pmatrix}
      -\tilde{\omega}_{-,p(m)}\sigma_0 & \Delta_-\\
      \Delta_-^\dag & \tilde{\omega}_{-,p(m)}\sigma_0
   \end{pmatrix}
   \begin{pmatrix}
      v_{-,p(m)}\\
      u_{-,p(m)}
   \end{pmatrix}
   &=
   E
   \begin{pmatrix}
      v_{-,p(m)}\\
      u_{-,p(m)}
   \end{pmatrix},
   \label{eq:Eigen_h}
\end{align}
where $\sigma_0$ is the $2 \times 2$ identity matrix and 
\begin{align}
   \Delta_\pm
   &=
   [D_\pm+\mathbf{d}_\pm\cdot\boldsymbol{\sigma}]i\sigma_2
   \nonumber
   \\
   \tilde{\omega}_{\pm,p}
   &=
   \sqrt{E^2-(|\mathbf{d}_\pm|^2+|D_\pm|^2+|\mathbf{J_\pm}|)},
   \label{eq:var_p}
   \\
   \tilde{\omega}_{\pm,m}
   &=
   \sqrt{E^2-(|\mathbf{d}_\pm|^2+|D_\pm|^2-|\mathbf{J_\pm}|)},
   \label{eq:var_m}
   \\
   \mathbf{J}_\pm
   &=
   \pm\mathbf{F}_\pm+\mathbf{q}_\pm
   \nonumber
   \\
   \mathbf{F}_\pm
   &=
   D_\pm\mathbf{d}_\pm^*+D_\pm^*\mathbf{d}_\pm,
   \nonumber
   \\
   \mathbf{q}_\pm
   &=
   i\mathbf{d}_\pm\times\mathbf{d}^*_\pm,
   \nonumber
\end{align}
with $D_+=D(\mathbf{k})$, $D_- = D(\tilde{\mathbf{k}})$,
$\mathbf{d}_+ = \mathbf{d}(\mathbf{k})$,
$\mathbf{d}_- = \mathbf{d}(\tilde{\mathbf{k}})$ and 
$\tilde{\mathbf{k}}=(k_x,k_y,-k_z)$.
$\mathbf{F}_\pm$ and $\mathbf{q}_\pm$ are real-valued functions and 
perpendicular to each other.
$D$ is a spin-singlet pair amplitude and $\mathbf{d}$ is a triplet one.

From Eqs.~(\ref{eq:Eigen_e}) and (\ref{eq:Eigen_h}),
we obtain following equations.
\begin{align}
   u_{+,p(m)}(E-\tilde{\omega}_{+,p(m)})(E+\tilde{\omega}_{+,p(m)})
   =&
   \Delta_+\Delta_+^\dag u_{+,p(m)},
   \label{eq:Eigen_up}
   \\
   u_{-,p(m)}(E-\tilde{\omega}_{-,p(m)})(E+\tilde{\omega}_{-,p(m)})
   =&
   \Delta_-^\dag\Delta_- u_{-,p(m)}.
   \label{eq:Eigen_um}
\end{align}
To simplify above equations, we define $2\times2$ matrices as
\begin{align}
   \check{\omega}_{\pm,pm}
   =&
   \begin{pmatrix}
      \tilde{\omega}_{\pm,p} &0 \\
      0 & \tilde{\omega}_{\pm,m}
   \end{pmatrix},
   \nonumber\\
   u_\pm
   =&
   \begin{pmatrix}
      u_{\pm,p} & u_{\pm,m}
   \end{pmatrix},
   \nonumber\\
   v_\pm
   =&
   \begin{pmatrix}
      v_{\pm,p} & v_{\pm,m}
   \end{pmatrix}.
   \nonumber
\end{align}
Eqs.~(\ref{eq:Eigen_up}) and (\ref{eq:Eigen_um}) become
\begin{align}
   u_{+}(E\sigma_0-\check{\omega}_{+,pm})(E\sigma_0+\check{\omega}_{+,pm})
   =&
   \Delta_+\Delta_+^\dag u_{+},
   \nonumber
   \\
   u_{-}(E\sigma_0-\check{\omega}_{-,pm})(E\sigma_0+\check{\omega}_{-,pm})
   =&
   \Delta_-^\dag\Delta_- u_{-}.
   \nonumber
\end{align}
$u_+$ and $u_-$ are given by \cite{Sigrist_RMP}
\begin{align}
   u_+
   =&
   \tilde{a}_+
   \left(
      |\mathbf{J}_+|\sigma_0 
      + 
      \mathbf{J}_+\cdot\boldsymbol{\sigma} 
   \right)
   (\sigma_0 + \sigma_3)
   + 
   \tilde{b}_+
   \left(
      |\mathbf{J}_+|\sigma_0 
      - 
      \mathbf{J}_+\cdot\boldsymbol{\sigma} 
   \right)
   (\sigma_0 - \sigma_3),
   \\
   u_-
   =&
   \tilde{a}_-
   \left(
      |\mathbf{J}_-|\sigma_0 
      + 
      \mathbf{J}_-\cdot\boldsymbol{\sigma}^*
   \right)
   (\sigma_0 + \sigma_3)
   + 
   \tilde{b}_-
   \left(
      |\mathbf{J}_-|\sigma_0 
      - 
      \mathbf{J}_-\cdot\boldsymbol{\sigma}^*
   \right)
   (\sigma_0 - \sigma_3),
\end{align}
with
\begin{align}
   \tilde{a}_\pm(\tilde{b}_\pm)
   &=
   \sqrt{%
   \frac{1}
   {16|\mathbf{J_\pm}|[|\mathbf{J_\pm}|+(\mathbf{J_\pm})_3]}
   \frac{E+\tilde{\omega}_{\pm,p(m)}}{E}
   }.
   \nonumber
\end{align}
By using $u_\pm$, $v_\pm$ can be expressed as 
\begin{align}
   v_+
   =&\Delta_+^\dag u_+(E\sigma_0+\check{\omega}_{+,pm})^{-1},
   \label{eq:vp}
   \\
   v_-
   =&\Delta_- u_-(E\sigma_0+\check{\omega}_{-,pm})^{-1}.
   \label{eq:vm}
\end{align}
\subsection{\label{sec:App_cond3}Conductance}
Tunneling 
conductance for general pair potential 
is obtained by solving following boundary conditions 
[Eq.~(\ref{eq:bound1}) and Eq.~(\ref{eq:bound2})].
\begin{align}
   e_\sigma
   +
   \begin{pmatrix}
      b_\sigma\\
      a_\sigma
   \end{pmatrix}
   =&
   \begin{pmatrix}
      u_+ & v_-\\
      v_+ & u_-
   \end{pmatrix}
   \begin{pmatrix}
      c\\
      d
   \end{pmatrix},
   \label{eq:bound_g1}
   \\
   ik_z
   \begin{pmatrix}
      u_+ & v_-\\
      v_+ & u_-
   \end{pmatrix}
   \begin{pmatrix}
      \sigma_0 & 0\\
      0 & -\sigma_0
   \end{pmatrix}
   \begin{pmatrix}
      c\\
      d
   \end{pmatrix}
   -
   ik_z
   e_\sigma
   +
   ik_z
   \begin{pmatrix}
      \sigma_0 & 0\\
      0 & -\sigma_0
   \end{pmatrix}
   \begin{pmatrix}
      b_\sigma\\
      a_\sigma
   \end{pmatrix}
   =&
   \frac{2mU_0}{\hbar^2}
   \left[
      e_\sigma
      +
      \begin{pmatrix}
         b_\sigma\\
         a_\sigma
      \end{pmatrix}	
   \right],
   \label{eq:bound_g2}
\end{align}
with
\begin{align}
   e_\uparrow
   =
   (1,0,0,0)^\mathrm{T},\:\:
   e_\downarrow
   =&
   (0,1,0,0)^\mathrm{T},\:\:
   a_\sigma
   =
   (a_{\sigma,\uparrow},a_{\sigma,\downarrow})^\mathrm{T},\:\:
   b_\sigma
   =
   (b_{\sigma,\uparrow},b_{\sigma,\downarrow})^\mathrm{T},\:\:
   c
   =
   (c_{\uparrow},c_{\downarrow})^\mathrm{T},\:\:
   d
   =
   (d_{\uparrow},d_{\downarrow})^\mathrm{T}.
   \nonumber
\end{align}

We define
\begin{align}
   \hat{\Theta}_\pm
   =&
   v_\pm(u_\pm)^{-1},
   \nonumber
   \\
   Z'
   =&
   \frac{2mU_0}{k_z\hbar^2},
   \nonumber
   \\
   G
   =&
   \begin{pmatrix}
      u_+ & v_-\\
      v_+ & u_-
   \end{pmatrix}.
   \nonumber
\end{align}
From Eqs.~(\ref{eq:vp}) and (\ref{eq:vm}), $\hat{\Theta}_+$
and $\hat{\Theta}_-$ satisfy, 
\begin{align}
   \hat{\Theta}_+
   =&
   \Delta_+^\dag u_+(E\sigma_0 + \check{\omega}_{+,pm})^{-1}u_+^{-1},
   \nonumber
   \\
   \hat{\Theta}_-
   =&
   \Delta_- u_-(E\sigma_0 + \check{\omega}_{-,pm})^{-1}u_-^{-1}.
   \nonumber
\end{align}
We can check that following $u_\pm^{-1}$ given by 
Eqs.~(\ref{eq:App_cond_u_inv1}) and (\ref{eq:App_cond_u_inv2}) satisfy $u_\pm u_\pm^{-1}=\sigma_0$.
\begin{align}
   u_+^{-1} 
   =&
   \check{a}_+
   (\sigma_0 + \sigma_3)
   \left(
      |\mathbf{J}_+|\sigma_0 + \mathbf{J}_+\cdot\boldsymbol{\sigma} 
   \right)
   + 
   \check{b}_+
   (\sigma_0 - \sigma_3)
   \left(
      |\mathbf{J}_+|\sigma_0 - \mathbf{J}_+\cdot\boldsymbol{\sigma} 
   \right) 
   \label{eq:App_cond_u_inv1}
   \\
   u_-^{-1} 
   =&
   \check{a}_-
   (\sigma_0 + \sigma_3)
   \left(
      |\mathbf{J}_-|\sigma_0 + \mathbf{J}_-\cdot\boldsymbol{\sigma}^*
   \right)
   + 
   \check{b}_-
   (\sigma_0 - \sigma_3)
   \left(
      |\mathbf{J}_-|\sigma_0 - \mathbf{J}_-\cdot\boldsymbol{\sigma}^*
   \right) 
   \label{eq:App_cond_u_inv2}
   \\
   \check{a}_\pm
   = &
   \sqrt{\frac{1}{4|\mathbf{J}_\pm|[|\mathbf{J}_\pm| + (\mathbf{J}_\pm)_3]}
   \frac{E}{E  + \tilde{\omega}_{\pm,p}}}
   \nonumber
   \\
   \check{b}_\pm
   = &
   \sqrt{\frac{1}{4|\mathbf{J}_\pm|[|\mathbf{J}_\pm| + (\mathbf{J}_\pm)_3]}
   \frac{E}{E  + \tilde{\omega}_{\pm,m}}}
   \nonumber
\end{align}

From Eqs.~(\ref{eq:App_cond_u_inv1}) and (\ref{eq:App_cond_u_inv2}), 
$\hat{\Theta}_+$ and $\hat{\Theta}_-$ are obtained as 
\begin{align}
   \hat{\Theta}_+
   =&
   \frac{\Delta_+^\dag}{2}
   \left[
      \left(
         \frac{1}{E+\tilde{\omega}_{+,p}}
         +
         \frac{1}{E+\tilde{\omega}_{+,m}}
      \right)
      +
      \left(
         \frac{1}{E+\tilde{\omega}_{+,p}}
         -
         \frac{1}{E+\tilde{\omega}_{+,m}}
      \right)
      \frac{\mathbf{J_+}}{|\mathbf{J_+}|}\cdot\boldsymbol{\sigma}
   \right],
   \nonumber
   \\
   \hat{\Theta}_-
   =&
   \frac{\Delta_-}{2}
   \left[
      \left(
         \frac{1}{E+\tilde{\omega}_{-,p}}
         +
         \frac{1}{E+\tilde{\omega}_{-,m}}
      \right)
      +
      \left(
         \frac{1}{E+\tilde{\omega}_{-,p}}
         -
         \frac{1}{E+\tilde{\omega}_{-,m}}
      \right)
      \frac{\mathbf{J_-}}{|\mathbf{J_-}|}
      \cdot\boldsymbol{\sigma}^*
   \right].
   \nonumber
\end{align}
If $\mathbf{F_\pm+q_\pm}=0$ is satisfied, 
from Eqs.~(\ref{eq:var_p}), (\ref{eq:var_m}), (\ref{eq:vp}), and (\ref{eq:vm}), 
we obtain 
$\tilde{\omega}_{\pm,p}=\tilde{\omega}_{\pm,m}$ and
\begin{align}
   \hat{\Theta}_+
   =&
   \frac{\Delta_+^\dag}{E+\tilde{\omega}_{+,p}},
   \nonumber
   \\
   \hat{\Theta}_-
   =&
   \frac{\Delta_-}{E+\tilde{\omega}_{-,p}}.
   \nonumber
\end{align}

We obtain $a_\sigma$ and $b_\sigma$ from Eqs.~(\ref{eq:bound_g1}) and 
(\ref{eq:bound_g2}).
\begin{align}
   \begin{pmatrix}
      b_\sigma\\
      a_\sigma
   \end{pmatrix}
   = &
   \left[
      I_4 + G
      \begin{pmatrix}
         \sigma_0&0\\
         0& -\sigma_0 
      \end{pmatrix}
      G^{-1}
      \begin{pmatrix}
         (1 + iZ')\sigma_0&0\\
         0& (- 1 + iZ')\sigma_0
      \end{pmatrix}
   \right]^{-1}
   \left[
      - I_4 + (1 - iZ')G
      \begin{pmatrix}
         \sigma_0&0\\
         0& -\sigma_0 
      \end{pmatrix}
      G^{-1}
   \right]
   e_\sigma,
   \nonumber
\end{align}
where $I_4$ is the $4\times4$ identity matrix.
From the general relation of $2n$ $\times$ $2n$ matrix, 
\begin{align}
   &
   \begin{pmatrix}
      A&B\\
      C&D
   \end{pmatrix}^{-1}
   \nonumber\\
   =&
   \begin{pmatrix}
      \left(
         A - BD^{-1}C
      \right)^{-1}
      &
      \left(
         C - DB^{-1}A
      \right)^{-1}
      \\
      \left(
         B - AC^{-1}D
      \right)^{-1}
      &
      \left(
         D - CA^{-1}B
      \right)^{-1}
   \end{pmatrix},
   \label{eq:rel_Inv_1}
\end{align}
where $A$, $B$, $C$ and $D$ are 
 $n\times n$ regular matrices, we obtain the following relation 

\begin{align}
   G
   \begin{pmatrix}
      \sigma_0 & 0\\
      0 & -\sigma_0
   \end{pmatrix}
   G^{-1}
   =&
   \begin{pmatrix}
      \left(
         \sigma_0-\hat{\Theta}_-\hat{\Theta}_+
      \right)^{-1}
      \left(
         \sigma_0+\hat{\Theta}_-\hat{\Theta}_+
      \right) 
      & 
      -2
      \left(
         \sigma_0-\hat{\Theta}_-\hat{\Theta}_+
      \right)^{-1}\hat{\Theta}_-
      \\
      2
      \left(
         \sigma_0-\hat{\Theta}_+\hat{\Theta}_-
      \right)^{-1}\hat{\Theta}_+
      &
      -
      \left(
         \sigma_0-\hat{\Theta}_+\hat{\Theta}_-
      \right)^{-1}
      \left(
         \sigma_0+\hat{\Theta}_+\hat{\Theta}_-
      \right) 
   \end{pmatrix}.
   \nonumber
\end{align}
We define $2\times2$ matrices $\alpha_{ij}$ as
\begin{align}
   \begin{pmatrix}
      \alpha_{11} & \alpha_{12}\\
      \alpha_{21} & \alpha_{22}
   \end{pmatrix}
   =
   \left[
      I_4 + G
      \begin{pmatrix}
         \sigma_0&0\\
         0& - \sigma_0
      \end{pmatrix}
      G^{-1}
      \begin{pmatrix}
         (1 + iZ')\sigma_0&0\\
         0& (- 1 + iZ')\sigma_0
      \end{pmatrix}
   \right]^{-1}.
   \nonumber
\end{align}
From Eq.~(\ref{eq:rel_Inv_1}), 
$\alpha_{ij}$ are obtained, 
\begin{align}
   \alpha_{11}
   =&
   (2-iZ')
   \left[
      Z'^2\hat{\Theta}_-\hat{\Theta}_+-(4+Z'^2)\sigma_0
   \right]^{-1}
   \left(
      \hat{\Theta}_-\hat{\Theta}_+ - \sigma_0
   \right)^{-1}
   \left(
      \sigma_0+\frac{-iZ'}{2-iZ'}\hat{\Theta}_-\hat{\Theta}_+
   \right)
   \left(
      \sigma_0 - \hat{\Theta}_-\hat{\Theta}_+
   \right),
   \nonumber\\
   \alpha_{22}
   =&
   (2+iZ')
   \left[
      Z'^2\hat{\Theta}_+\hat{\Theta}_- - (4+Z'^2)\sigma_0
   \right]^{-1}
   \left(
      \hat{\Theta}_+\hat{\Theta}_- - \sigma_0
   \right)^{-1}
   \left(
      \sigma_0 +  \frac{iZ'}{2+iZ'}\hat{\Theta}_+\hat{\Theta}_-
   \right)
   \left(
      \sigma_0 -\hat{\Theta}_+\hat{\Theta}_-
   \right),
   \nonumber\\
   \alpha_{12}
   =&
   2(-1+iZ')
   \left[
      (4+Z'^2)\sigma_0-Z'^2\hat{\Theta}_-\hat{\Theta}_+
   \right]^{-1}
   \left(
      \sigma_0-\hat{\Theta}_-\hat{\Theta}_+
   \right)^{-1}
   \hat{\Theta}_-
   \left(
      \sigma_0 -\hat{\Theta}_+\hat{\Theta}_-
   \right),
   \nonumber\\
   \alpha_{21}
   =&
   -2(1 + iZ')
   \left[
      (4+Z'^2)\sigma_0 - Z'^2\hat{\Theta}_+\hat{\Theta}_-
   \right]^{-1}
   \left(
      \sigma_0 - \hat{\Theta}_+\hat{\Theta}_-
   \right)^{-1}
   \hat{\Theta}_+
   \left(
      \sigma_0 -\hat{\Theta}_-\hat{\Theta}_+
   \right).
   \nonumber
\end{align}
Finally, $b_\sigma$ and $a_\sigma$ are obtained 
\begin{align}
   b_\sigma
   =&
   \frac{-iZ'}{2+iZ'}
   \left(
      \sigma_0-\hat{\Theta}_-\hat{\Theta}_+
   \right)
   \left[
      \sigma_0-(1-\sigma_\mathrm{N})\hat{\Theta}_-\hat{\Theta}_+
   \right]^{-1}
   \tilde{e}_\sigma,
   \nonumber
   \\
   a_\sigma
   =&
   \sigma_\mathrm{N}
   \hat{\Theta}_+
   \left[
      \sigma_0 - (1-\sigma_\mathrm{N})\hat{\Theta}_-\hat{\Theta}_+
   \right]^{-1}
   \tilde{e}_\sigma,
   \nonumber
\end{align}
where $\tilde{e}_\sigma$ is defined by 
$e_\sigma^\mathrm{T}=(\tilde{e}_\sigma^\mathrm{T},0,0)$.
Then a general formula of tunneling conductance is expressed compactly  by 
using $\hat{\Theta}_{+}$ and $\hat{\Theta}_{-}$.
\begin{align}
   \sigma_\mathrm{S}
   =&
   \frac{1}{2}\sum_{\sigma}
   \left[
      1+a_\sigma^\dag a_\sigma-b_\sigma^\dag b_\sigma
   \right]
   \nonumber\\
   =&
   \frac{\sigma_\mathrm{N}}{2}\mathrm{Tr}
   \left\{
      \left[
         \sigma_0-(1-\sigma_\mathrm{N})\hat{\Theta}_+^\dag\hat{\Theta}_-^\dag
      \right]^{-1}
      \left[
         \sigma_0
         +
         \sigma_\mathrm{N}\hat{\Theta}_+^\dag\hat{\Theta}_+
         +
         (\sigma_\mathrm{N}-1)
         \hat{\Theta}_+^\dag\hat{\Theta}_-^\dag\hat{\Theta}_-\hat{\Theta}_+
      \right]
      \left[
         \sigma_0-(1-\sigma_\mathrm{N})\hat{\Theta}_-\hat{\Theta}_+
      \right]^{-1}
   \right\}.
   \label{eq:cond_gen}
\end{align}

\subsection{\label{sec:App_cond4}Spin rotation}
In this subsection, we explain that the conductance is invariant under
the spin rotation.
Under the spin rotation ($\Psi\rightarrow U\Psi$), 
\begin{align}
   U
   =&
   \begin{pmatrix}
      R &0 \\
      0 & R^*\\
   \end{pmatrix},
   \nonumber
   \\
   R
   =&
   e^{-i\boldsymbol{\theta}\cdot\boldsymbol{\sigma}/2},
   \nonumber
\end{align}
pair potential $\Delta_\pm$ 
and 
$\mathbf{J}_\pm$ 
are transformed as
\begin{align}
   \Delta_\pm\rightarrow& R\Delta_\pm R^\mathrm{T}
   \nonumber
   \\
   \mathbf{J}_+\cdot\boldsymbol{\sigma}
   \rightarrow& 
   R\mathbf{J}_+\cdot\boldsymbol{\sigma} R^\dag,
   \nonumber
   \\
   \mathbf{J}_-\cdot\boldsymbol{\sigma}^*
   \rightarrow& 
   R^*\mathbf{J}_-\cdot\boldsymbol{\sigma}^* R^\mathrm{T}.
   \nonumber
\end{align}

$\hat{\Theta}_\pm$ is transformed as
\begin{align}
   \hat{\Theta}_+
   \rightarrow&
   \frac{R^*\Delta_+^\dag R^\dag}{2}
   \left[
      \left(
         \frac{1}{E+\omega_{+,p}}
         +
         \frac{1}{E+\omega_{+,m}}
      \right)
      +
      \left(
         \frac{1}{E+\omega_{+,p}}
         -
         \frac{1}{E+\omega_{+,m}}
      \right)
      \frac{R\mathbf{J}_+\cdot\boldsymbol{\sigma}R^\dag}{|\mathbf{J}_+|}
   \right]
   \nonumber\\
   &=
   R^*\hat{\Theta}_+ R^\dag,
   \nonumber
   \\
   \hat{\Theta}_-
   \rightarrow&
   \frac{R\Delta_- R^\mathrm{T}}{2}
   \left[
      \left(
         \frac{1}{E+\omega_{-,p}}
         +
         \frac{1}{E+\omega_{-,m}}
      \right)
      +
      \left(
         \frac{1}{E+\omega_{-,p}}
         -
         \frac{1}{E+\omega_{-,m}}
      \right)
      \frac{R^*\mathbf{J}_-\cdot\boldsymbol{\sigma}^*R^\mathrm{T}}{|\mathbf{J}_-|}
   \right]
   \nonumber\\
   &=
   R\hat{\Theta}_- R^\mathrm{T}.
   \nonumber
\end{align}

Then the conductance is invariant under the spin rotation.
\begin{align}
   \sigma_\mathrm{S}
   \rightarrow&
   \mathrm{Tr}
   \left[
      \sigma_0-(1-\sigma_\mathrm{N})R\hat{\Theta}_+^\dag R^\mathrm{T} 
      R^* \hat{\Theta}_-^\dag R^\dag
   \right]^{-1}
   \left[1+\sigma_\mathrm{N}R\hat{\Theta}_+^\dag R^\mathrm{T}R^*\hat{\Theta}_+R^\dag
      +(\sigma_\mathrm{N}-1)R\hat{\Theta}_+^\dag R^\mathrm{T}R^*\hat{\Theta}_-^\dag R^\dag
      R\hat{\Theta}_- R^\mathrm{T}R^*\hat{\Theta}_+R^\dag
   \right]\nonumber\\
   &
   \left[
      \sigma_0-(1-\sigma_\mathrm{N})R\hat{\Theta}_- R^\mathrm{T}R^*\hat{\Theta}_+R^\dag
   \right]^{-1}\nonumber\\
   =&
   \mathrm{Tr}
   R
   \left[
      \sigma_0-(1-\sigma_\mathrm{N})\hat{\Theta}_+^\dag\hat{\Theta}_-^\dag
   \right]^{-1}
   R^\dag R
   \left[1+\sigma_\mathrm{N}\hat{\Theta}_+^\dag\hat{\Theta}_+
      +(\sigma_\mathrm{N}-1)
      \hat{\Theta}_+^\dag\hat{\Theta}_-^\dag\hat{\Theta}_-\hat{\Theta}_+
   \right]
   R^\dag R
   \left[
      \sigma_0-(1-\sigma_\mathrm{N})\hat{\Theta}_-\hat{\Theta}_+
   \right]^{-1}
   R^\dag\nonumber\\
   =&\sigma_\mathrm{S}.
   \nonumber
\end{align}

\end{widetext}
\end{document}